\documentclass[a4paper,13pt,twoside,openright]{book}

\usepackage[latin1]{inputenc}
\usepackage{feynarts}
\usepackage[english]{babel}
\usepackage{amsmath}
\usepackage{amssymb}
\usepackage[dvips]{graphicx}
\usepackage{epsfig}
\usepackage{mcite}
\DeclareGraphicsExtensions{.eps}
\author{Alessandra Repetto}
\def\la{\lambda}
\def\s{\sigma}

\def\a{\alpha}

\def\c{\gamma}
\def\b{\beta}
\def\d{\delta}
\def\eps{\epsilon}

\def\C{\Gamma}

\def\T{\Theta}

\def\n{\eta}

\def\.{\cdot}

\def\+{\bigoplus}
\def\D{\Delta}
\def\({\left(}
\def\){\right)}
\def\[{\left[}
\def\]{\right]}
\def\l.{\left.}
\def\r.{\right.}

\def\<{\left\langle}
\def\r|{\right|}
\def\>{\right\rangle}
\def\l|{\left|}

\def\beq{\begin{equation}}
\def\eeq{\end{equation}}
\def\bea{\begin{eqnarray}}
\def\eea{\end{eqnarray}}

\def\ber{\begin{array}}
\def\eer{\end{array}}

\newcommand{\pasl}{ \mkern-6mu \not \mkern-3mu \partial}
\newcommand{\psl}{ \mkern-6mu \not \mkern-3mu p}
\newcommand{\qsl}{ \mkern-6mu \not \mkern-3mu q}
\newcommand{\ksl}{ \mkern-6mu \not \mkern-3mu k}
\newcommand{\Dsl}{ \mkern-6mu \not \mkern-3mu D}
\begin{document}
\clearpage{\pagestyle{empty}\cleardoublepage \thispagestyle{empty}
\begin{center}
\includegraphics[width=0.1\textwidth]{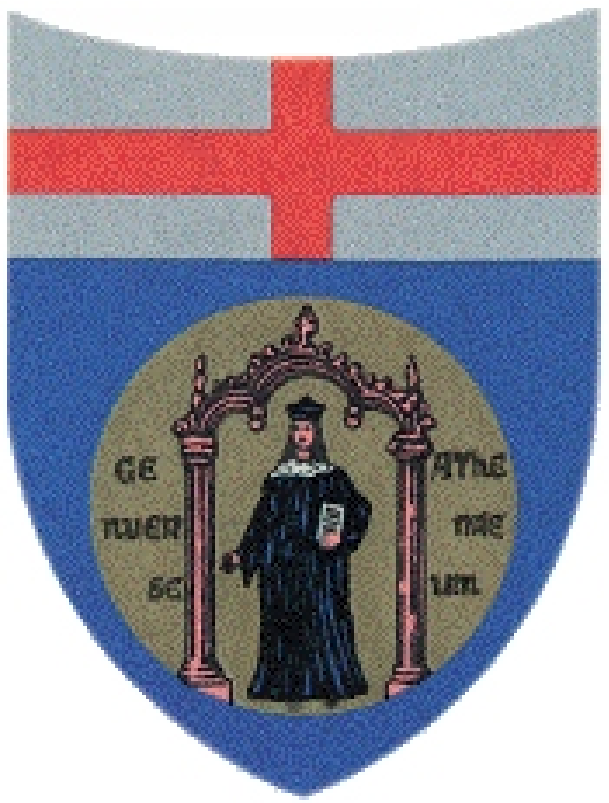}\\
\Large \textsc{Universit\`a degli studi di Genova}\\
\textit{Facolt\`a di Scienze Matematiche, Fisiche e Naturali}
\end{center}
\vspace{1.2cm}
\begin{center}
\huge \textsc{Tesi di Dottorato in Fisica}
\end{center}
\vspace{2cm}
\begin{center}
\Huge \textbf{A strategy for the off-shell analysis of collinear singularities in Feynman diagrams}
\end{center}
\vspace{1.5cm}

\newlength{\spacebetween}
\setlength{\spacebetween}{\textwidth}
\addtolength{\spacebetween}{-7.3cm}
\addtolength{\spacebetween}{-3.8cm}
\addtolength{\spacebetween}{-2cm}

\begin{center}

{\Large Alessandra Repetto}\\
\vspace{50pt}
{\Large Supervisor: Prof. Carlo Maria Becchi}





\end{center}

\newpage \thispagestyle{empty} 
\pagestyle{plain}
\newpage

\chapter*{Introduction}
The success of Feynman's parton model \cite{Feynman:1969,*Bjorken:1969ja} has pushed the search of a field theory formulation of the parton hypothesis. The discovery of asymptotic freedom of unbroken non-abelian gauge theories, which implies that the corresponding particle interaction weakens at short distances, has clearly indicated the field content of the Feynman model. Partons are fermions, quarks, and vector bosons interacting through exchanges of vector bosons, gluons.

Asymptotic freedom \cite{Gross:1973ju,*Gross:1974cs,*unp:1972,*Politzer:1973fx} makes more precise Feynman's idea that, whenever one has high energy and momentum transfer, partons behave as free particles. This is indeed what happens in Deep Inelastic Scattering (DIS) and in quark anti-quark annihilations into heavy bosons.

However, partons, therefore quarks and gluons, should be bound in a hadronic initial state. Therefore the partonic process should be characterized by two scales, the high energy momentum transfer scale and the binding energy scale, which should be less than few hundred MeV.

Therefore if one tries to exploit perturbative field theory in order to compute corrections to the naive free parton model, one finds typically logarithmic factors in the ratio of the two scales, which is about one thousand. These logarithmic factors appear as multipliers of the strong structure constant $\a_s$. Therefore the validity of perturbation theory becomes less obvious in much the same way as the idea of parton independence, that is, factorization. Feynman amplitudes in these calculations must be computed in the neighbourhood of singularities which are typically of the collinear type. The only simplification induced by the presence of the two above mentioned scales is that at least for the quarks up and down, whose mass is few MeV, one can consider quarks as massless as gluons are. Therefore the field model should be considered in first (free) approximation scale invariant and the radiative corrections should violate Bjorken scale invariance due to renormalization. This is indeed what happens and the corrections to Bjorken scaling \cite{Bjorken:1968dy} give a strong evidence in favour of the non-abelian QCD model. 

However the calculations of radiative corrections together with the presence of the above mentioned scales, one of which is associated with the process, the other appearing in the hadronic wave function, have posed non-trivial technical difficulties which have been overcome thanks to the discovery of dimensional regularization \cite{'tHooft:1972fi,*Bollini:1995pp,Breitenlohner:1977hr} and of the corresponding minimal renormalization schemes.

The simplicity of computation is only one of the advantages which have strongly favoured the use of the dimensional method. A second very important advantage is that it doesn't break gauge invariance and applies without problems to the quark and gluon mass-shell amplitudes. We shall see that this is very important in view of parton factorization hypothesis. The method is consistent with the description of the hadron as a gas of free quarks and gluons whose distribution function has to be computed on the basis of consistency conditions of different scale choices.

Even if the free gas scheme is not physically consistent, otherwise the hadronic density of state should be exceedingly large, the presence of a consistent computational method for high energy processes has given solid technical foundations to the parton model.

In this framework a huge amount of work has been done aiming at the analysis of the singularities of Feynman diagrams as functions of the kinematic invariants, that is, of the invariant squared partial sum of the momenta carried by the external vertices. In particular, in the study of DIS, the behaviour of a mass-shell QCD amplitude when partons are considered massless is related to collinear singularities of the amplitude. The study of these singularities gives information through the Altarelli-Parisi approach \cite{Altarelli:1977zs,*Altarelli:1978id,*Curci:1980uw} on the parton distribution function of e.g. a target baryon.

Collinear singularities appear in dimensional regularization of a mass-shell massless parton amplitude when the dimensional regularization parameter $\eps=\frac{d-4}{2}$ goes to zero; that is collinear singularities appear as poles in the origin of the $\eps$ complex plane in much the same way as UV singularities do. This implies that one has to avoid IR-UV singularity mixings by first subtracting UV divergences. 

An important result concerning the Feynman graph singularities is that in the Schwinger parametric representation \cite{Speer:1970ss} the divergent parts of a diagram are confined in particular sectors of the parametric space. 

In fact in a mass-shell amplitude, using dimensional regularization, one finds that $\eps\rightarrow 0$ singularities associated with collinear divergences appear only in the restriction of the parametric representation to particular sectors which can be identified using the results published by Speer in 1975 \cite{Speer:1975dc}. This induces a remarkable simplification in the singularity analysis but until recently it has not been exploited in calculations.(See however \cite{Carter:2010hi} and references therein). 

On account of these results and of more recent analogous formulations \cite{Smirnov:2008aw}, it is reasonable to verify the simplifying power of Speer's results in the calculation of collinear singularities of parton processes. However, beyond purely technical aspects, an easy tool for the computation of the infrared singularities of the Feynman amplitudes allows a better physical insight into Parton Physics. From the point of view of the physical interpretation, one should take into account that partons are identified with bound, and hence off-shell quarks and gluons. However single parton amplitudes with off-shell initial particle states present a further technical difficulty. Indeed it is fairly well known since the seminal Yennie-Frautschi-Suura work \cite{1961AnPhy..13..379Y} that, already in QED, off-shell electron amplitudes are not "gauge invariant" and that this has consequences on the compensation of infrared (soft) divergences of the single graphs. Here gauge invariance should be given two different meanings, the most common one being the vanishing of space-time divergence of any conserved current vertex; there is however a second meaning concerning the independence of the gauge fixing of quantization choice. As a matter of fact strictly speaking the single parton contributions are not physically meaningful and hence parton factorization, which is the basis of the parton model, should be taken "cum grano salis".

As a matter of fact hadrons are colour neutral and hence the cross section of a fully inclusive deep inelastic process is free of infrared singularities. Indeed this is a consequence of the famous Kinoshita-Lee-Nauenberg theorem \cite{Kinoshita:1962ur,*Lee:1964is}, which can be easily understood noticing that the cross section is the absorptive part of a virtual forward Compton scattering whose amplitude should not be affected by infrared divergences.

There are however two obstructions in pushing forward this point of view. The first one is that this would bring the analysis far away from the independent parton model, the second one is the need of knowledge of the hadron-parton vertex which cannot be deduced from perturbative theory.

One could try to overcome these difficulties assuming that the partons be "almost" on-shell and the collinear singularities be induced by the "almost" on-shellness. However the analysis should also consider the singularities due to the exact on-shell nature of the final state partons.

In QED, when the initial state is neutral, final state singularities come from the coupling of real and virtual soft photons to the final charged particles and cancel in the inclusive cross sections. This happens also in first order corrections to QCD amplitudes and suggests that the final parton on-shell singularities should be forgotten keeping only the contributions which are singular when the initial parton goes on-shell.

Notice that considering off-shell amplitudes is a kind of regularization since the infrared singularities are regularized by a suitable off-shell (Euclidean) choice of external momenta and are therefore well separated from the (dimensionally regularized) UV singularities. Indeed studying a massless theory for space-like external momenta \cite{Speer:1975dc,Zimmermann:1975gm}, one shows that if the external momenta are non-exceptional, i.e. no partial sum vanishes, the diagram has no infrared divergences if the field have positive mass dimension and all the internal vertices have dimension four, as in QCD.



Hence the idea is to regularize the parton amplitudes choosing off-shell initial momenta \cite{Gabrieli:1997rx,*Dorn:2008dz,*Carlitz:1988ab} and to select the contributions which are singular when the parton goes on-shell. Notice that contrary to the off-shell amplitude, its mass-shell singular part should be gauge invariant and independent.

We hope that applying the present idea to a physical situation, such as Deep Inelastic Scattering, will help us to make this point clear.

In this work we consider the difficult off-shell analysis in the light of few, more or less, recent progresses in Feynman graph computations. These are essentially based on the extended use of Mellin-Barnes transform \cite{librosm,*Freitas:2010nx,*Allendes:2009bd,*Smirnov:2009up,*Smirnov:2008tz,*Aguilar:2008qj,*Bierenbaum:2007zza,*Gluza:2007bd,*Gluza:2007rt,*Czakon:2005rk,*Borisov:2005ka,*Friot:2009fw,*Friot:2005gh,*Friot:2005cu,*Suzuki:2003jn,*Smirnov:2004ip,*Kharchev:2000ug,*Kowalenko:1998qm,*Passare:1996db,*Elizalde:1994dm,*Bytsenko:1993cf,*Anastasiou:2005cb,*Bolzoni:2009ye,*Bolzoni:2010sp,*Drummond:2006rz}, aiming at the singularity analysis, and the sector decomposition of parametric Feynman integrals, mentioned above.

It is clear that a generic not renormalized amplitude could present UV singularities which could combine with mass singularities. The presence of UV and mass singular contributions could appear in the case of multiloop diagrams with UV divergent subdiagrams. However, once these ultraviolet divergences are subtracted using the Breitenlohner-Maison scheme \cite{Breitenlohner:1975hg}, they do not interfere with the mass singularities \cite{mia:tesi}.

The best way of describing a computational algorithm consists in applying it to a simple, however not trivial case. What we are going to show in this thesis is how the combined sectorialization-Mellin Barnes transform technique works in the study of collinear singularities in the first order  in the strong constant $\alpha_s$ structure functions of DIS.

\chapter{Massless QCD}

\textit{Currently Particle Physics is successfully described by the Standard Model \cite{Weinberg:1996kr,*Weinberg:1995mt,Becchi:2006sk}, i.e. the theory of strong and electroweak interactions, based on gauge group $SU(3)\otimes SU(2)\otimes U(1)_Y$.} \textit{The corresponding gauge fields are gluons, which are the mediators of the strong interaction, and the electroweak bosons, the photon,} $W^\pm$ \textit{and} $Z_0$. 

\textit{The Standard Model contains Quantum Chromodynamics or QCD, the non-abelian gauge theory which describes strong interactions based on the colour group $SU(3)$.} 

\textit{$SU(2)\otimes U(1)_Y$ is the Glashow-Weinberg-Salam  electroweak symmetry group. The $U(1)_Y$ factor is the abelian phase group generated by the weak hyperchage operator $Y$, which is related to the electric charge $Q$ and to the weak isospin third component $T_3$, i.e. one of the $SU(2)$ generators, by means of $Q=T_3+Y$.}

\textit{$W^\pm$ and $Z_0$ bosons, differently from the photon and the gluon, are massive; Lagrangian mass terms are induced by the partial spontaneous breakdown of the symmetry group $SU(2)\times U(1)_Y$ according to the Higgs mechanism.}

\textit{The unbroken gauge symmetry group is $SU(3)\otimes U(1)_Q$. It corresponds to the exact symmetry of Nature; $U(1)_Q$ is the phase group generated by the electric charge. Its gauge boson is the photon. $SU(3)$ gauge bosons are gluons: the eight gluons mediate the strong interactions between quarks, antiquarks and gluons, commonly called "the partons". As a matter of fact quarks are massive, however, in particular up and down quarks have masses of few MeV. If one renormalizes the theory at the light quark mass scale $\mu$ in the physical amplitudes the coupling $\a_s$ appears multiplied by $\log\(\frac{E}{\mu}\)$, where $E$ corresponds to the energy scale of the process. If e.g. $E$ is about $100$ GeV, $\log(\frac{E}{\mu})\approx 12$ and hence the perturbative expansion parameter is too large. If, on the contrary, one chooses renormalization scale corresponding to $E$, the universally subtracted off-shell amplitudes become quark mass independent and hence, in general, they can be identified with the massless quark amplitudes}

\textit{In presence of massless particles, Feynman field theory, i.e. diagrammatic perturbative theory, produces infrared divergences, which necessarily have a physical nature: they must be carefully studied in the light of the physical process.}

\textit{In conclusion: fundamental physics must deal with infrared divergence problem because of the existence of massless particles. Infrared divergences arise particularly in QCD. In this chapter we give a brief summary of perturbative QCD, focusing on applications in which infrared divergences appear; in this way we want to give the physical background of our work.}

\section{Introduction to QCD}
 
$QCD$ or Quantum Chromodynamics \cite{Itzykson:1980rh,Ellis:1991qj,Peskin:1995ev} is a gauge theory associated with the colour group $SU(3)$, i.e. the group of $3\times 3$ complex unitary matrices for which the determinant equals unity; they satisfy
\beq U^+U=1 \text{, }\quad det(U)=1. \eeq
QCD is the mathematical model which describes strong interactions, that are generated by exchanges of gluons and quark-antiquark pairs in the hadrons.  The $SU(3)$ group acts on the fields of the theory, quarks, which correspond to colour triplets and gluons, corresponding to octets.

The corresponding Lagrangian density is: 

\beq
{\cal L}=-\frac{1}{4}\sum^{8}_{A=1}F^{\mu\nu}_A F^A_{\mu\nu}+\sum^{n_f}_{f=1}\overline{\psi}^f_i(i \Dsl_{ij}-m_f)\psi^f_j,
\label{Lqcd}
\eeq
where:
\begin{itemize}
	\item $\psi^f_j$ is the spinor field associated  with the $j$ colour component of a quark with flavour $f$ and mass $m_{f}$.
	\item $\Dsl=D_{\mu} \c^{\mu}$, where $\c^{\mu}$ are Dirac matrices and $D_{\mu}$ is the covariant derivative:
	\beq D_{\mu,ij}=\partial_{\mu}\d_{ij}-ie_s A_{\mu,ij}. \eeq
In this last expression the gauge constant $e_s$ has been introduced. In general one refers to the strong coupling constant $\a_s=\frac{e_s^2}{4\pi}$, the QCD constant in analogy with the fine structure constant in QED. 
	
One has introduced also $\textbf{A}_{\mu,ij}=\sum_A t_{A\,ij} \cdot A_{\mu}^A$, where $A=[1,8]$: $A_{\mu}^A$ are the eight gluonic massless fields and $t^A$ are the eight $SU(3)$ generators of the quark tridimensional representation; therefore they are a basis of the linear space of the $3\times 3$ traceless matrices acting on quark $\psi_j$ and they satisfy $Tr\ ({t_At_B})=\delta_{AB}/2$. Structure constants $f_{ABC}$ are defined by the commutation relations $[t_A,t_B]=if_{ABC}t_C$. 

\item $F^A_{\mu \nu}$ is an antisymmetric tensor that transforms under the gauge adjoint representation
\beq \textbf{F'}_{\mu \nu}=U \textbf{F}_{\mu \nu} U^{-1}\quad \text{with}\quad \textbf{F}_{\mu \nu}=\sum_A F^A_{\mu \nu}t_A \eeq
and in components:
\beq F^A_{\mu \nu}=\partial_{\mu}A_{\nu}^A-\partial_{\nu}A_{\mu}^A-e_s f_{ABC} A_{\mu}^B A_{\mu}^C.\eeq

\end{itemize}
It is believed that the very simple Lagrangian (\ref{Lqcd}) describes the whole strong interaction physics: from the hadronic spectrum to confinement properties, to high energy processes.
In ordinary situations quarks can't be free, because of confinement, i.e. the existence of a quark-antiquark potential that rises linearly with distance; free particles correspond only to states that are invariant by colour transformations.
On the contrary high energy phenomena are characterized by asymptotic freedom: at short distances (and at high energies) the renormalized coupling constant becomes small, in contrast to what one observes in QED: therefore quarks inside hadrons appear nearly free.

Asymptotic freedom is a property of utmost importance because it allows perturbative calculation in QCD processes characterized by short distances or by large energy or momentum transfer, such as "hard" processes. The perturbative method is very predictive as far these processes are concerned, particularly in electron-positron high energy annihilation into quark-antiquark pairs and in deep inelastic scattering. In these processes the comparison between theory and experiments provides a remarkable confirmation of QCD predictions.

There is another method which allows the study of QCD in a non-perturbative way: this is lattice QCD, i.e. QCD on a finite and discrete space-time. Lattice calculations allow to compute physical observables by means of numerical simulations and to study QCD also at large coupling constant. In our work we consider only the perturbative approach, because it is the most used analytical method in experimental data analysis in the physical conditions when IR-singularities become important.

Starting from Lagrangian density given in equation (\ref{Lqcd}), which describes classical QCD, one builds up Feynman rules corresponding to diagram vertices and propagators. QCD Feynman diagrams have different kinds of vertices since QCD is a non-abelian theory: in addition to the gluon-quark-antiquark vertex (similar to the QED photon-electron-positron vertex) one has also three and four gluons vertices and ghost field vertices; the ghost fields, which are unphysical scalar fields, are introduced into the theory in the Faddeev-Popov quantization procedure.

In figure (\ref{feyqcd}) QCD Feynman rules are shown.

\begin{figure}[htbp]
\begin{picture}(0,0)%
\includegraphics[scale=0.7]{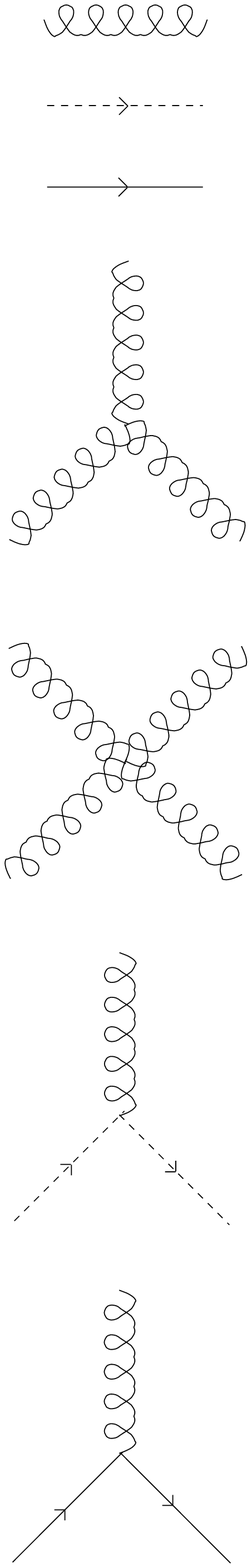}%
\end{picture}%
\setlength{\unitlength}{0.0200bp}%
\begin{picture}(38000,25000)(0,0)%
\fontsize{12}{\baselineskip}\selectfont
\put(7000,28000){\makebox(10000,200)[l]{$=\quad\delta^{a b} \frac{-i\,g^{\a\b}}{p^{2}+i\eps}$}}%
\put(7000,26500){\makebox(10000,200)[l]{$=\quad\delta^{a b} \frac{i}{p^{2}+i\eps}$}}%
\put(7000,25000){\makebox(10000,200)[l]{\strut{}$=\quad\delta^{i k} \left(\frac{i}{\psl-m+i\eps}\right)_{mn}\text{ e se m=0 si ha}\quad\delta^{i k} \left(\frac{i}{\psl+i\eps}\right)_{mn}$}}%
\put(7000,21000){\makebox(10000,200)[l]{\strut{}$=\quad e_sf^{abc}\left[g^{\a \b}(p-q)^{\c}+g^{\b \c}(q-r)^{\a}+g^{\c \a}(r-p)^{\b}\right]$}}%

\put(7000,15000){\makebox(10000,200)[l]{$=\quad-ie_s^2f^{xac}f^{xbd}(g^{\a\b}g^{\c\d}-g^{\a\d}g^{\b\c})+$}}
\put(7000,14000){\makebox(10000,200)[l]{$\phantom{=\quad}-ie_s^2f^{xad}f^{xbc}(g^{\a\b}g^{\c\d}-g^{\a\c}g^{\b\d})+$}}
\put(7000,13000){\makebox(10000,200)[l]{$\phantom{=\quad}-ie_s^2f^{xab}f^{xcd}(g^{\a\c}g^{\b\d}-g^{\a\d}g^{\b\c})$}}

\put(7000,8500){\makebox(10000,200)[l]{\strut{}$=\quad -e_sf^{abc}q^\a$}}%
\put(7000,2000){\makebox(10000,200)[l]{\strut{}$=\quad ie_s t^a_{ki}\c^{\a}_{mn}$}}%
\fontsize{10}{\baselineskip}\selectfont
\put(900,28500){\makebox(0,0)[r]{$a,\a$}}%
\put(4200,28500){\makebox(0,0)[r]{$b,\b$}}%
\put(2350,28500){\makebox(0,0)[r]{$p$}}%
\put(800,26900){\makebox(0,0)[r]{$a$}}%
\put(4000,26900){\makebox(0,0)[r]{$b$}}%
\put(2350,26900){\makebox(0,0)[r]{$p$}}%
\put(900,25400){\makebox(0,0)[r]{$i,n$}}%
\put(4200,25400){\makebox(0,0)[r]{$k,m$}}%
\put(2350,25400){\makebox(0,0)[r]{$p$}}%

\put(2550,24000){\makebox(0,0)[r]{$b,\beta$}}%
\put(600,18100){\makebox(0,0)[r]{$a,\alpha$}}%
\put(4600,18100){\makebox(0,0)[r]{$c,\gamma$}}%
\put(2900,22100){\makebox(0,0)[r]{$q$}}%
\put(800,19900){\makebox(0,0)[r]{$p$}}%
\put(3900,19900){\makebox(0,0)[r]{$r$}}%

\put(600,17000){\makebox(0,0)[r]{$a,\alpha$}}%
\put(4600,17000){\makebox(0,0)[r]{$b,\beta$}}%
\put(600,12000){\makebox(0,0)[r]{$c,\gamma$}}%
\put(4600,12000){\makebox(0,0)[r]{$d,\delta$}}%

\put(2550,11400){\makebox(0,0)[r]{$a,\alpha$}}%
\put(200,5800){\makebox(0,0)[r]{$b$}}%
\put(4200,5800){\makebox(0,0)[r]{$c$}}%
\put(3500,7300){\makebox(0,0)[r]{$q$}}%

\put(2550,5350){\makebox(0,0)[r]{$a,\alpha$}}%
\put(500,-300){\makebox(0,0)[r]{$i,n$}}%
\put(4600,-300){\makebox(0,0)[r]{$k,m$}}%

\end{picture}
\caption{\small{QCD Feynman rules. From the top to the bottom one has: gluon propagator (expressed in Feynman gauge), ghost propagator and quark propagator. Then one has three gluon vertex, four gluon vertex, non-physical gluon-ghost-antighost vertex and gluon-quark-antiquark vertex.}}
\label{feyqcd}
\end{figure}

Using Feynman rules, one builds up the perturbative expansion of transition amplitudes. One knows that Feynman diagrams with closed meshes, loops, correspond to integrals on the momentum space; such integrals could produce ultraviolet divergences in the high momentum region. In presence of massless particles and in the small momentum region, also infrared divergences appear. A regularization process is introduced to remove ultraviolet divergences preserving gauge symmetry and Lorentz invariance; this process yields UV-finite amplitudes.

In general one computes cross sections which are related to the massless square of amplitudes which however are in general linear combinations of difficult terms. More frequently one is interested in inclusive cross sections which can be obtained from the imaginary (absorptive) part of forward amplitudes. These are typically, when the final state is a many particle state, loop amplitudes, that is, amplitudes corresponding to diagrams with loops. 

Inclusive cross sections are in general less singular in their dependence of kinematic parameters than exclusive ones, those associated with a restricted selection of possible final states, since many potentially singular contributions compensate each other.

There are many ways of proceeding. The main one is the \textit{dimensional regularization} \cite{'tHooft:1972fi,*Bollini:1995pp,Breitenlohner:1977hr}.

Integrals are made finite varying space-time dimension from $d=4$ to $d=4-2\eps$, where $\eps$ is infinitesimal.
Feynman amplitudes become meromorphic functions of the variable $d$ and therefore of $\eps$. So ultraviolet divergences appear in the amplitude as poles in $\eps=0$. Finite parts are obtained by subtracting Maclaurin expansion terms in $\epsilon$.

Formally this subtraction is obtained by means of field and parameter redefinition (for example of $e_s$) which depend on $\eps$ and have poles that cancel diagram divergences. This process is called \textit{renormalization}.

The quantum theory is represented by a Lagrangian density, called renormalized, which has the same structure as the classical one.
The redefinition of the starting Lagrangian parameters, called bare,  in terms of physical fields and parameters and of $\eps$, allows to obtain amplitudes at every order of perturbation theory without ultraviolet divergences and to express amplitudes by means of physical parameters measured in experiments.

\section{Massless QCD and infrared divergences}

Equation (\ref{Lqcd}), as we said, is the QCD classical Lagrangian density. Quark masses are the only parameters with physical dimensions, in particular with energy dimension using the natural system $\hbar=c=1$.

Therefore one expects that massless QCD is scale invariant. Scale invariance is the property for which observables are not dependent on the energy scale, but only depend on ratios of dimensional quantities.

In reality QCD is scale invariant only classically and not after quantization because scale symmetry is broken by regularization and renormalization processes. Going from $4$ dimensions to $4-2\epsilon$ dimensions, it is necessary to introduce a new dimensional parameter in the quantum version of the theory which introduces a new mass scale into the theory. In other words scale invariance is broken by regularization.
Anyway deviations from scaling are asymptotically small, logarithmic and computable.

In massive QCD there are other scaling corrections due to the presence of quark masses. However mass corrections decrease as powers of the ratios between the masses and the energy scale in  high energy processes and in particular in hard processes.

Hard processes are characterized by energy and momentum exchanges much larger than $\Lambda_{QCD}$, the characteristic scale introduced by the breakdown of the scale invariance:
\beq
E_i\simeq E, \quad E>>\Lambda_{QCD}>>m_i \ .
\eeq 

As we said,  hard process amplitudes, when masses are null or negligible, are affected by infrared divergences (IR). These must be distinguished between soft infrared divergences and collinear ones, also called mass singularities.

The first ones arise when the four-momentum of a real or virtual particle become soft, i.e. its components tend to zero. Finally one has mass or collinear singularities  when one has transitions from a state of a massless particle to a state of two massless particles with four-momentum $p^{\mu}$ and $p'^{\mu'}$ moving  parallel to each other. The invariant mass of the system becomes zero, although none of the momenta are soft:
\beq 
k^2=(p+p')^2=2EE'(1-\cos\theta)\rightarrow 0
\eeq

In literature one defines collinear the divergences due to a term $\frac{1}{(1-\cos\theta)}$ inside phase space integral (in cross section calculation); if instead the divergence appears inside a loop integral, it is called mass singularity. In this work we use two terms refer for the same thing.

\begin{figure} 
\centering
\includegraphics[scale=0.5]{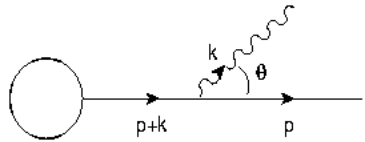}
\caption{\small{Diagram containing both a soft divergence and a mass singularity.}}
\label{sing}
\end{figure}

For example we consider a virtual quark (figure $\ref{sing}$) that transforms into a real quark emitting a gluon at an angle $\theta$  with respect to its momentum.
The virtual quark propagator, which carries four-momentum $p+k$ and which corresponds to the amplitude of the diagram in figure, is:

\beq
	\frac{\psl+\ksl}{(p+k)^2-m^2}=\frac{\psl+\ksl}{2(p\cdot k)}=\frac{1}{2E_k E_p}\cdot \frac{\psl+\ksl}{1-\b_p \cdot \cos\theta},
\eeq
where $E_k$ corresponds to the photon energy, $E_p$ and $m$ are respectively quark energy and mass and one has $\b_p=\sqrt{1-\frac{m^2}{E_p^2}}$.
Since the gluon is massless, $E_k$ could vanish giving a soft infrared divergence.
But when quark mass is zero, one has a second type of divergence, the collinear singularity. Indeed for  $m\rightarrow 0$ one has $\b_p \rightarrow 1$, therefore $(1-\b_p \cdot \cos\theta)$ is null in the forward direction. Due to the presence of an helicity constraint which forbids the transition in the forward directions since angular momentum conservation requires a fermion helicity flip which is forbidden by the collinear helicity conservation, the transition probability turns out to be proportional to $\frac{1}{1-\cos\theta}$ and hence its phase space integral of the cross section,  diverges as \beq \int \frac{d\cos\theta}{1-\cos\theta}.\eeq This is a collinear divergence.

A renormalization theory, which is able to remove infrared divergences, as in the case of ultraviolet singularities, doesn't exist. For example if one  assigned a mass to the gluon, such that the amplitude would converge, one could not remove the mass dependence from the amplitude redefining the Lagrangian parameter.

Nevertheless there are two theorems \cite{Kinoshita:1962ur,*Lee:1964is,jauch:r} which in certain conditions guarantee the cancellation of the infrared divergences in physical processes.

This suggests the idea that infrared divergences are due to an improper definition of initial and final states of the process.

The first theorem is \textit{Bloch-Nordsieck}'s theorem, which is valid in $QED$ and asserts that summing up over the final states in a little energy interval,  soft infrared divergences disappear because there is a cancellation between the virtual diagram divergences and the real ones.
One understands this deletion, observing that a particular scattering process, which consists in a transition from an initial state $| a>$  to a final state $| b>$, is  experimentally indistinguishable from the one with an arbitrary number of photons in the final state, provided that the photon energy is small enough.

In QCD, this theorem is violated, but this happens because of subdominant terms, as it has been shown in \cite{Doria:1980ak} and in \cite{Di'Lieto:1980dt}.

The second theorem is the \textit{Kinoshita-Lee-Nauenberg (KLN)}'s one, which is valid in perturbation theory:  collinear divergences, originating from massless particles in initial and/or final states, cancel up if one sums the transition probabilities between all of the states that are approximately mass degenerate with respect to the considered ones.

Whereas the sum on degenerate final states is motivated by the finite resolution of measure instruments and is an example of the already mentioned cancellation of singularities in inclusive cross sections, it is more difficult to explain the sum on initial states: indeed one must first define the distribution of degenerate states upon which the sum must be carried. In $QCD$ the condition which allows the cancellation of mass singularities coming from initial states generates Altarelli-Parisi equations. Mass singularities due to initial states are absorbed into the initial partonic densities, that correspond to the probability densities of finding the given parton into the initial hadron. Partons are the fundamental constituents, i.e. quarks and gluons.

Considering hard processes in a more detailed way, one describes as an example the process of hadrons production coming out from electron-position annihilation. The reference is figure (\ref{adroni}).\begin{figure} 
\centering
\includegraphics[scale=0.8]{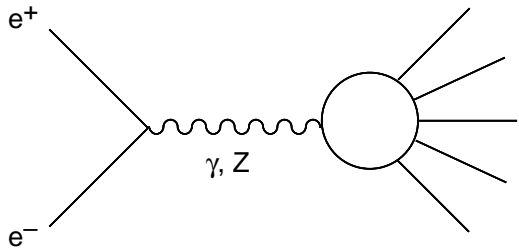}
\caption{\small{ Electron-positron annihilation with hadrons production by means of an exchanged virtual gauge boson.}}
\label{adroni}
\end{figure}
\beq R=\frac{\s(e^+e^-\rightarrow hadrons)}{\s_{point}(e^+e^-\rightarrow \mu^+\mu^-)}. \label{R}\eeq 
Historically the measure of this observable has confirmed the number of QCD colours, which is equal to $3$, as we know.
In (\ref{R}) the point cross section is given by $\s_{point}(e^+e^-\rightarrow \mu^+\mu^-)=\frac{4\pi\a_s^2}{3s}$, where $s=Q^2=4E^2$ is the square of the center of mass (the invariant square of the total 4-momentum).

\section {An example of hard process with infrared singularities: the Deep Inelastic Scattering (DIS)}
\vspace{0.8 cm}
\begin{figure}[h]
\centering
\begin{picture}(0,0)
\includegraphics[scale=0.8]{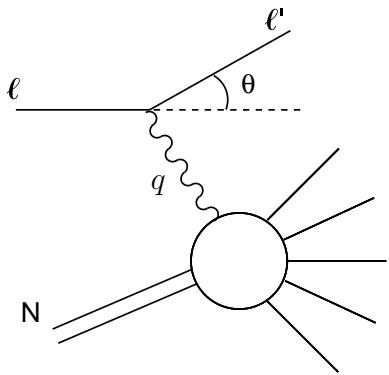}
\end{picture}
\setlength{\unitlength}{0.0200bp}%
\begin{picture}(10000,5000)(0,0)%
\fontsize{12}{\baselineskip}\selectfont
\put(2550,3500){\makebox(100,100)[l]{$q$}}%
\end{picture}
\caption{\small{Deep Inelastic Scattering Diagram.}}
\label{DIS}
\end{figure}
Another important class of hard processes is the one known as Deep Inelastic Scattering (DIS) \cite{Ellis:1991qj,Ridolfi:2002}, which groups processes of the type:
\beq l+N\rightarrow l'+X, \quad l=e^\pm,\ \mu^\pm,\ \nu,\ \overline{\nu}.
\eeq
DIS processes have a great importance in physics history because they clearly confirmed QCD predictions on the structure of nucleons, which are considered composite states of elementary particles: quarks and gluons.\footnote{The parton model treats hadrons as particles made of elementary constituents, said partons. Together with the study of the hadronic spectrum and of QCD itself, one assumes that partons are just quarks and gluons.}

In figure (\ref{DIS}) the DIS Feynman diagram is shown.

Introducing the four-momentum transferred by the electron
\beq 
q=k-k',  
\nonumber
\eeq one defines kinematic invariants $ \nu=\frac{p\cdot q }{m},$ and $Q^2=-q^2$. Then one defines the Bjorken variable:
\beq x_B=\frac{Q^2}{2m\nu} ,\eeq which is adimensional and whose values  belong to the interval between $0$ and $1$. In the case of elastic scattering one has $x_B=1$; that is, $x_B=1$ corresponds to a single parton final state.
In the laboratory system, where the nucleon with mass $m$ is at rest, lepton and nucleon four-momenta are:
\bea
k_{\mu}=(E,\vec{k})\qquad k'_{\mu}=(E',\vec{k'})\qquad p_{\mu}=(m,0,0,0),
\nonumber
\eea
therefore the energy transferred from the electron to the nucleon is \beq \nu=E-E'.\eeq 

In covariant form the Bjorken variable is: \beq x_B=\frac{Q^2}{2p\cdot q}.\eeq 
In case of deep inelastic scattering, the process duration, as it is seen by partons, is of order $(1/\nu)$ and it is much less than the interaction time of partons, whose order of magnitude is the inverse nucleon mass:
\beq m<<\nu .\eeq
In such conditions one can neglect quark interactions.

In DIS experiments physicists measure gluon densities and quark densities as function of $x_B$ and of $\frac{Q^2}{\mu^2}$, where $\mu$ is the reference energy, they measure also $\a_s(\frac{Q^2}{\mu^2})$, i.e. how the coupling constant depends on the momentum transfer.

The cross section of the reaction is given by the product of a leptonic tensor, easy to evaluate because the lepton is point-like, and a hadronic tensor:
\beq
\s=L^{\a \b}W_{\a \b}.
\nonumber
\eeq

The hadronic tensor $W_{\a \b}$ is given by the Fourier transform of the invariant expectation value of the product of the two electroweak currents in the suitably normalized nucleon state:
\beq
W_{\a \b}(p,q)\approx\int dx\quad \exp(iqx) <p\left|J^+_{\mu}(x)J_{\nu}(0)\right|p>.
\eeq

This expectation value is directly related to the total (inclusive) cross section of the interaction of a nucleon with a virtual photon and hence, due to the optical theorem, to the imaginary part of the forward virtual photon-nucleon Compton scattering. 

In the case of an unpolarized nucleon  current conservation and  CP invariance imply that this tensor has the form:
 {\small\beq
W_{\a\b}(p,q)=-(g_{\a\b}+\frac{q_{\a}q_{\b}}{Q^2})F_1(x_B,Q^2)+(p_{\a}+\frac{q_{\a}}{2 x_B})(p_{\b}+\frac{q_{\b}}{2 x_B})\frac{F_2(x_B,Q^2)}{p\cdot q}.
\eeq}
Here the coefficient functions $F_i(x_B,Q^2)$ are called electromagnetic structure functions of the nucleon and are related to the hadronic tensor in the following way:
{\small
\bea
g^{\a \b}  W_{\a,\b}(p,q)&=&-3 F_1(x_B,Q^2)+\frac{F_2(x_B,Q^2)}{2x_B}=\frac{2\,F_2(x_B,Q^2)}{x_B}-\frac{3\,F_L(x_B,Q^2)}{2 x_B}\nonumber\\
p^\a p^\b W_{\a,\b}(p,q)&=&\frac{p\cdot q}{4 x_B^2}\(F_2(x_B,Q^2)- 2 x_B\,F_1(x_B,Q^2)\)=\frac{p\cdot q}{4 x_B^2}F_L(x_B,Q^2),
\label{c}
\eea}
where $F_L(x_B,Q^2)$ is known in literature as \textit{longitudinal structure function}.

In the limit of very high energies $Q^2>>m^2$, i.e. in the limit in which the hadronic masses are negligible, the structure functions approximatively obey Bjorken scaling:
\bea 
&F_1(x_B,Q^2)&\rightarrow F_1(x_B)\\
&F_2(x_B,Q^2)&\rightarrow F_2(x_B).\nonumber
\eea
In other words in the above limit, i.e. for large values of $Q^2$, the structure functions become scale invariant, they depend only on the dimensionless variable $x_B$.

The physical meaning of the structure functions is described by Feynman parton model which considers the nucleon seen from the scattering lepton as a gas of non-interacting, massless elementary particles, called partons. A field theoretic foundation of the parton model has been given by QCD thanks to its asymptotic freedom. At short distances and during short interaction times, QCD particles interact weakly and hence, if the nucleon is a bound state of relativistic QCD particles, these appear in a DIS experiment as Feynman partons. Therefore the process should be described as a direct quark-virtual-photon interaction in the framework of QCD. Even if gluons should be considered partons together with quarks, they are not considered as partons in DIS since they do not directly interact with photon.  However QCD Feynman model should present perturbative corrections, since quarks are not exactly free; as we mentioned in the previous section, perturbation theory induces logarithmic corrections to the pure scaling because of renormalization, but these corrections are computable in QCD.

Neglecting QCD correction, one has the naive parton model in which:
\beq g^{\a,\b}W_{\a,\b}(p,q)=-\int^{1}_{x_B^-}\,\s_{point}(x_B/y)\,q_0(y)\,\frac{dy}{y},\label{F}\eeq
where $q_0(y)$ is the parton distribution function 
\beq
\s_{point}(x_B/y)=e^2\delta(x_B/y-1).\eeq  
From this relation \cite{Altarelli:2002wg} one immediately has \beq -W_{\a}^{\a}(x_B)=\frac{F_2(x_B)}{x_B}=2\,F_1(x_B)=e^2q_0(x_B).\footnote{$e$ is the fractional electric charge of the considered quark}
\eeq The equality $F_2(x_B)=2x_B\,F_1(x_B)$, called Callan Gross equality, is experimentally verified.
\begin{figure} 
\centering
\includegraphics[scale=0.7]{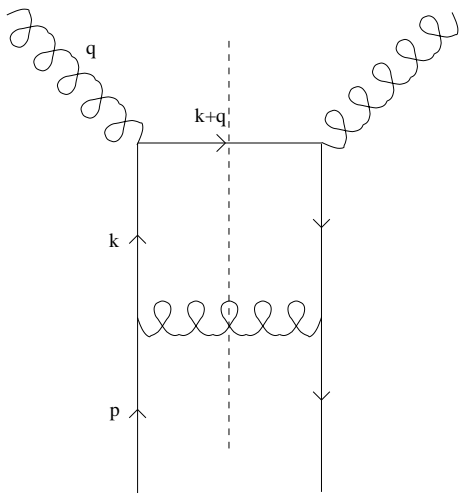}
\label{quad}
\caption{\small{Box diagram associated with the forward Scattering Compton amplitude in first order of $\a_s$.}}
\end{figure}
If one considers the radiative correction in QCD, and in particular diagrams with loops like the box (1.5), which has three fermionic lines and a gluonic one, one meets a divergence in the limit $p^2\rightarrow 0$.  This divergence arises in regions of momentum space, in which the gluon momentum is parallel to that of the massless quark. \\
Considering all the first order in $\a_s$ radiative corrections the cross section becomes \cite{Altarelli:2002wg}: 
\beq \s_{point}(z)=e^2\.\[\delta(z-1)+\frac{\a_s}{2\pi}\(\log(Q^2/m^2)P(z)+f(z)\)\],\label{G}\eeq
where $P(z)$ and $f(z)$ are finite and $\log(Q^2/m^2)$ contains a mass singularity in the limit $m\to 0$.

The infrared problem is solved by the Altarelli-Parisi equations \cite{Altarelli:1977zs,*Altarelli:1978id,*Curci:1980uw}, which allow the inclusion of such a singularity in the effective density of the quark. Indeed the effective parton density depends by the scale $\mu$:
\beq q_0(y)\rightarrow q(y,\log(\mu^2/m^2))\equiv q_0(y)+\Delta q(y,\log(\mu^2/m^2)),\label{sost}\eeq
where  $\Delta q(y,\log(\mu^2/m^2))$ contains the mass singularity of the divergent diagram:
\beq 
\Delta q(y,\log(\mu^2/m^2))=\frac{\a_s}{2\pi}\log(\mu^2/m^2)\int_{x_B^-}^1 dy \frac{q_0(y)}{y}\. P(x_B/y).
\eeq
The parton density given in $(\ref{sost})$ satisfies the following evolution equation:
\beq  \frac{d}{dt}q(x_B,t)=\frac{\a_s(t)}{2\pi}\int^{1}_{x_B^-}dy \frac{q(y,t)}{y}\.P(x_B/y)+O(\a_s(t)^2),
\eeq whose solution depends on the initial condition at $t=0$ ($q_0(x)$).
The trace of the hadronic tensor $(\ref{F})$  in terms of the effective parton density becomes:
\beq
\begin{array}{rl}
W_{\a}^{\a}(x_B,t)&=- \int^{1}_{x_B^-}dy \frac{q(y,t)}{y} \.e^2\[\delta(x_B/y-1)+\frac{\a_s}{2\pi}\(tP(x_B/y)+f(x_B/y)\)\]=\nonumber\\ 
 &=-e^2q(x_B,t)+O(\a_s(t))
\end{array}\eeq
The function $P(x)$ is known as the splitting function and it is evaluated summing on the radiative corrections of the diagrams which contribute to the first order in $\a_s$ of the cross section, isolating the collinear divergent part. We need a regularization in order to find the collinear divergent part of the diagrams. In literature the dimensional regularization is used and the collinear singularities appear as poles in $\epsilon$ of mass-shell massless quark amplitudes in much the same way as UV-divergences do. In other words the dimensional regularization parameter $\epsilon$ has a universal meaning. In order not to mix divergences of different nature, the UV-divergences must be subtracted first. 

In our approach the quark is considered weakly bound inside the hadrons and hence off-shell. 
Let $\d^2$ measure the off-shellness of the parton initial state. This regularizes collinear divergences but not the infrared singularities associated with the single parton final states.
This choice, which allows the absence of UV-divergence mixing, combined with the use of the Speer Sectors will allow us isolate the collinear divergences, which appear as divergent function of $\d^2$. The use of an off-shell amplitude together with the Speer sectors representation provides us an alternative method for the study of collinear singularities in QCD, which will be presented in this work. 

We consider this new approach worth a careful analysis, at least for physical and formal reasons, even if it is possibly more  cumbersome than the standard general one.
\chapter{Schwinger representation of Feynman amplitudes.}
\textit{In perturbative field theory the Feynman amplitudes are generally represented either in coordinate or in momentum space.} 

\textit{There is another representation which is very important in practical calculation and will be systematically used in this work. This is the Schwinger parameter representation, that we shall introduce in this section.}

\textit{In the framework of this representation one can divide the parameter measure in Speer sectors and distinguish the singularities of the amplitudes more easily.}

\section{Proper diagrams}
Scattering amplitudes and operator matrix elements within scattering amplitude states are related by L.S.Z. reduction formulae \cite{Lehmann:1954rq} to the Fourier transformed time-ordered Green functions, that is, to the Fourier transformed vacuum expectation values of time ordered products of fields and local composite operators, such as currents. 

In Feynman $\hbar$ expansion Green functions are related to series of {\large \textbf{connected}} Feynman diagrams. These correspond to sets of vertices connected by lines forming a connected polygon. 

Among the vertices one distinguishes those associated with the fields appearing in the Green functions; in the diagrams they appear as end-points of single lines, the "legs" of the diagram. Computing physical quantities the lines connecting the field vertices to the rest of the diagram are omitted. The diagram is "amputated" of its legs. 

Considering a diagram contributing to a Fourier transformed amputated Green function, one has a product of terms some of which are propagators and some others appear as a momentum multi-integral of a product of vertex and line factors (propagators). Each of these terms corresponds to a subdiagram, which is {\large \textbf{one-particle irreducible}} ({\large \textbf{1-P.I}}.), also called {\large \textbf{proper}}, meaning that it is not only connected, but remains connected if one of its lines is omitted. The computation of the 1-P.I. diagrams is the technically difficult part of the calculations. Therefore we shall concentrate our study on 1-P.I. diagrams. 

A 1-P.I. subdiagram has three kinds of vertices, all of them are end-points of at least two lines of the diagram. The vertices of the first kind correspond to local composite operators. Those of second kind correspond to Lagrangian interaction terms, which are connected through one line either to the other parts of the whole diagram or to a field vertex. The vertices of the third kind are only joined by lines of the same subdiagram. We call the vertices of the first two kinds {\large \textbf{external vertices}} since they carry momentum flowing into the diagram from outside. {\large \textbf{Internal vertices}} are those of the third kind, they do not carry any external momentum flow.
\begin{figure}[h]
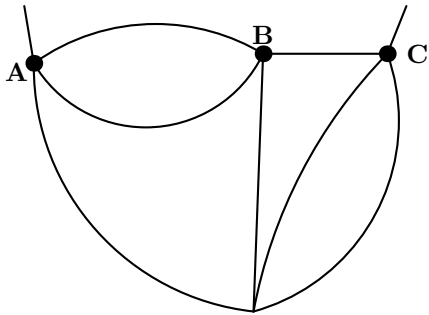

\centering
\begin{feynartspicture}(159,159)(1,1)
\FADiagram{}
\FAProp(0.5,14.)(12.5,14.5)(-0.3,){/Straight}{0}
\FAProp(12.5,14.5)(19.,14.5)(0.,){/Straight}{0}
\FAProp(0.5,14.)(12.5,14.5)(0.6,){/Straight}{0}
\FAProp(19.,14.5)(20.,17.)(0.,){/Straight}{0}
\FAProp(0.5,14.)(-0.,17.)(0.,){/Straight}{0}
\FAProp(12.5,14.5)(12.,1.)(0.,){/Straight}{0}
\FAProp(12.,1.)(19.,14.5)(-0.1483,){/Straight}{0}
\FAProp(12.,1.)(19.,14.5)(0.4361,){/Straight}{0}
\FAProp(0.5,14.)(12.,1.)(0.3862,){/Straight}{0}
\FAVert(0.5,14.){0}
\FALabel(0.2,14)[tr]{\textbf{A}}
\FAVert(12.5,14.5){0}
\FALabel(12.5,15)[b]{\textbf{B}}
\FALabel(20,14.5)[l]{\textbf{C}}
\FAVert(19.,14.5){0}
\end{feynartspicture}
\caption{Example of diagram with an internal vertex, two external vertices of the second kind, $A$ and $C$, and an external of the first kind, $B$.}
\label{4loop}
\end{figure}

If a connected diagram has $I$ lines and $V$ vertices, the number of {\large \textbf{loops}} $L$, i.e. meshes present in the diagram, is given by the following topological relation:
	\beq L=I-V+1 \label{L}. \eeq
A {\large \textbf{tree}} (sub)-diagram is a connected diagram without loop

Defining $P_v$ as the total four-momentum flowing into the diagram through the vertex $v$, the four-momentum taken by internal lines will be determined 
identifying in the diagram $L$ different loop subdiagrams, that is, closed not self intersecting polygons, and selecting a line in each loop (polygon). Notice that even if there are lines which belong to more than one loop, one has to avoid selecting the same line for two or more loop subdiagrams or lines whose momenta differ for an external momentum. Asking for momentum conservation in all the vertices, the line momenta are determined in terms of the external momenta and of momenta carried by the loop lines, that we call loop momenta. 
\begin{figure}[ht]
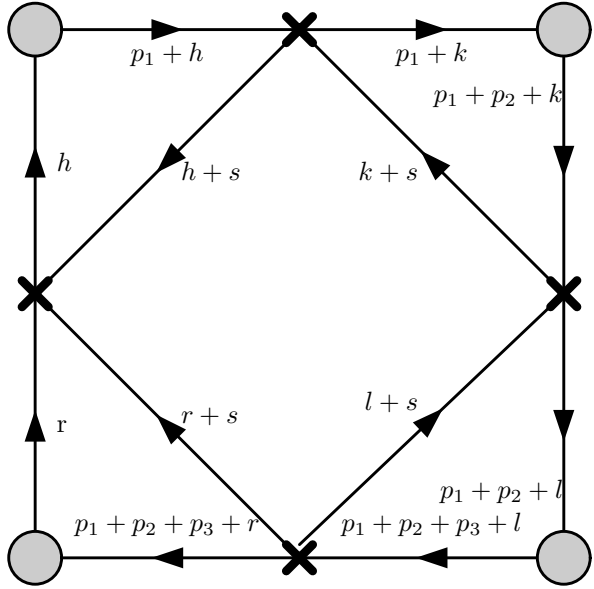

\centering
\begin{feynartspicture}(220,220)(1,1)
\FADiagram{}
\FAProp(0.,20.)(10.,20.)(0.,){/Straight}{1}
\FALabel(5.,19.48)[t]{$p_1+h$}
\FAProp(10.,20.)(20.,20.)(0.,){/Straight}{1}
\FALabel(15.,19.48)[t]{$p_1+k$}
\FAProp(20.,20.)(20.,10.)(0.,){/Straight}{1}
\FALabel(19.98,17.5)[r]{$p_1+p_2+k$}
\FAProp(20.,10.)(20.,0.)(0.,){/Straight}{1}
\FALabel(19.98,2.5)[r]{$p_1+p_2+l$}
\FAProp(20.,0.)(10.,0.)(0.,){/Straight}{1}
\FALabel(15.,0.8199)[b]{$p_1+p_2+p_3+l$}
\FAProp(10.,0.)(0.,-0.)(0.,){/Straight}{1}
\FALabel(5.,0.8199)[b]{$p_1+p_2+p_3+r$}
\FAProp(0.,-0.)(-0.,10.)(0.,){/Straight}{1}
\FALabel(0.8199,5.)[l]{r}
\FAProp(-0.,10.)(0.,20.)(0.,){/Straight}{1}
\FALabel(0.8199,15.)[l]{$h$}
\FAProp(-0.,10.)(10.,20.)(0.,){/Straight}{-1}
\FALabel(5.52,14.98)[tl]{$h+s$}
\FAProp(10.,20.)(20.,10.)(0.,){/Straight}{-1}
\FALabel(14.48,14.98)[tr]{$k+s$}
\FAProp(-0.,10.)(10.,0.)(0.,){/Straight}{-1}
\FALabel(5.52,5.02)[bl]{$r+s$}
\FAProp(10.,0.5)(20.,10.)(0.,){/Straight}{1}
\FALabel(14.48,5.52)[br]{$l+s$}
\FAVert(10.,20.){1}
\FAVert(20.,20.){-1}
\FAVert(20.,10.){1}
\FAVert(20.,0.){-1}
\FAVert(10.,0.){1}
\FAVert(0.,-0.){-1}
\FAVert(-0.,10.){1}
\FAVert(0.,20.){-1}\end{feynartspicture}
\caption{Example of a proper diagram with external (grey balls) and internal (crosses) vertices}
\label{proprioei}
\end{figure}	
A Feynman amplitude in d dimensional space-time is an integral of a product of vertex factors and propagators over the $L$ loop independent momentum variables, hence it is an integral in $L\,d$ variables.
We start our analysis in a scalar theory, characterized by couplings without derivatives and scalar propagators, calculating the Schwinger representation of 1-P.I. diagrams.
Then we shall extend the results to the more general and physical framework of theories characterized by non-scalar propagators and couplings, hence to theories with spinor and vector fields and with vertices containing derivatives.

We limit ourselves to massless theories.
\section{Schwinger parameter representation of Feynman amplitudes in $\lambda\,\phi^4$ theory.}
%

The scalar propagator corresponding to the $i$-th line, which carries a four-momentum $k_i$  in momentum representation is described by:
\beq \frac{i}{k^{2}_{i}+i\n}. \eeq 
The scalar amplitude of a proper\footnote{The relation counts also for a connected diagram, but we choose to only deal with proper diagrams} diagram $G$ with $I$ lines, $V$ vertices and $L=I-V+1$ loops is the following:
\bea
\label{ampimp}
&&\tilde{A}_G(P)=K\,\,(2\pi)^d\,\delta^d(\sum P)\,A_G(P)=\\
&&=K(-i\,\la)^{V}\mu^{\frac{(4-d)}{L}}\int \prod_{l=1}^I\left[\frac{d^d k_l}{(2\pi)^d}\frac{i}{k_l^2+i\eta}\right]\prod_{v=1}^V
\left[(2\pi)^d\,\delta^{(d)}(P_v-\sum_{l=1}^I \eps_{v\textit{l}}k_l)\right],\nonumber
\eea
with:
\begin{itemize}
\item $d=4-2\eps$ is the dimension of the space-time, with the $\eps$ dimensional regularization complex parameter.
\item $K$ is the symmetry factor of the diagram.
\item $\mu$ is the renormalization scale.
\item $\la$ is the coupling constant of the theory.
\item $P=\left\{P_1, ..., P_n\right\}$ is the set of the external momenta which satisfy
\beq \sum^{n}_{i=1} P_v=0 \label{impext}\eeq and $P_v$ is the total external momentum entering through the vertex $v$.
\item $\eps_{v\textit{l}}$ is the $V \times I$ incidence matrix, where $v$ refers to the vertex and $l$ to the line:
\beq
\eps_{v\textit{l}}=\left\{
\begin{array}{rl}
+1 & {\rm v\, is\, the\, starting\, point\, of\,  \textit{l}}\\
-1 & {\rm v\, is\, the\, ending\, point\,of\,\textit{l}}\\
0& {\rm otherwise}\\
\end{array}
\right.
\label{inc}
\eeq
\end{itemize}
In order to pass from momentum to Schwinger parameter representation \cite{Breitenlohner:1977hr,Speer:1970ss,Itzykson:1980rh} one has to make the following replacements:
\bea
\frac{i}{k_l^2+i\eta}&=&\int_0^\infty d\alpha_l\,e^{i\alpha_l(k_l^2+i\eta)}\nonumber
\\
(2\pi)^d\,\delta^{(d)}\left(P_v-\sum_{l=1}^I \epsilon_{vl}k_l\right)&=&
\int d^d y_v\,\exp
\left[-iy_v\left(P_v-\sum_{l=1}^I \epsilon_{vl}k_l\right)\right].
\label{iden}
\eea
Then for every loop one has a Gaussian integral in d-dimensions which is easily performed leading to an expression of the form:
\bea
A_G(P)=\mu^{-2L\eps}(-i\,\la)^{V} \frac{i^{L(1-\frac{d}{2})}}{(4\pi)^{\frac{Ld}{2}}}
\prod_{l=1}^I
\left[\int_0^\infty d\alpha_l\right]
\,\frac{\exp\left[i(Q_G(\alpha,P) \right]}{[P_G(\alpha)]^{d/2}}.
\label{PQ}
\eea
In this expression the Symanzik functions $P_G(\a)$ and $Q_G(\a,P)$ appear.

$P_G(\a)$ is a homogeneous polynomial  of degree $L$ in the $\a$ variables. It can be computed identifying the set ${\cal J}_T$ of all the {\large \textbf{tree}} subdiagrams of $G$ which are {\large \textbf{maximal}}, that is, such that one cannot add a line to the subdiagram without closing a loop. This implies that all maximal tree subdiagrams are obtained deleting in $G$ one line for every independent loop in all possible ways. Therefore the maximal tree subdiagrams of $G$, elements of ${\cal J}_T$, have $I-L$ lines. Notice that a maximal tree subdiagram contain all the external vertices of $G$.

Once one has identified all the elements of ${\cal J}_T$, one computes $P_G(\a)$ using the formula
\beq
P_G(\alpha)\equiv \sum_{T\in {\cal J}_T } \prod_{l\not\in T} \alpha_l,
\label{pg}
\eeq
where the sum is over all maximal tree diagrams $T$ contained in the set ${\cal J}_T$.

One also has:
\beq Q_G(\alpha,P)\equiv\frac{1}{P_G(\alpha)}\,D_G(\alpha,P),\label{Qg}\eeq where $D_G(\alpha,P)$ is a quadratic form in the vertex momenta $P_v$, whose coefficients are homogeneous polynomials of degree $L+1$ in the $\a$'s.

In order to select independent momentum variables, the set of the external vertices must exclude an arbitrarily chosen vertex, that we call $v_0$. Indeed $P_{v_0}$ is not independent of the other $P_{v}$'s since momentum conservation implies $P_{v_0}=-\sum_{v\neq v_0}P_v$. 

Therefore one has 
\beq D_G(\a,P)=\sum_{v,v'\neq v_0}C_{G}(\a)_{v,v'/v_0}P_vP_{v'}
\label{dgb}
\eeq
It remains to compute the coefficients $C_{G}(\a)_{v,v'/v_0}$. 

For this one has to identify ${\cal J}_{T_2}$, the set of the {\large \textbf{2-tree}} subdiagrams which can be obtained from the tree subdiagrams, deleting one of their lines. It follows that in general a 2-tree subdiagram is not connected and that, if it is connected, it does not contain all the  vertices of $G$; in other words in general a 2-tree has two disconnected components and, more important, divides the set of  vertices of $G$ in  two subsets of the  vertices contained in each component. If the 2-tree has a single component, one of the sets contains a single element, the vertex not belonging to the 2-tree.

Considering the set of the lines of $G$ not belonging to a 2-tree, these can be identified with the cut lines when one cuts the diagram in two components breaking all its loops. We call this cut a {\large \textbf{complete cut}}. From this point of view the above mentioned two subsets of vertices are those lying on opposite sides of the complete cut.\footnote{In the following with complete cut we will refer to a set $C$ of lines whose deletion breaks $G$ into two tree subdiagrams each of which contains at least an external vertex.}\\

\begin{tiny}
\begin{figure}[hc]
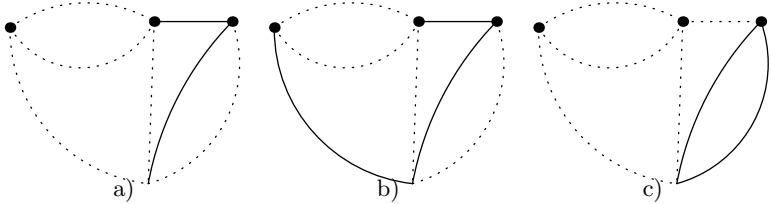

\centering
\begin{feynartspicture}(300,100)(3,1)
\FADiagram{a)}
\FAProp(0.5,14.)(12.5,14.5)(-0.3,){/GhostDash}{0}
\FAProp(12.5,14.5)(19.,14.5)(0.,){/Straight}{0}
\FAProp(0.5,14.)(12.5,14.5)(0.6,){/GhostDash}{0}
\FAProp(12.5,14.5)(12.,1.)(0.,){/GhostDash}{0}
\FAProp(12.,1.)(19.,14.5)(-0.1483,){/Straight}{0}
\FAProp(12.,1.)(19.,14.5)(0.4361,){/GhostDash}{0}
\FAProp(0.5,14.)(12.,1.)(0.3862,){/GhostDash}{0}
\FAVert(0.5,14.){0}
\FAVert(12.5,14.5){0}
\FAVert(19.,14.5){0}
\FADiagram{b)}
\FAProp(0.5,14.)(12.5,14.5)(-0.3,){/GhostDash}{0}
\FAProp(12.5,14.5)(19.,14.5)(0.,){/Straight}{0}
\FAProp(0.5,14.)(12.5,14.5)(0.6,){/GhostDash}{0}
\FAProp(12.5,14.5)(12.,1.)(0.,){/GhostDash}{0}
\FAProp(12.,1.)(19.,14.5)(-0.1483,){/Straight}{0}
\FAProp(12.,1.)(19.,14.5)(0.4361,){/GhostDash}{0}
\FAProp(0.5,14.)(12.,1.)(0.3862,){/Straight}{0}
\FAVert(0.5,14.){0}
\FAVert(12.5,14.5){0}
\FAVert(19.,14.5){0}
\FADiagram{c)}
\FAProp(0.5,14.)(12.5,14.5)(-0.3,){/GhostDash}{0}
\FAProp(12.5,14.5)(19.,14.5)(0.,){/GhostDash}{0}
\FAProp(0.5,14.)(12.5,14.5)(0.6,){/GhostDash}{0}
\FAProp(12.5,14.5)(12.,1.)(0.,){/GhostDash}{0}
\FAProp(12.,1.)(19.,14.5)(-0.1483,){/Straight}{0}
\FAProp(12.,1.)(19.,14.5)(0.4361,){/Straight}{0}
\FAProp(0.5,14.)(12.,1.)(0.3862,){/GhostDash}{0}
\FAVert(0.5,14.){0}
\FAVert(12.5,14.5){0}
\FAVert(19.,14.5){0}
\end{feynartspicture}
\caption{a) is an example of a \textit{2-tree} b) is an example of a \textit{tree} c) is neither a tree nor a 2-tree}
\end{figure}
\end{tiny}

Thus a 2-tree contains $I-L-1$ lines and is the complement in $G$ of  a complete cut of $L+1$ lines. Given two, possibly coinciding, external vertices $v$,  $v'$ and $v_0$, one identifies a subset of 2-trees ${\cal J}_{T_2\, v,v'/v_0}$ containing the 2-trees for which the pair $v,v'$ and $v_0$ lie on opposite sides of the corresponding complete cut.

One has: 
\beq
C_{G}(\a)_{v,v'/v_0}=\sum_{T_2\in {\cal J}_{T_2\, v,v'/v_0}}\prod_{l\notin T_2}\a_l.
\eeq
Alternatively if we call ${\cal C}$ the set of complete cuts C, understood as a subset of lines of $G$, and ${\cal C}_{v,v'/v_0}$ the set of complete cuts separating the pair $v, v'$ from $v_0$, one can write:
\beq
C_{G}(\a)_{v,v'/v_0}=\sum_{C\in {\cal C}_{v,v'/v_0}}\prod_{l\in C}\a_l.
\label{CG}
\eeq


The integral $(\ref{PQ})$ might diverge when one or more parameters $\a$ vanish or tend to infinity. The first case corresponds to ultraviolet (UV) divergences; the second one to infrared (IR) divergences. 


In order to evaluate equation $(\ref{PQ})$, one introduces the scale variable $t$ according to:
\beq
\a_l=t\ \b_l \ .
\label{s}\eeq

It is apparent that the definition of the new parameters $\b$ is not unique, since they are $I$ coordinates of a $I-1$-dimensional manifold which is identified with the quotient space of the positive sector of the $I$-dimensional Cartesian space and a positive semiline, whose points are identified by the scale factor $t$. This manifold is the positive sector of the $I-1$- dimensional projective space.

Therefore the integration measure becomes
\beq\prod_{l=1}^I d\a_l =dt \:t^{I-1} d\mu(\b)\label{demu},\eeq where $d\mu(\b)$ depends on the choice of independent coordinates $\b$ of the projective space and the polynomials $Q\ ,\ P\ $ e $D$ depend on $t$ in the following way:

\bea
&&Q_G(\a_l\cdot t, P)=t\cdot Q_G(\b_l,P),\\ \nonumber
&&P_G(\a_l\cdot t)=t^{L}\cdot P_G(\b_l), \\ \nonumber
&&D_G(\a_l\cdot t, P)=t^{L+1}\cdot D_G(\b_l,P).\nonumber \eea
In terms of the new parameters $\b_l$, one has:
\bea
A_G(P)&=&\mu^{2L\eps}(-i\,\la)^{V}\frac{i^{L-dL/2}}{(4\pi)^{dL/2}}
\int \frac{dt}{t} \ t^{I-dL/2} \int d\mu(\b)
\frac{\exp\left[it\ Q_G(\b,P) \right]}{[P_G(\b)]^{d/2}}\nonumber\\
&=&\mu^{2L\eps}\frac{i^{(2-d)L-1}\la^{V}}{(4\pi)^{dL/2}}
 \int d\mu(\b)
\frac{\Gamma(I-dL/2)}{[P_G(\b)]^{\frac{d(L+1)}{2}-I}
\left[D_G(\b,P)\right]^{I-\frac{dL}{2}}}.\qquad
\label{princ}
\eea
The argument of the Gamma-function is $(-D/2)$, where $D=dL-2I$ is the diagram \textit{superficial divergence degree}. In a four dimensional scalar theory D is an even integer. When $D$ is a positive integer or null the function $\Gamma(-D/2)$ has a singularity corresponding to the overall UV divergence of the diagram. It is important to notice that the amplitude can be divergent also when $D$ is negative, since the diagram might contain divergent subdiagrams.\\
As said above, after the integration over the scale parameter the Feynman integral is transformed into an integral over a positive sector of $(I-1)$-dimensional projective space. One can choose among different possible parametrizations of the projective space.

These are the most relevant for our purposes:

\begin{enumerate}
    \item  the \textit{hypercubic parametrization}, which satisfies the constraint $\sum_{i=1}^I\b_i=1$, whose integration measure is \beq d\mu(\b)=\(\prod_{i=1}^I d\b_i\)\delta(1-\sum_{i=1}^I\b_i);\eeq 
    \item the \textit{parametrization based on Speer-Smirnov sectors}, that is, on the division of the integration domain into subdomains. This will be described in the following. In each  subdomain one of the $\b$ is assigned the value $1$ and the others take positive values less than $1$.
\end{enumerate}
    
The great advantage of Speer-Smirnov choice is that UV and the IR divergent parts of a diagram appear only in few sectors and hence if one is interested in singularities, one can limit the number of considered subsectors.


\section{Schwinger representation of tensorial Feynman amplitudes.}
If one considers a generic theory, like QCD, one deals with tensorial amplitudes in loop momenta because of the presence of fermion propagators, which differ from the scalar ones, since they have a line momentum dependent numerator $\ksl$, and derivative and hence momentum dependent interactions terms: the vertex with three gluons in QCD, shown in figure $(\ref{feyqcd})$, introduces a function of the line momenta in the numerator of the amplitude.


Therefore the generic amplitude in momentum representation, differently from the scalar amplitude, has as a numerator consisting in a sum of different rank tensors in loop momenta, which are contracted with tensors depending on covariant objects, like external momenta, gamma matrices and the metric tensor. Let's call the numerator $N(P,k)$. 

The momentum representation amplitude for a proper diagram with $I$ massless lines and $V$ vertices is\footnote{$C_V$ is a factor according to Feynman rules}:
\beq
\tilde{A}_G(P)=\mu^{2L\eps} C_V\int \prod_{i=1}^I
\left[\frac{d^d k_i}{(2\pi)^d}\,\frac{i}{k_i^2+i\eta}\right]\,N(P,k)
\prod_{v=1}^V 
\left[(2\pi)^d\,\delta^{(d)}(P_v-\sum_{i=1}^I \epsilon_{vi}k_i)\right].
\label{ampimp_n}
\eeq
In order to give a parametric representation of this integral, one has to proceed in the following way:
\begin{itemize}
	\item One chooses a line $l$ for every loop of the diagram (the line which carries the independent loop momentum) and multiplies its propagator $\frac{i}{k^2_l+i \n}$ by the factor $e^{ik_l\cdot u_l}$, where $u_l$ is a vectorial parameter assigned to line $l$.
	\item One replaces $N(P,k)$ with $N(P,-i\cdot\partial_u)$, which depends on the differential operators $\partial_u$ and acts on the scalar amplitude.
\end{itemize}
The amplitude is \cite{Breitenlohner:1977hr}:  
	\bea 
&&\mu^{2L\eps}\,N(P,-i\partial_u) \int \prod_{l=1}^I
\left[\frac{d^d k_l}{(2\pi)^d}\cdot\frac{i\cdot e^{ik_l\cdot u_l}}{k_l^2+i\eta}\right]\nonumber\\
&&\qquad\qquad\qquad\qquad \prod_{v=1}^V
\left[(2\pi)^d\,\delta^{(d)}(P_v-\sum_{l=1}^I \epsilon_{vl}k_l)\right]\vert_{u=0}\\ &&\equiv \mu^{2L\eps}(2\pi)^d\,\delta^d(\sum P)\,N(-i\partial_u) A_G(P,u)\vert_{u=0}\ .\nonumber\label{nume}
\eea
After performing the $L$ gaussian integrals in d-dimensions, one obtains for $A_G(P,u)$ the following relation:


\beq
A_G(P,u)=\mu^{2L\eps}\, C_V \,\frac{i^{L(1-\frac{d}{2})}}{(4\pi)^{\frac{L d}{2}}}
\prod_{l=1}^I
\left[\int_0^\infty d\alpha_l\right]
\,\frac{\exp\left[iQ_G(\alpha,P,u)\right]}{[P_G(\alpha)]^{d/2}}|_{u=0},
\label{PQ_n}
\eeq
with
\beq
Q_G(\alpha,P,u)\equiv\frac{1}{P_G(\alpha)}\,D_G(\alpha,P,u)=Q_G(\alpha_l,P_v+\sum_l \epsilon_{l,v}\frac{u_l} {2\alpha_l})-{\frac{1}{4}}\sum_l \frac{u_l^2}{\alpha_l}.
\label{Qu}
\eeq
Equation $(\ref{PQ_n})$, that one commonly finds in the literature, is not sufficient  for our purposes. Indeed in order  to profit completely of the Speer-Smirnov sector decomposition, it is convenient to interchange in equations $(\ref{dgb})$ and $(\ref{CG})$, which defines the quadratic form $D_G$, the sum over vertices with that over 2-trees.

Indeed, a given 2-tree $T_2$, or better, the corresponding complete cut $C$, contributes in the vertex sum to the coefficients $C_{G}(\a)_{v,v'/v_0}$, when $v$ and $v'$ lie on the opposite side to $v_0$ of the cut diagram; let us call ${\cal V}_{T_2,v_0}$ the set of vertices opposite to $v_0$. Then the contribution to $D_G$ from $T_2$ is equal to
\beq
\prod_{l\in C}\a_l\sum_{v,v'\in{\cal V}_{T_2,v_0}}P_v\,P_{v'}=\prod_{l\in C}\a_l\(\sum_{v}P_v\)^2.
\nonumber
\eeq
It is therefore natural to consider the kinematic invariant variable $\(\sum_{v}P_v\)^2=P_{T_2}^2$, where $P_{T_2}$ is the total momentum crossing the complete cut $C$.

After this choice one has
\beq D_G(\a,P)=\sum_{T_2\in {\cal J}_{T_2}} P_{T_2}^2\prod_{l\notin T_2} \a_l=\sum_{C\in {\cal C}} P_{C}^2\prod_{l\in C} \a_l ,\label{dg}\eeq
where $P_{T_2}=P_C$ is the total momentum crossing the complete cut $C$.

Let us now consider $D_{G}(\a,P,u)$, given in equation $(\ref{Qu})$, and single out the contribution of a given 2-tree $T_2$ and the corresponding cut $C$. This is given by:
\beq
\prod_{l\in C}\a_l\(\sum_{v}(P_v+\sum_{l'}\epsilon_{v,l'}\frac{u_{l'}^2}{2\a_{l'}})\)^2.
\nonumber
\eeq

In the sum over $l'$, appearing in this equation, it is clear that the lines connecting pair of vertices on the same side of the cut $ C$ do not contribute, since the sign of $\epsilon_{v,l'}$ is opposite for the two vertices of the pair. Therefore the only lines contributing are the elements of the complete cut $C$ whose end points belong to opposite connected components of $G$. These lines do not belong to $T_2$ and are such that the union $T_2\cup l'$ is a tree subdiagram. Therefore the linear term in $u$ in $(\ref{Qu})$ is given by
\beq
\sum_{T_2\in {\cal J}_{T_2}}\sum_{\substack{l'\notin T_2\\ T_2\cup l'\in{\cal J}_{T}}}\eps_{l',T_2}\,P_{T_2}\cdot u_{l'}\prod_{l\notin T_2}\a_l
\label{lin2}
\eeq
$\eps_{l',T_2}$ is positive when $u$ crosses $C$ in the same direction of $P_{T_2}$; otherwise it has the opposite sign.

This formula and the following ones can be written in terms of the complete cuts and the {\large \textbf{reduced cuts}} $\hat{C}$, which are obtained omitting a line from complete cuts and are identified with the complements in $G$ of the maximal trees. Therefore, together with the set of maximal trees ${\cal J}_{T}$, we have the set of reduced cuts $\hat{{\cal C}}$, such that for any $\hat{C}\in \hat{{\cal C}}$ one has a $T\in {\cal J}_{T}$ satisfying: 
\beq
\hat{C}=G/T.
\nonumber
\eeq
Now the term in equation $(\ref{lin2})$ is written:
\beq
\sum_{C\in {\cal C}}\sum_{\substack{l'\in C\\C/l'\in\hat{{\cal C}}}}\eps_{l',C}\,P_{C}\cdot u_{l'}\prod_{l\in C}\a_l,
\label{lin}
\eeq
where we have identified $\eps_{l',T_2}$ with $\eps_{l',C}$, if $C$ is the complement of $T_2$.

Now we come to the quadratic part in $u$ of $D_G(\a,P,u)$, for which the contribution of a given complete cut is \beq\prod_{l\in C}\a_l\(\sum_{v,l'}\epsilon_{v,l'}\frac{u_{l'}^2}{2\a_{l'}}\)^2.\nonumber\eeq
Once again the lines contributing to this term are those belonging to $C$ and connecting vertices lying in opposite sides of the cut diagram. Let us consider first of all the coefficients $u_{l'}\cdot u_{l''}$ for $l'\neq l''$. As above $l'$ and $l''$ are restricted by the same condition appearing in $(\ref{lin})$ and hence one has for the corresponding term in $D_G(\a,P,u)$ in terms of the 2-trees and the complete cuts:
\bea
&&\frac{1}{4}\sum_{T_2\in {\cal J}_{T_2}}\sum_{\substack{l'\neq l''\\l'\cup T_2\in{\cal J}_T\\T_2\cup l''\in{\cal J}_{T}}}\eps_{l',T_2}\,\eps_{l'',T_2}\,u_{l'}\cdot u_{l''}\prod_{l\notin T_2\cup l\cup l'}\a_l=\nonumber\\
&&\frac{1}{4}\sum_{C\in {\cal C}}\sum_{\substack{ l'\neq l''\\C/l'\in\hat{{\cal C}}\\C/l''\in\hat{{\cal C}}}}\eps_{l',C}\,\eps_{l'',C}\,u_{l'}\cdot u_{l''}\prod_{l\in C/(l\cup l')}\a_l.
\label{uiuj}
\eea
Considering instead the terms in $u_l^2$, one has two contributions of opposite signs in $(\ref{Qu})$, the quadratic part in $Q_G$ and the last term. 
The first contribution is positive and, following the analysis of the other terms, turn out to be equal to
\beq
\frac{1}{4}\sum_{T_2\in {\cal J}_{T_2}}\sum_{\substack{l'\notin T_2\\ l'\cup T_2\in{\cal J}_{T}}}\prod_{l\notin T_2\cup l'}\a_l \frac{u_{l'}^2}{\a_{l'}}.
\label{u}
\eeq
The second contribution is 
\beq
-\frac{1}{4}P_G(\a)\sum_{l'}\frac{u_{l'}^2}{\a_{l'}}=-\frac{1}{4}\sum_{T\in {\cal J}_{T}}\prod_{l\in T}\a_l\sum_{l'}\frac{u_{l'}^2}{\a_{l'}}.
\label{v}
\eeq
Combining these terms we find
\beq
-\frac{1}{4}\sum_{T\in {\cal J}_{T}}\sum_{l'\notin T}u_{l'}^2\prod_{l\notin T\cup l'}\a_l. 
\label{u2}
\eeq
Indeed the first term given in $(\ref{u})$ selects the maximal trees which coincide with $T_2\cup l'$ for some $T_2$. These are all the maximal trees containing $l'$. Since the expression given in $(\ref{v})$ refers to all the elements of ${\cal J}_{T}$, the difference corresponds to maximal trees which don't contain $l'$.

The formula in terms of the reduced and complete cuts becomes
\beq
-\frac{1}{4}\sum_{\hat{C}\in \hat{{\cal C}}}\sum_{l'\in \hat{C}}u_{l'}^2\prod_{l\neq l',\,l\in \hat{C}}\a_l. 
\label{u2b}
\eeq
In conclusion we have this expression in terms of  maximal trees and  2-trees:
\bea
&&D_G(\a,P,u)=\sum_{T_2\in {\cal J}_{T_2}} P_{T_2}^2\prod_{l\notin T_2} \a_l+\sum_{\substack{T_2\in {\cal J}_{T_2},\,i,\\( l_i\notin T_2)\\(T_2\cup l_i\in{\cal J}_{T})}}\eps_{i,T_2}\,P_{T_2}\cdot u_i\prod_{l\notin T_2}\a_l+\qquad \\
&&-\frac{1}{4}\( \sum_{\substack{T\in{\cal J}_{T},\,i\\(l_i\notin T)}}u_i^2\prod_{l\notin T\cup l_i}\a_l-\sum_{\substack{T_2\in{\cal J}_{T_2},\,i\neq j\\(l_i\cup T_2\in{\cal J}_{T})\\(l_j\cup T_2\in{\cal J}_{T})}}u_i\cdot u_j\, \eps_{i,T_2}\,\eps_{j,T_2}\prod_{l\notin T_2\cup l_i\cup l_j}\a_l\)\nonumber
\label{DG_p}.
\eea
While in terms of the complete and the reduced cuts one has the alternative formulation:
\bea
&&D_G(\a,P,u)=\sum_{C\in {\cal C}} P_{C}^2\prod_{l\in C} \a_l+\sum_{\substack{C\in {\cal C},\,i,\\( l_i\in C)\\(C/l_i\in\hat{{\cal C}})}}\eps_{i,C}\,P_{C}\cdot u_i\prod_{l\in C}\a_l+\\
&&-\frac{1}{4}\( \sum_{\substack{\hat{C}\in\hat{{\cal C}},\,i\\(l_i\in \hat{C})}}u_i^2\prod_{l\in \hat{C}/ l_i}\a_l-\sum_{\substack{C\in {\cal C},\,i\neq j\\(C/l_i\in\hat{\cal C})\\(C/l_j\in\hat{\cal C})}}u_i\cdot u_j\, \eps_{i,C}\eps_{j,C}\prod_{l\in C/(l_i\cup l_j)}\a_l\)\nonumber
\label{DG_p2}.\eea
We recall the equivalent form for $(\ref{dg})$,
\beq
D_G(\a,P)=\sum_{C\in {\cal C}} P_{C}^2\prod_{l\in C} \a_l,
\label{dgc}
\eeq
the other Symanzik function being:
\beq
P_G(\a)=\sum_{\hat{C}\in\hat{\cal C}}\prod_{l\in\hat{C}}\a_l.
\label{pgc}
\eeq 
Going on we shall use this alternative form.\\

Considering equation $(\ref{ampimp_n})$, one can write a generic amplitude as a linear combination of homogeneous polynomials of the components of the external momenta whose coefficient are given by suitable integrals in the Schwinger parametric space. Indeed let us decompose $N(P,k)$ into a sum of products of homogeneous terms in the external momenta $P$ and internal loop momenta $k$:
\beq
N(P,k)=\sum_n N_n^l(P)\cdot N_n^i(k).
\label{N}
\eeq
Let $d_{N_n^l}$ and $d_{N_n^i}$ be the mass dimensions of the factors.

Correspondingly in equation $(\ref{nume})$ one has $N(P,-i\partial_u)=\sum_n N_n^l(P)\cdot N_n^i(-i\partial_u)$. The action of $N_n^i(-i\partial_u)$ on $\exp\left[i\frac{D_G(\alpha,P,u)}{P_G(\a)}\right]\left|_{u=0}\right.$ generates a sum of contributions corresponding to the possible actions of the $u$-derivative either on the linear or on the quadratic part in $u$ of $D_G(\a,P,u)$. The number of times the $u$-derivatives in $N_n^i$ act on the $u$-quadratic part of $D_G(\a,P,u)$ must be an even number, $2a$, since the result is computed at $u=0$.
Therefore $N_n^i$ acts on $d_{N_n^i}-2a$ times on the $n$-linear part of $D_G(\a,P,u)$.

One has correspondingly a factor $\frac{R(\a)}{P_G(\a)^a}$ with $R(\a)$ homogeneous of degree $(L-1)a$ from the quadratic part of $D_G(\a,P,u)$ and a factor $\frac{S(\a,P)}{P_G(\a)^{d_{N_n^i}-2a}}$ with $S(\a,P)$ homogeneous of degree $d_{N_n^i}-2a$ in the momentum components and of $(d_{N_n^i}-2a)L$ in  $\a$ from the linear part. Therefore one has a global factor of the type:
\beq
\sum_{a=0}^{\[\frac{d_{N_n^i}}{2}\]}\frac{P^i_{n,a}(\a,P)}{P_G(\a)^{d_{N_n^i}-a}},
\eeq
where we denote by $\[X\]$ the integer part of the positive number $X$, and $P^i_{n,a}(\a,P)$ is a homogeneous polynomial of degree $d_{N_n^i}-2a$ in the external momentum components and of degree
\beq
(d_{N_n^i}-2a)L+(L-1)a=L\,d_{N_n^i}-(L+1)a
\eeq
in $\a$.

The $\a$-degree is apparently non-negative. Therefore $P^i_{n,a}(\a,P)$ is a linear combination of $\a$-monomials with momentum-dependent coefficient:
\beq
P^i_{n,a}(\a,P)=\sum_{\tau}C^i_{n,a,\tau}(P)\,\prod_{l=1}^I\a^{\lambda_{l,\tau}}.
\eeq
Consequently in the parametric integral the numerator $N$, whose dimension $d_N$ is directly related to the global degree of the amplitude $G$,
\beq
d_G=d_N+d\,L-2I,
\eeq
corresponds to the factor :
\beq
N=\sum_{b=0}^{d_{N}}\sum_{a=0}^{\frac{d_N-b}{2}}\sum_{\tau}\Theta_{a,b,\tau}(P)\frac{\prod_{l=1}^I \(d\a_i \a_i^{\lambda_{l,\tau}}\)}{P_G(\a)^{b-a}},
\eeq
where we have denoted by $b$ the degree in $k$ of the homogeneous polynomials $N_n^i(k)$ in the decomposition $(\ref{N})$.
Furthermore $\Theta_{a,b,\tau}$ is a homogeneous polynomial in the momentum components of degree \beq d_{\Theta}=d_N-2a\nonumber\eeq and one has, as shown above,
\beq
\sum_i^I\lambda_{i,\tau}=L(b-a)-a.
\eeq

In a scalar theory, $(b=0,\ a=0)$, this quantity is null as expected.

At this point in much the same way as in scalar theory, one introduces the scale variable $t$ according to $(\ref{s})$ and integrates on it.

One obtains the analogous generic relation to the scalar case, $(\ref{princ})$:
\beq
\tilde{A_G}(P)=\frac{i^{I+L-Ld}}{(4\pi)^{Ld/2}}\,C_V \sum_{\tau}\sum_{a=0}^{\frac{d_N}{2}}\sum_{b=0}^{d_N-2a}\Theta_{a,b,\tau}(P) \,I_{G,(a,b,\tau)}(P),
\label{princ2}
\eeq
where
\beq
I_{G,(a,b,\tau)}(P)= \mu^{2L\eps}\Gamma(\mu_{\tau,a}) \int \frac{d\mu(\b)\,\prod_{i=1}^I \b_i^{\lambda_{i,\tau}}}{P_G(\b)^{\frac{d}{2}+b-a-\mu_{\tau,a}}D_G(\b,P)^{\mu_{\tau,a}}}
\label{Ib}
\eeq
and \beq \mu_{\tau,a}=I-\frac{Ld}{2}-a .\eeq

In chapter $4$ we shall study the ultraviolet and infrared divergences related to $I_{G,(a,b,\tau)}(P)$ decomposing the integral into the Speer sectors. 



\subsection{An example: the vertex correction}
\begin{center}
\begin{feynartspicture}(200,150)(1,1)
\FADiagram{}
\FAProp(0.,20.)(2.,15.)(0.,){/Sine}{0}
\FALabel(1.52,18.)[l]{q,\ $\mu$}
\FAProp(2.,15.)(2.,5.)(0.,){/Straight}{-1}
\FALabel(1.48,10.)[r]{k\ \ \textit{1}}
\FAProp(2.,5.)(0.,0.)(0.,){/Straight}{-1}
\FALabel(1.52,2.)[l]{p}
\FAProp(2.,15.)(15.,15.)(0.,){/Straight}{1}
\FALabel(8.5,16.02)[b]{\textit{2}\ \ q+k}
\FAProp(15.,15.)(20.,15.)(0.,){/Straight}{1}
\FALabel(17.5,16.02)[b]{q+p}
\FAProp(15.,15.)(2.,5.)(0.,){/Cycles}{0}
\FALabel(9.02,8.98)[tl]{\textit{3}\ \ k-p}
\end{feynartspicture}
\end{center}
One computes the off-shell, dimensionally regularized amplitude of the diagram in figure using the expressions $(\ref{PQ_n})$ and $(\ref{DG_p})$. The amplitude in momentum representation is
{\small\beq \Delta_{\eps}^{\mu}=\mu^{2\eps} C_V\,i^3\int \frac{d^dk}{(2\pi)^d}\,\frac{\c^{\rho}(\qsl+\ksl)\c^{\mu}\ksl\c_{\rho}}{k^2(q+k)^2(p-k)^2}=-2\mu^{2\eps} C_V\,i^3\int \frac{d^dk}{(2\pi)^d}\,\frac{\ksl \c^{\mu}\qsl+\ksl \c^{\mu}\ksl}{k^2(q+k)^2(p-k)^2},\eeq}
where the numerator depending on k is due to the two fermionic internal lines, and one has $C_V=-ie_s^2\,e\,c_F$\footnote{$t^at^a=c_F\textit{1}$}, since there are two QCD vertices gluon-fermion-fermion and a QED one.

One assigns the fourvector $u$ to line $1$ and from $(\ref{pg})$ and $(\ref{DG_p})$ one has: 
{\small \bea
&&P_G(\a)=\a_1+\a_2+\a_3;\\
&&Q_G(\a,P,u)=\frac{D_G(\a,P,u)}{P_G(\a)}=\frac{1}{P_G(\a)}\(D_G(\a,P)+\alpha_3 \,p\cdot u-\alpha_2 \,q\cdot u-\frac{u^2}{4}\)\nonumber
\eea}
with $D_G(\a,P)=p^2\,\a_1\a_3+q^2\,\a_1\a_2+S\,\a_2\a_3$ and $S=(q+p)^2.$

Passing from momentum to parameter representation, the amplitude becomes:
{\small\bea
&&\Delta_{\eps}^{\mu}=-\mu^{2\eps} C_V \,\frac{2i}{(4\pi)^{d/2}}
\int_0^\infty\prod_{l=1}^Id\alpha_l(i\pasl_u\c^{\mu}\qsl+\pasl_u\c^{\mu}\pasl_u)\frac{\exp\left[iQ_G(\alpha,P,u)\right]}{[P_G(\alpha)]^{d/2}}|_{u=0}=\nonumber\\
&& \mu^{2L\eps} \frac{2iC_V}{(4\pi)^{d/2}}\int_0^\infty \frac{\prod_{l=1}^Id\alpha_l}{(P_G(\a))^{d/2}}\[\frac{\(\ \a_3\ \psl-\a_2\ \qsl\ \)\c^{\mu}\(\ \qsl\ \(\a_1+\a_3\)+\ \psl\ \a_3\)}{P_G(\a)^{2}}-\frac{i\c^{\mu}}{P_G(\a)}\]\nonumber
\eea}
At this point one scales the parameters with respect to the variable $t$, like in $(\ref{s})$, and, after $t$-integration, one obtains
{\small \bea
&&\Delta_{\eps}^{\mu}=-\frac{2\,C_V}{(4\pi)^{2}}\int_0^\infty \frac{\prod_{l}^3d\b_l\d(1-\sum_{i}^3\b_i)}{(P_G(\b))^{3}}\[\frac{\(\ \b_3\ \psl-\b_2\ \qsl\ \)\c^{\mu}\(\ \qsl\ \(\b_1+\b_3\)+\ \psl\ \b_3\)}{(p^2\,\b_1\b_3+q^2\,\b_1\b_2+S\,\b_2\b_3)}\right.\nonumber\\
&&-\Gamma[\eps]\c^{\mu} \(\frac{\mu^{2}P_G(\b)^{2}}{p^2\,\b_1\b_3+q^2\,\b_1\b_2+S\,\b_2\b_3}\)^{\eps}\left.\].
\label{Deltal}
\eea}
Here the second term is apparently UV-divergent since it diverges in the limit $\eps\rightarrow 0$.

The $\overline{MS}$ renormalized amplitude is given by
\beq
\Delta^{\mu}=lim_{\eps\rightarrow 0}\(\Delta_{\eps}^{\mu}-\frac{\c^{\mu}}{\eps}\frac{C_V}{(4\pi)^{2}}\),
\eeq
where we have subtracted the UV-divergence due to the pole of $\Gamma(\eps)=\frac{1}{\eps}+\zeta$.

The result is:
{\small\bea
&\Delta^{\mu}=-\frac{2\,C_V}{(4\pi)^{2}}\int_0^\infty &\frac{\prod_{l}^3d\b_l\d(1-\sum_{i}^3\b_i)}{(P_G(\b))^{3}}\\
&&\[\frac{\(\ \b_3\ \psl-\b_2\ \qsl\ \)\c^{\mu}\(\ \qsl\ \(\b_1+\b_3\)+\ \psl\ \b_3\)}{(p^2\,\b_1\b_3+q^2\,\b_1\b_2+S\,\b_2\b_3)}+\right.\nonumber\\
&&-\left.\c^{\mu}\, \log\(\frac{P_G(\b)^2\mu^2}{p^2\,\b_1\b_3+q^2\,\b_1\b_2+S\,\b_2\b_3}\)\]\nonumber.
\label{Delta}
\eea}
We will study the infrared divergences of this amplitude once introduced the Speer sector parametrization.

\chapter{Singularity families and Speer sectors}
\textit{In my degree thesis\cite{mia:tesi} I have studied the scalar Feynman amplitudes in the space of Schwinger parameters, divided into Speer sectors, showing that this division separates the amplitudes in the sum of terms with different singular behaviour.  I have identified and classified the infrared divergences for scalar amplitudes. The construction of the Speer sectors was proposed by Speer in the article \cite{Speer:1975dc} in 1975. In this chapter we present the construction of the Speer Sectors due to Smirnov in \cite{Smirnov:2008aw}. The sectors one obtains are slightly different from the those identified by Speer, but related to them. They are more numerous than the Speer ones and the range values of the parameters are limited all between 0 and 1, instead of taking different range values.}

\textit{The advantage of Smirnov's choice is the availability of power counting formulae}
\section{Preliminary definitions}
\begin{itemize}
\item An {\large \textbf{irreducible}} diagram is a diagram which either is 1-P.I., but it is not the union of two 1-P.I. diagrams with a single vertex in common, or it consists of a single line.

\begin{figure}[ht]
\centering
\includegraphics[scale=0.4]{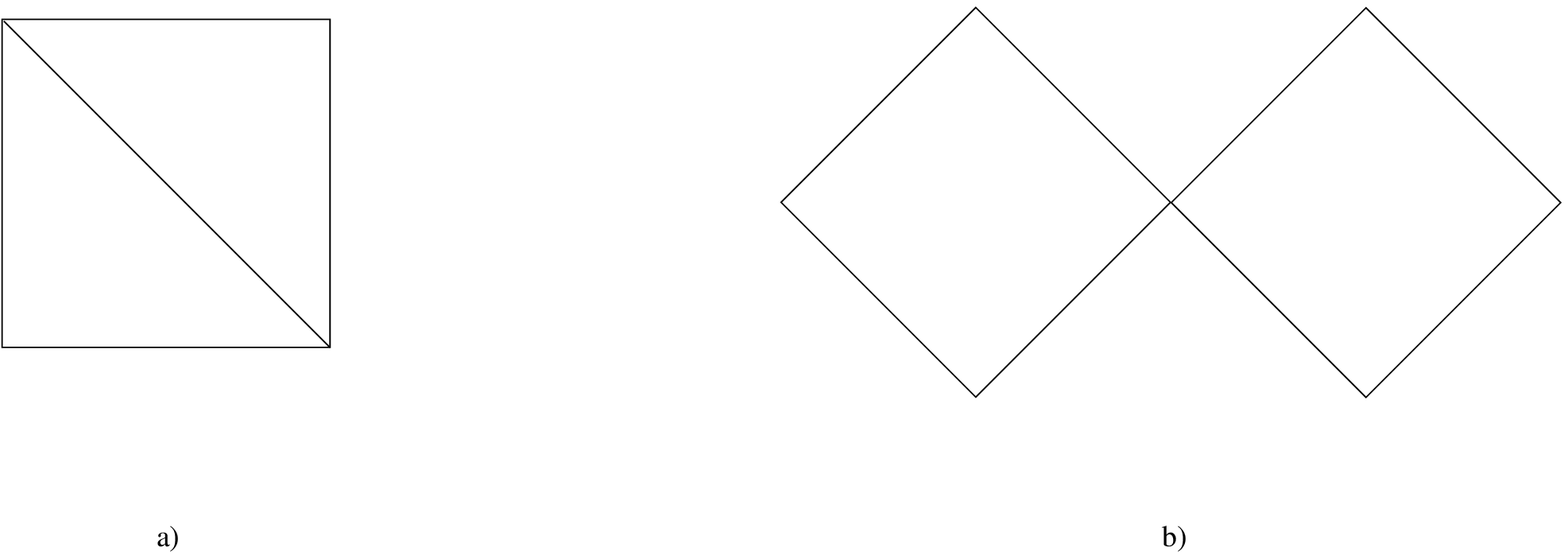}
\caption{\small{$a)$ is irreducible. $b)$ is 1-P.I., but not irreducible.}}
\label{propri}
\end{figure}

\item The {\large \textbf{parts}} of a diagram are the maximal irreducible subgraphs of the original diagram. 
\item A {\large \textbf{link}} \cite{Speer:1975dc} in a diagram in a massless theory is a connected diagram which contains all vertices of the diagram and is minimal: any subgraph obtained after the deletion of any part of the diagram is not anymore a link.\\

\begin{tiny}
\begin{figure}[hc]
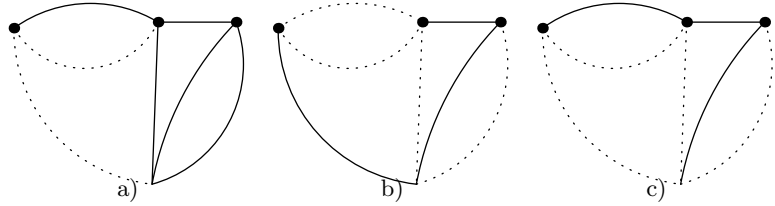

\centering
\begin{feynartspicture}(300,100)(3,1)
\FADiagram{a)}
\FAProp(0.5,14.)(12.5,14.5)(-0.3,){/Straight}{0}
\FAProp(12.5,14.5)(19.,14.5)(0.,){/Straight}{0}
\FAProp(0.5,14.)(12.5,14.5)(0.6,){/GhostDash}{0}
\FAProp(12.5,14.5)(12.,1.)(0.,){/Straight}{0}
\FAProp(12.,1.)(19.,14.5)(-0.1483,){/Straight}{0}
\FAProp(12.,1.)(19.,14.5)(0.4361,){/Straight}{0}
\FAProp(0.5,14.)(12.,1.)(0.3862,){/GhostDash}{0}
\FAVert(0.5,14.){0}
\FAVert(12.5,14.5){0}
\FAVert(19.,14.5){0}
\FADiagram{b)}
\FAProp(0.5,14.)(12.5,14.5)(-0.3,){/GhostDash}{0}
\FAProp(12.5,14.5)(19.,14.5)(0.,){/Straight}{0}
\FAProp(0.5,14.)(12.5,14.5)(0.6,){/GhostDash}{0}
\FAProp(12.5,14.5)(12.,1.)(0.,){/GhostDash}{0}
\FAProp(12.,1.)(19.,14.5)(-0.1483,){/Straight}{0}
\FAProp(12.,1.)(19.,14.5)(0.4361,){/GhostDash}{0}
\FAProp(0.5,14.)(12.,1.)(0.3862,){/Straight}{0}
\FAVert(0.5,14.){0}
\FAVert(12.5,14.5){0}
\FAVert(19.,14.5){0}
\FADiagram{c)}
\FAProp(0.5,14.)(12.5,14.5)(-0.3,){/Straight}{0}
\FAProp(12.5,14.5)(19.,14.5)(0.,){/Straight}{0}
\FAProp(0.5,14.)(12.5,14.5)(0.6,){/GhostDash}{0}
\FAProp(12.5,14.5)(12.,1.)(0.,){/GhostDash}{0}
\FAProp(12.,1.)(19.,14.5)(-0.1483,){/Straight}{0}
\FAProp(12.,1.)(19.,14.5)(0.4361,){/GhostDash}{0}
\FAProp(0.5,14.)(12.,1.)(0.3862,){/GhostDash}{0}
\FAVert(0.5,14.){0}
\FAVert(12.5,14.5){0}
\FAVert(19.,14.5){0}
\end{feynartspicture}
\caption{a) is a \textit{link} b) is a \textit{link} c) is not a link since the subdiagram is not minimal}
\end{figure}
\end{tiny}

\end{itemize}

\section{Construction of a singularity-family}
Let $G$ be an irreducible Feynman diagram with $I$ internal lines and $L$ loops. A {\large\textbf{(singularity)-family}} ${\cal F}$  in $G$ is a set of subgraphs of $G$, which are either links or irreducible subgraph of $G$, and which do not {\large \textbf{overlap}}, that is, either they have no lines  in common or are contained into one another. This set of subgraphs is also called a {\large\textbf{forest}}.

Since the diagram $G$ is a link, we include it as the first element of the family. Then we delete any line $l_1$ in $G$. Let us denote by $G/l_1$ the subdiagram of $G$, obtained deleting the line $l_1$ from $G$. $G/l_1$ contains all the external vertices of $G$, however in general it is not a link since it is not necessarily minimal; $G/l_1$ is the union of parts, some of which form a link. This link, that we call the second link of the forest and we denote by $L_{1,\cal{F}}$ ($L_{0,\cal{F}}$ coinciding with $G$) and the other parts are new elements of the singularity-family.  

Then we delete a further line $l_2$ in $L_{1,\cal{F}}$. 
If some parts of $L_{1,\cal{F}}/l_2$ form a link, $L_{2,\cal{F}}$,  we add $L_{2,\cal{F}}$ and the other irreducible parts of $L_{1,\cal{F}}/l_2$ to $\cal{F}$ and continue deleting a further line, $l_3$, in $L_{2,\cal{F}}$. We repeat the same procedure until the last link, $L_{k,\cal{F}}$ is broken by a further deletion. Let us notice that each of the first $k$ deletions has opened a loop in the corresponding link whereas the $k+1$-th deletion has broken $L_{k,{\cal F}}$ into two disconnected subdiagrams, each containing external vertices. At this stage the family is composed of $k+1$ links contained into one another and a set of trivial or non-trivial irreducible subdiagrams which do not intersect, that is, which have no lines in common. A {\large\textbf{trivial}} subgraph contains a single line.  We choose an arbitrary non-trivial irreducible element of ${\cal F}$ and delete one of its lines, $l_{\c}$, adding the parts of $\c/l_{\c}$ to ${\cal F}$. This procedure is repeated considering at each step the set of non-trivial irreducible elements of ${\cal F}$, which are minimal, i.e., which do not contain other elements of ${\cal F}$, choosing an arbitrary element of this set and deleting a line in it. 

The procedure stops when one is left with only trivial elements. Now the construction  of the family is complete. Since the deletions of the first set  have opened $k$ loops contained in the links $L_{i,{\cal F}}$, whereas each deletion of the second set has opened a loop in a minimal non-trivial irreducible element of ${\cal F}$ and the sequence of deletions stops when there are no unbroken loops left, the number of deletions is equal to $L+1$.

Furthermore in every non-trivial element of ${\cal F}$ we have selected a line. The set of selected lines corresponds to a complete cut. It is apparent from the construction procedure that the unselected lines belong to the trivial elements of the family and are identified with them. Therefore the total number of trivial and non-trivial elements of ${\cal F}$ is $I$, the number of lines in $G$. There is a one-to-one correspondence between the elements of ${\cal F}$ and the lines in $G$, $l_{\gamma,{\cal F}}$. 

What is worth noticing is the fact that the sequence of deleted lines identifies a partially ordered complete cut in the diagram and hence the set of undeleted lines identifies a 2-tree. Indeed the first set of deletions including the one which breaks the last link, the $k+1$-th deletion, forms a cut which separates the set of external vertices in two subsets. Let's call it the {\large \textbf{breaking cut}} $C_{{\cal F}}^s$, corresponding to the family ${\cal F}$. This is completely ordered. The number of lines contained in the breaking cut equals the number of the links in ${\cal F}$. However the cut is not complete; indeed non-trivial irreducible subdiagrams remain after the considered deletions.\\

\begin{small}
We show an example of construction of a family for the five loop diagram of the figure: 
\begin{center}
\begin{scriptsize}
\begin{feynartspicture}(150,100)(1,1)
\FADiagram{$\Gamma$}
\FAProp(0.,20.)(10.,20.)(0.,){/Straight}{0}
\FALabel(5.,19.18)[t]{5}
\FAProp(10.,20.)(20.,20.)(0.,){/Straight}{0}
\FALabel(15.,19.18)[t]{4}
\FAProp(20.,20.)(20.,10.)(0.,){/Straight}{0}
\FALabel(19.18,15.)[r]{7}
\FAProp(20.,10.)(20.,0.)(0.,){/Straight}{0}
\FALabel(19.18,5.)[r]{6}
\FAProp(20.,0.)(10.,0.)(0.,){/Straight}{0}
\FALabel(15.,0.82)[b]{8}
\FAProp(10.,0.)(0.,-0.)(0.,){/Straight}{0}
\FALabel(5.,0.82)[b]{10}
\FAProp(0.,-0.)(-0.,10.)(0.,){/Straight}{0}
\FALabel(0.82,5.)[l]{1}
\FAProp(-0.,10.)(0.,20.)(0.,){/Straight}{0}
\FALabel(0.82,15.)[l]{11}
\FAProp(-0.,10.)(10.,20.)(0.,){/Straight}{0}
\FALabel(5.52,14.98)[tl]{12}
\FAProp(10.,20.)(20.,10.)(0.,){/Straight}{0}
\FALabel(14.48,14.98)[tr]{3}
\FAProp(-0.,10.)(10.,0.)(0.,){/Straight}{0}
\FALabel(5.52,5.02)[bl]{2}
\FAProp(10.,0.5)(20.,10.)(0.,){/Straight}{0}
\FALabel(14.48,5.52)[br]{9}
\FAVert(20.,20.){0}
\FAVert(20.,0.){0}
\FAVert(0.,-0.){0}
\FAVert(0.,20.){0}
\label{boxinbox}
\end{feynartspicture}
\end{scriptsize}
\end{center}
If one deletes in  order the lines $1$, $2$ and $3$ in $\Gamma$, the first element, three links are formed: $L(2,3,4,5,6,7,8,9,10,11,12)$,  $L(3,4,5,6,7,8,9,10,11,12)$,  and $L(4,5,6,7,8,9,10,11,12)$.  Deleting the line $4$, there are not links anymore; so one includes in the family the parts of the subdiagram: $H(7),\ H(10),\ H(5,11,12),\ \text{and}\ H(6,8,9)$. Deleting the lines $5$ and $6$ in the not trivial parts, one finally has $H(8),H(9), H(11),H(12)$ and the procedure stops with a family of $12$ elements:
\bea
&&\Gamma \oplus L(\Gamma/ \{1\})\oplus L(\Gamma/ \{1,2\})\oplus L(\Gamma/ \{1,2,3\})\oplus H(7)\oplus H(10)\oplus\\
&& \oplus H(5,11,12)\oplus H(6,8,9)\oplus H(8)\oplus H(9)\oplus H(11)\oplus H(12)\nonumber
\label{fam}
\eea

The 2-tree associated with this family is formed by the lines of the trivial parts of the families, that is, $\{7,8,9,10,11,12\}$ and the breaking cut is $\{1,2,3,4\}$.
\end{small}\\

In summary we have met three types of cuts:
\begin{itemize}
\item The \textit{complete cut}, that is the set of $L+1$ not ordered deleted lines, is the complement in a diagram $G$ of a 2-tree; each of two tree subdiagrams contains at least an external vertex.
\item The \textit{reduced cut} is the complement of a maximal tree in a diagram $G$ and contains $L$ not ordered lines.
\item The \textit{breaking cut}, corresponding to a family ${\cal F}$, is an ordered set of lines, whose number depends on the singularity family.
\end{itemize}

Given a family ${\cal F}$, generated as described above, one has a corresponding complete cut $C_{\cal F}$, which is identified with the set of deleted lines.

We want to show that, given any family ${\cal F}$ and any complete cut $C$, $C$ intersects all the non-trivial elements of ${\cal F}$; more precisely, for every $\c$, non-trivial element of ${\cal F}$, there is a \textit{distinct} line in $C$, which is contained in $\c$, meaning that one can set a one-to-one correspondence between the $L+1$ lines in $C$ and the non-trivial elements of ${\cal F}$.

In order to prove this, let us notice that, given the set of $L+1$ non-trivial elements $\c$ of ${\cal F}$, it is possible to identify, in general not uniquely, $L$ independent loops in $G$, asking that every loop is contained in a $\c$. This is possible since the elements $\c$ do not overlap. The complete cut $C$ contains at least a line in every loop, otherwise it should not be a complete cut; however there is a one more line in $C$ and one more non-trivial element in ${\cal F}$ than there are loops in $G$. One has to select a line $\widehat{l}_k$ in $C$ which is contained in the smaller link $L_k$ and such that $C/\hat{l}_k$ is a reduced cut. This is always possible at least in one way. Indeed, first, the cut $C$ intersects $L_k$ since it separates the external vertices, while $L_k$ joins them.  Second, at least for one line $\widehat{l}_k$ in the intersection of $C$ with $L_k$, $C/\widehat{l}_k$ is a reduced cut. Indeed $T_{2,C}$, the complement of $C$,   separates the external vertices while $L_k$ joins them: this means that there is at least one line in $C\cap L_k$, which added to $T_2$ gives a $T\in {\cal J}_T$, since this line joins two disjoint subgraphs of $T_2$.

Once the reduced cut $C/\widehat{l}_k$ is extracted from $C$ and $\widehat{l}_k$ is associated with $L_k$, which contains it, $C/\widehat{l}_k$ cuts all the loops and hence it contains one distinct line for every loop. Since there is a correspondence between loops and non-trivial elements of ${\cal F}$ containing them, we have identified for every non-trivial element of ${\cal F}$ a distinct line of $C$ contained in it.\\
An example can make this argument clearer.\\


 


\small
Let's consider the diagram $G$ of figure $(\ref{4loop})$ and two different sets of non-trivial elements contained in two singularity-families, ${\cal F}_a$ and ${\cal F}_b$, related to the diagram $G$.

Deleting  the lines $\{4,6,2,3\}$ respectively one obtains the four links shown in figure $(\ref{3a})$. 

\begin{tiny}
\begin{figure}[hc]
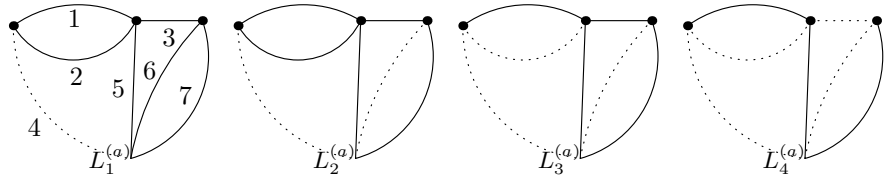

\centering
\begin{feynartspicture}(340,100)(4,1)
\FADiagram{$L_1^{(a)}$}
\FAProp(0.5,14.)(12.5,14.5)(-0.3,){/Straight}{0}
\FALabel(6.4666,15.5308)[t]{1}
\FAProp(12.5,14.5)(19.,14.5)(0.,){/Straight}{0}
\FALabel(15.75,13.68)[t]{3}
\FAProp(0.5,14.)(12.5,14.5)(0.6,){/Straight}{0}
\FALabel(6.7041,9.8311)[t]{2}
\FAProp(12.5,14.5)(12.,1.)(0.,){/Straight}{0}
\FALabel(11.4308,7.7981)[r]{5}
\FAProp(12.,1.)(19.,14.5)(-0.1483,){/Straight}{0}
\FALabel(14.5315,8.7766)[br]{6}
\FAProp(12.,1.)(19.,14.5)(0.4361,){/Straight}{0}
\FALabel(17.9986,6.3328)[br]{7}
\FAProp(0.5,14.)(12.,1.)(0.3862,){/GhostDash}{0}
\FALabel(3.2454,4.8975)[tr]{4}
\FAVert(0.5,14.){0}
\FAVert(12.5,14.5){0}
\FAVert(19.,14.5){0}
\FADiagram{$L_2^{(a)}$}
\FAProp(0.5,14.)(12.5,14.5)(-0.3,){/Straight}{0}
\FAProp(12.5,14.5)(19.,14.5)(0.,){/Straight}{0}
\FAProp(0.5,14.)(12.5,14.5)(0.6,){/Straight}{0}
\FAProp(12.5,14.5)(12.,1.)(0.,){/Straight}{0}
\FAProp(12.,1.)(19.,14.5)(-0.1483,){/GhostDash}{0}
\FAProp(12.,1.)(19.,14.5)(0.4361,){/Straight}{0}
\FAProp(0.5,14.)(12.,1.)(0.3862,){/GhostDash}{0}
\FAVert(0.5,14.){0}
\FAVert(12.5,14.5){0}
\FAVert(19.,14.5){0}
\FADiagram{$L_3^{(a)}$}
\FAProp(0.5,14.)(12.5,14.5)(-0.3,){/Straight}{0}
\FAProp(12.5,14.5)(19.,14.5)(0.,){/Straight}{0}
\FAProp(0.5,14.)(12.5,14.5)(0.6,){/GhostDash}{0}
\FAProp(12.5,14.5)(12.,1.)(0.,){/Straight}{0}
\FAProp(12.,1.)(19.,14.5)(-0.1483,){/GhostDash}{0}
\FAProp(12.,1.)(19.,14.5)(0.4361,){/Straight}{0}
\FAProp(0.5,14.)(12.,1.)(0.3862,){/GhostDash}{0}
\FAVert(0.5,14.){0}
\FAVert(12.5,14.5){0}
\FAVert(19.,14.5){0}
\FADiagram{$L_4^{(a)}$}
\FAProp(0.5,14.)(12.5,14.5)(-0.3,){/Straight}{0}
\FAProp(12.5,14.5)(19.,14.5)(0.,){/GhostDash}{0}
\FAProp(0.5,14.)(12.5,14.5)(0.6,){/GhostDash}{0}
\FAProp(12.5,14.5)(12.,1.)(0.,){/Straight}{0}
\FAProp(12.,1.)(19.,14.5)(-0.1483,){/GhostDash}{0}
\FAProp(12.,1.)(19.,14.5)(0.4361,){/Straight}{0}
\FAProp(0.5,14.)(12.,1.)(0.3862,){/GhostDash}{0}
\FAVert(0.5,14.){0}
\FAVert(12.5,14.5){0}
\FAVert(19.,14.5){0}
\end{feynartspicture}
\caption{{\scriptsize These are, together with the diagram itself, the non-trivial elements of a family ${\cal F}_a$, produced by the first $L=4$ choices. In this case they are all links. The remaining two elements of the family, which are trivial, are obtained by deletion of any line of the last link.}}
\label{3a}
\end{figure}
\end{tiny}
We see that each of the above deletions produces a link. So one has as many links as there are loops, if we don't consider $G$. In order to complete the family one has to choose a line between the ones of the last link $L_4^{(a)}$, for example line $5$, so that the diagram breaks  into two trivial parts. In this example the breaking cut coincides with the complete cut $C_{{\cal F}_a}$: $\{4,6,2,3,5\}$

Alternatively one can delete from the diagram $G$ the lines $\{4,5,3\}$. One has the non-trivial elements in figure $(\ref{3b})$, showing links together with irreducible diagrams.

\begin{tiny}
\begin{figure}[hhhh]
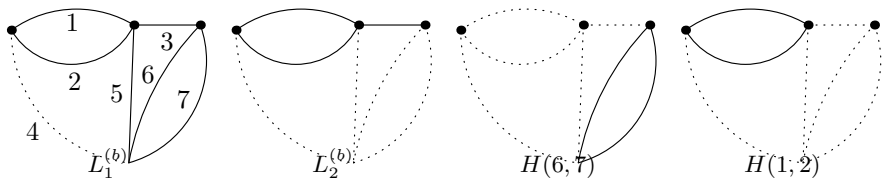

\centering
\begin{feynartspicture}(340,100)(4,1)
\FADiagram{$L_1^{(b)}$}
\FAProp(0.5,14.)(12.5,14.5)(-0.3,){/Straight}{0}
\FALabel(6.4666,15.5308)[t]{1}
\FAProp(12.5,14.5)(19.,14.5)(0.,){/Straight}{0}
\FALabel(15.75,13.68)[t]{3}
\FAProp(0.5,14.)(12.5,14.5)(0.6,){/Straight}{0}
\FALabel(6.7041,9.8311)[t]{2}
\FAProp(12.5,14.5)(12.,1.)(0.,){/Straight}{0}
\FALabel(11.4308,7.7981)[r]{5}
\FAProp(12.,1.)(19.,14.5)(-0.1483,){/Straight}{0}
\FALabel(14.5315,8.7766)[br]{6}
\FAProp(12.,1.)(19.,14.5)(0.4361,){/Straight}{0}
\FALabel(17.9986,6.3328)[br]{7}
\FAProp(0.5,14.)(12.,1.)(0.3862,){/GhostDash}{0}
\FALabel(3.2454,4.8975)[tr]{4}
\FAVert(0.5,14.){0}
\FAVert(12.5,14.5){0}
\FAVert(19.,14.5){0}
\FADiagram{$L_2^{(b)}$}
\FAProp(0.5,14.)(12.5,14.5)(-0.3,){/Straight}{0}
\FAProp(12.5,14.5)(19.,14.5)(0.,){/Straight}{0}
\FAProp(0.5,14.)(12.5,14.5)(0.6,){/Straight}{0}
\FAProp(12.5,14.5)(12.,1.)(0.,){/GhostDash}{0}
\FAProp(12.,1.)(19.,14.5)(-0.1483,){/GhostDash}{0}
\FAProp(12.,1.)(19.,14.5)(0.4361,){/GhostDash}{0}
\FAProp(0.5,14.)(12.,1.)(0.3862,){/GhostDash}{0}
\FAVert(0.5,14.){0}
\FAVert(12.5,14.5){0}
\FAVert(19.,14.5){0}
\FADiagram{$H(6,7)$}
\FAProp(0.5,14.)(12.5,14.5)(-0.3,){/GhostDash}{0}
\FAProp(12.5,14.5)(19.,14.5)(0.,){/GhostDash}{0}
\FAProp(0.5,14.)(12.5,14.5)(0.6,){/GhostDash}{0}
\FAProp(12.5,14.5)(12.,1.)(0.,){/GhostDash}{0}
\FAProp(12.,1.)(19.,14.5)(-0.1483,){/Straight}{0}
\FAProp(12.,1.)(19.,14.5)(0.4361,){/Straight}{0}
\FAProp(0.5,14.)(12.,1.)(0.3862,){/GhostDash}{0}
\FAVert(0.5,14.){0}
\FAVert(12.5,14.5){0}
\FAVert(19.,14.5){0}
\FADiagram{$H(1,2)$}
\FAProp(0.5,14.)(12.5,14.5)(-0.3,){/Straight}{0}
\FAProp(12.5,14.5)(19.,14.5)(0.,){/GhostDash}{0}
\FAProp(0.5,14.)(12.5,14.5)(0.6,){/Straight}{0}
\FAProp(12.5,14.5)(12.,1.)(0.,){/GhostDash}{0}
\FAProp(12.,1.)(19.,14.5)(-0.1483,){/GhostDash}{0}
\FAProp(12.,1.)(19.,14.5)(0.4361,){/GhostDash}{0}
\FAProp(0.5,14.)(12.,1.)(0.3862,){/GhostDash}{0}
\FAVert(0.5,14.){0}
\FAVert(12.5,14.5){0}
\FAVert(19.,14.5){0}
\end{feynartspicture}
\caption{{\scriptsize These are the non-trivial elements of the family ${\cal F}_b$. The deletion of the lines $\{4,5\}$ produces two links. From the second deletion one has also the part $H(6,7)$. From the choice of the line $3$ one has another irreducible element, $H(1,2)$.}}
\label{3b}
\end{figure}
\end{tiny}
The breaking cut contains the lines $\{4,5,3\}$, since the choice of line $3$ breaks the last link. In order to complete the family one has to delete one of the lines $\{1,2\}$ of $H(1,2)$ and one of the lines in $H(6,7)$. 

Now we verify that it is possible to create a correspondence (not unique) between $L$ loops and $L$ non-trivial elements of a family, by selecting in every non-trivial element an independent loop. Indeed for the family ${\cal F}_a$ one can establish for example this correspondence
\beq
G\leftrightarrow \{2,4,5\};\ L_1^{a} \leftrightarrow\{6,7\};\ L_2^{a}\leftrightarrow\{1,2\};\ L_3^{a}\leftrightarrow\{3,5,7\} 
\label{L1}
\eeq
and for the family ${\cal F}_b$:
\beq
G\leftrightarrow \{1,4,5\};\ L_1^{b} \leftrightarrow\{3,5,7\};\ H(1,2)\leftrightarrow\{1,2\};\ H(6,7)\leftrightarrow\{6,7\}. 
\label{L2}
\eeq
In this way we have created a correspondence between  $4$ loops and $4$ non-trivial elements of the two families.

Now we consider a complete cut $C=\{1,3,4,5,7\}$. It contains $5>4$ lines. As explained above, we first construct from $C$ two reduced cuts  related to the two families. 
One has to take the last link of the family and consider the lines $\widehat{l}$, which are contained in the intersection of the last link  and the cut. For the two families one has
\beq
\begin{array}{rl}
 & L^{a}_4\cap C=\{1,5,7\}\\
 & L^{b}_2\cap C=\{1,3\}.
\end{array}
\label{set}
\eeq
In case $a$ one has to exclude both lines $1$ and $7$ since neither $\{3,4,5,7\}$ nor $\{1,3,4,5\}$ are reduced cuts; their complements contain loops. Hence one must choose $\widehat{l}_a=5$. In case $b$ one must choose $\widehat{l}_a=3$.  

Therefore, given $C$, in this example one has only one reduced cut in both families:
\beq
\begin{array}{rl}
 &\hat{C}^a=C/\widehat{l}^a=\{1,3,4,7\}\\
 &\hat{C}^b=C/\widehat{l}^b=\{1,4,5,7\}.
\end{array}
\eeq
Now, taking into account of $(\ref{L1})$ and $({\ref{L2}})$, we construct one of the possible correspondences between the lines of the complete cut and the non-trivial elements of the family:
\beq
\begin{array}{rl}
 &G\leftrightarrow 4;\ L_1^{a} \leftrightarrow 7;\ L_2^{a}\leftrightarrow 1;\ L_3^{a}\leftrightarrow 3;\ L_4^{a}\leftrightarrow 5\\
 &G\leftrightarrow 4;\ L_1^{b} \leftrightarrow 5;\ L_2^{b} \leftrightarrow 3;\ H(1,2)\leftrightarrow 1;\ H(6,7)\leftrightarrow 7,
\end{array}
\eeq
where we have associated $\widehat{l}^a_4=5$ with the last link of the family ${\cal F}_a$ and $\widehat{l}^b_2=3$ with the last link of the family ${\cal F}_b$.
This shows how we create a correspondence between distinct lines of a complete cut and  non-trivial elements of a family.
\normalsize
\section{Speer-Smirnov sectors}
The crucial step in our program is the identification of the decomposition of the positive sector of the projective parameter space seen in the previous chapter into subsectors, each associated with a particular parametrization of the projective space. 

This is accomplished referring to the singularity families ${\cal F}$. We have seen that a singularity family is a partially ordered set of non-overlapping subdiagrams of $G$ including $G$ itself and that there is a corresponding complete cut $C_{{\cal F}}$ which singles out a line in each not trivial element of the family. 

The partial order of the elements of ${\cal F}$ is determined by their inclusion properties: every subdiagram precedes in the order the elements of ${\cal F}$ contained in it. 
With each element $\gamma$ of ${\cal F}$ we have associated a unique line $l_{\gamma,{\cal F}}$ which is contained in $\gamma$ and in no element of ${\cal F}$ contained in $\gamma$. If $\c$ is non-trivial this is a line of the complete cut associated with the family, $C_{{\cal F}}$. The set of the lines associated with the links, which are elements of the breaking cut, have a natural complete order since the links are contained into one another. The lines associated with other irreducible elements are partially ordered, as already said, and they are ordered after those of the link. 

Therefore the partial order in ${\cal F}$ corresponds to a partial order in the complete cut, which consists in the ordered lines of the links and in the partial ordered lines of the non-trivial irreducible elements. Since every line of $G$ is associated with a particular element of ${\cal F}$,  the family contains as many elements as there are lines in $G$, as we have said. 

Notice that the lines of the 2-tree, which is the complement in $G$ of $C_{{\cal F}}$, are elements of the trivial subdiagrams in ${\cal F}$. Having recalled these points, we come to the parametrization associated with a family ${\cal F}$. One associates a parameter $t_{\c}$, which takes values between $0$ and $1$, with every element $\c$ di ${\cal F}$, while one sets $t_{G}$ equal to $1$.

The parameter $\b_l$, appearing in the scalar amplitude $(\ref{princ})$ and in the generic one $(\ref{princ2})$, is set equal to the product of the $t_{\c}$ corresponding to the elements of ${\cal F}$ containing the line $l$. That is one sets:
\beq
\b_l=\prod_{\c\in{\cal F},l\in \c}t_{\c}.
\label{Se}
\eeq
The range of values taken by the $\b$'s after this prescription defines a Speer sector. It is apparent that, due to the range $0\leq t_{\gamma}\leq 1$, the partial ordering of the elements of ${\cal F}$ corresponds to a partial ordering of the $\b$'s.  
In a given non-trivial element of the family ${\cal F}$ the line with the largest $\b_l$ parameter is the line $l_{\c,{\cal F}}$.\\


\begin{small}
Coming back to the first example, we see that the  Speer sector corresponding to the family $(\ref{fam})$ is:
\beq
\a_1>\a_2>\a_3>\a_4>\left\{
\begin{array}{ll}
&\a_5>\a_{11},\,\a_{12}\\
&\a_6>\a_8,\,\a_9\\
&\a_7,\,\a_{10}
\end{array}\right.
\eeq

The biggest parameter is assigned to line $1$: $\a_1=t_{\Gamma}=1$. Calling $t_i$ the variable corresponding to the $i$-th element of the family, starting from $L(\Gamma/\{1\})$, the parameters have the following explicit form:
\bea
&&\a_1=t_{\Gamma}=1,\a_2=t_1,\,\a_3=t_1t_2,\,\a_4=t_1t_2t_3,\,\a_5=t_1t_2t_3t_6,\,\a_6=t_1t_2t_3t_7,\qquad\nonumber\\
&&\a_7=t_1t_2t_3t_4,\,\a_{10}=t_1t_2t_3t_5,\,\a_{11}=t_1t_2t_3t_6t_{10},\,\a_{12}=t_1t_2t_3t_6t_{11},\\
&& \a_8=t_1t_2t_3t_7t_8,\,\a_9=t_1t_2t_3t_7t_9.
\nonumber
\eea
\end{small}

It is not difficult to verify that different sectors do not intersect, since different families correspond to different sectors. Indeed considering the family construction procedure, there is necessarily an irreducible diagram, e.g. $G$ itself, with which different families associate different lines. This corresponds to a different ordering of the lines of the irreducible diagram.  

Furthermore one verifies that the union of all the sectors covers the whole positive sector of the projective space. 

In order to show this second point, let us consider any ordered choice of the Schwinger parameters 
$ \a_{i_1}>\a_{i_2}>...>\a_{i_I}>0 $. Let us identify $\a_{i_1}$ with the scale parameter $t$, and hence choose:
\beq \b_{i_1}=1>\b_{i_2}>...>\b_{i_I} .\eeq We have to show that this corresponds to a Speer-Smirnov sector.

From the construction procedure it is clear that $l_{i_1}$ is $l_G$. The situation with $l_{i_2}$ is less simple; indeed $l_{i_2}$ could either belong to a link contained in $G/l_{i_1}$ or to a non-trivial irreducible part of $G/l_1$ or else to a trivial irreducible part. In any case $l_{i_2}$ is identified with the line of either the link or the irreducible part.

The above chosen order is compatible with this identification. Let us now consider $l_{i_3}$; once again it must be contained in one of the minimal elements introduced into the family after deletion of lines $l_{i_1}$ and $l_{i_2}$, let $l_{i_3}$ be the line associated with this element; let us furthermore reduce what remains of this element after deletion of $l_{i_3}$ in its irreducible parts, we can now consider $l_{i_4}$ repeating the same procedure. Once reached $l_{i_I}$, we have identified a family ${\cal F}_{i1,i2,..,i_I}$, and hence a Speer-Smirnov sector, compatible with the considered order of parameters. This proves that the sector decomposition exhausts the whole parameter space.\\

In order to make the construction clear we discuss a particular example. One can then verify our claims concerning the disjointness and completeness of sector decomposition.
\subsection*{Example of construction of the complete set of singularity-families for a two loop diagram}

\begin{figure}[ht]
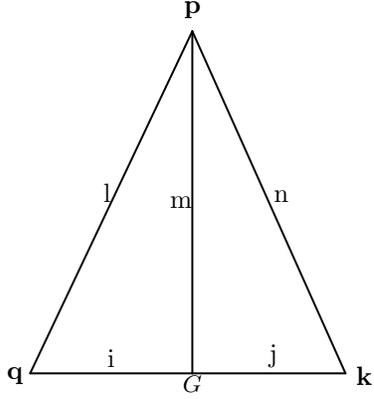

\begin{center}
\begin{feynartspicture}(150,150)(1,1)
\FADiagram{$G$}
\FAProp(1.,0.5)(10.,19.5)(0.,){/Straight}{0}
\FALabel(5.48,10.02)[br]{l}
\FAProp(10.,19.5)(18.5,0.5)(0.,){/Straight}{0}
\FALabel(14.52,10.02)[bl]{n}
\FAProp(10.,19.5)(10.,0.5)(0.,){/Straight}{0}
\FALabel(9.98,10.)[r]{m}
\FAProp(1.,0.5)(10.,0.5)(0.,){/Straight}{0}
\FALabel(5.5,0.85)[b]{i}
\FAProp(10.,0.5)(18.5,0.5)(0.,){/Straight}{0}
\FALabel(14.5,0.85)[b]{j}
\FALabel(0.6,0.35)[r]{\textbf{q}}
\FALabel(10.,20.2)[b]{\textbf{p}}
\FALabel(19.1,0.37)[l]{\textbf{k}}
\end{feynartspicture}
\end{center}
\label{Stl}
\caption{Scalar two loop diagram with three external vertices and an internal one}
\end{figure}

Let's consider the two loop diagram of the figure with three external vertices and an internal one and let's build the entire set of singularity families following the procedure above. 

From the initial diagram every choice of the first line produces a link after its deletion. So the links which are formed are so many as the the lines of the diagram, which are five: 
\beq L(l,m,i,j),\ L(m,n,i,j),\ L(l,m,n,j),\ L(l,m,n,i)\ \text{and}\ L(l,n,i,j).\nonumber\eeq
 Let's start considering the first link $L(l,m,i,j)$. In order to obtain another link, one has to delete anyone of the lines except for $j$; in this way the links $L(m,i,j)$, $L(l,m,j)$ or $L(l,i,j)$ are formed. The construction stops choosing anyone of the lines of the link and one obtains different families with the same structure which we call of type "A". If otherwise one chooses line $j$ in the initial link what remains is a non-trivial irreducible subgraph, $H(l,m,i)$, which has to be included into the family. The procedure stops choosing one of the lines contained in $H(l,m,i)$. These families have a different structure, since they contain a not banal irreducible element. We call them of type "B".

Thanks to the symmetry of the diagram the families which contain the second link are obtained from the  above described ones exchanging the lines $i\longleftrightarrow j$ and $l\longleftrightarrow n$  .

From the third link $L(l,m,n,j)$ only the deletion of line $n$ produces families of type A, since the link $L(l,m,j)$ is formed. Otherwise from the deletion of line $l$, which produces $H(m,n,j)$, one has three families of type B. Finally deleting a further line, either $j$ or $m$, one adds to the family the new link $L(l,n)$ and a further trivial element, either $m$ or $j$. The procedure stops choosing  a line in $L(l,n)$. We call this set of families type "C". We distinguish this set from set $A$  since its elements contain a two line link.

The families which contain the fourth link are obtained for symmetry  from the ones which  contain the third one.

Finally from the fifth link one can alternatively delete line $l/n$ or line $i/j$. In the first case the link $L(n,i,j)/L(l,i,j)$ is formed and so one has families of the type $A$; in the second we have families of type $C$, since the deletion produces $L(l,n)\oplus j/i$.

In the end the number of the families is 54: 30 of type $A$, 12 of type $B$ and 12 of type $C$.

They are shown in the table $(\ref{tabella})$.

\begin{small}
\begin{table}[ht] \caption{Singularity families}  
\centering 
\begin{tabular}{llll} 
\hline\hline 
A & Singularity Family & &Singularity Family \\ [0.5ex] 
\hline\hline 
1&$L(l,m,i,j)\oplus L(m,i,j)\oplus m \oplus i$&16&$L(l,n,i,j)\oplus L(l,i,j)\oplus i \oplus j$\\
2&$L(l,m,i,j)\oplus L(m,i,j)\oplus m \oplus j$&17&$L(l,n,i,j)\oplus L(l,i,j)\oplus j\oplus l$\\
3&$L(l,m,i,j)\oplus L(m,i,j)\oplus i \oplus j$&18&$L(l,n,i,j)\oplus L(l,i,j)\oplus i \oplus l$\\ 
4&$L(l,m,i,j)\oplus L(l,i,j)\oplus l \oplus i$&19&$L(l,n,i,j)\oplus L(n,i,j)\oplus n \oplus i$\\
5&$L(l,m,i,j)\oplus L(l,i,j)\oplus l \oplus j$&20&$L(l,n,i,j)\oplus L(n,i,j)\oplus n \oplus j$\\
6&$L(l,m,i,j)\oplus L(l,i,j)\oplus i \oplus j$&21&$L(l,n,i,j)\oplus L(n,i,j)\oplus i \oplus j$\\
7&$L(l,m,i,j)\oplus L(l,m,j)\oplus l \oplus m$&22&$L(m,n,i,j)\oplus L(m,n,i)\oplus m \oplus n$\\
8&$L(l,m,i,j)\oplus L(l,m,j)\oplus l \oplus j$&23&$L(m,n,i,j)\oplus L(m,n,i)\oplus m \oplus i$\\
9&$L(l,m,i,j)\oplus L(l,m,j)\oplus m \oplus j$&24&$L(m,n,i,j)\oplus L(m,n,i)\oplus n \oplus i$\\
10&$L(l,m,n,j)\oplus L(l,m,j)\oplus l \oplus j$&25&$L(m,n,i,j)\oplus L(n,i,j)\oplus n \oplus i$\\
11&$L(l,m,n,j)\oplus L(l,m,j)\oplus l \oplus m$&26&$L(m,n,i,j)\oplus L(n,i,j)\oplus n \oplus j$\\
12&$L(l,m,n,j)\oplus L(l,m,j)\oplus m \oplus j$&27&$L(m,n,i,j)\oplus L(n,i,j)\oplus i \oplus j$\\
13&$L(l,m,n,i)\oplus L(m,n,i)\oplus n \oplus i$&28&$L(m,n,i,j)\oplus L(m,i,j)\oplus m \oplus i$\\
14&$L(l,m,n,i)\oplus L(m,n,i)\oplus m \oplus i$&29&$L(l,n,i,j)\oplus L(m,i,j)\oplus i \oplus j$\\
15&$L(l,m,n,i)\oplus L(m,n,i)\oplus m \oplus n$&30&$L(m,n,i,j)\oplus L(m,i,j)\oplus m \oplus j$\\[1ex]
\hline\hline 
B & Singularity Family & &Singularity Family  \\ [0.5ex] 
\hline\hline 
1&$L(m,n,i,j)\oplus H(m,n,j)\oplus m \oplus n$&7&$L(l,m,n,i)\oplus H(l,m,i)\oplus l \oplus m$\\
2&$L(m,n,i,j)\oplus H(m,n,j)\oplus m \oplus j$&8&$L(l,m,n,i)\oplus H(l,m,i)\oplus l \oplus i$\\
3&$L(m,n,i,j)\oplus H(m,n,j)\oplus n \oplus j$&9&$L(l,m,n,i)\oplus H(l,m,i)\oplus m \oplus i$\\
4&$L(l,m,n,j)\oplus H(m,n,j)\oplus m \oplus n$&10&$L(l,m,i,j)\oplus H(l,m,i)\oplus l \oplus m$\\
5&$L(l,m,n,j)\oplus H(m,n,j)\oplus m \oplus j$&11&$L(l,m,i,j)\oplus H(l,m,i)\oplus l \oplus i$\\
6&$L(l,m,n,j)\oplus H(m,n,j)\oplus n \oplus j$&12&$L(l,m,i,j)\oplus H(l,m,i)\oplus m \oplus i$\\ [1ex]
\hline\hline 
C & Singularity Family &  &Singularity Family \\ [0.5ex] 
\hline\hline 
1&$L(l,n,i,j)\oplus L(l,n)\oplus i\oplus n$&7&$L(l,m,n,j)\oplus L(l,n)\oplus j\oplus n$\\
2&$L(l,n,i,j)\oplus L(l,n)\oplus j\oplus n$&8&$L(l,m,n,j)\oplus L(l,n)\oplus m\oplus n$\\
3&$L(l,n,i,j)\oplus L(l,n)\oplus i\oplus l$&9&$L(l,m,n,i)\oplus L(l,n)\oplus m\oplus l$\\
4&$L(l,n,i,j)\oplus L(l,n)\oplus j\oplus l$&10&$L(l,m,n,i)\oplus L(l,n)\oplus i\oplus l$\\
5&$L(l,m,n,j)\oplus L(l,n)\oplus m\oplus l$&11&$L(l,m,n,i)\oplus L(l,n)\oplus i\oplus n$\\
6&$L(l,m,n,j)\oplus L(l,n)\oplus j\oplus l$&12&$L(l,m,n,i)\oplus L(l,n)\oplus m\oplus n$\\ [1ex]
\hline 
\end{tabular} 
\label{tabella} 
\caption{All the singularity families related to the diagram in figure $(3.5)$ are shown. Every family contains also the diagram itself, which in this table is omitted.}
\end{table}
\end{small}
\chapter{The Symanzik functions in the Speer-Smirnov sectors}
\textit{In this chapter we will show that the Speer-Smirnov sector division plays an important role in the study of the asymptotic behaviour of the amplitudes.  In every sector a power counting is possible for the study of the infrared divergences. This is shown for a generic theory.}

\textit{In order to evaluate exactly the divergent part of the amplitude, we will use the Mellin-Barnes transform, which will be described in the following chapter.}

\section{Symanzik functions in a Speer-Smirnov sector}
We have seen in chaper $2$ the  expression $(\ref{princ2})$ of the amplitude $\tilde{A_G}(P)$ in Schwinger parameter representation, given by a sum over the indices $a,b,\tau$. 

We want to focus on a single term of the sum, in particular we recall 
\beq
I_{G,(a,b,\tau)}(P)=\mu^{2L\eps}\,\Gamma(I-\frac{Ld}{2}-a)\int \frac{d\mu(\b)\,\prod_{i=1}^I \b_i^{\lambda_{i,\tau}}}{P_G(\b)^{\frac{d(L+1)}{2}-I+b}D_G(\b,P)^{I-\frac{Ld}{2}-a}}
\label{st}
\eeq 
and express it in the Speer-Smirnov parameter representation. 

In this representation $I_{G,(a,b,\tau)}(P)$ decomposes into the sum of different contributions $I_{G,(a,b,\tau)}^S(P)$, each of which is evaluated in a Speer-Smirnov sector:
\beq
I_{G,(a,b,\tau)}(P)=\sum_S I^S_{G,(a,b,\tau)}(P)
\eeq
with
\beq I^S_{G,(a,b,\tau)}(P)=\mu^{2L\eps}\,\Gamma(I-\frac{Ld}{2}-a)\int \frac{d\mu_S\,\prod_{i=1}^I \b_i^{\lambda_{i,\tau}}}{P_S^{\frac{d(L+1)}{2}-I+b}D_S^{I-\frac{Ld}{2}-a}}.
\label{I_Sb}
\eeq
In the last expression the measure $d\mu_S$ and the Symanzik functions $P_S$ and $D_S$ are labelled by $S$, since they are different in different sectors.


From $(\ref{Se})$, specifying the sector parametrization, we can deduce the corresponding integration measure of the sector amplitude, which is 
\beq
d\mu_S=\prod_{l=2}^I d\b_{i_l}=\prod_{\substack{\c\in{\cal F}\\ \c\neq G }}t_{\c}^{I(\c)-1}\,dt_{\c},
\label{mu}
\eeq
where the product runs on all the elements of the family, except for the diagram $G$ itself,  and $I(\c)$ is the number of lines contained in the element $\c$ of the family.

From the relation which defines the Symanzik functions $(\ref{pgc})$ and $(\ref{dgc})$, changing the variables from $\b$ to $t_{\gamma}$ according to the definition of the sector, one has the general expressions for the Symanzik functions in a sector:   
\bea
P_S&=&\sum_{\hat{C}\in\hat{\cal C}}\prod_{\c\in{\cal F}}t_{\c}^{I(\c\cap \hat{C})};\nonumber\\
D_{S}&=&\sum_{C\in {\cal C}}P_{C}^2\prod_{\c\in{\cal F}}t_{\c}^{I(\c\cap C)},
\label{PsDs}
\eea
where we recall that $C$ is a complete cut, $\hat{C}$ is the reduced cut, which contains $L$ lines (one line less than the complete cut), $I(\c\cap \hat{C})$ is the number of the lines in the intersection between an element of the family $\c$ and a cut $\hat{C}$ and
$I(\c\cap C)$ is the number of the lines in the intersection between an element of the family $\c$ and a complete cut $C$.
\\ 

In the previous chapter we have seen that a singularity family ${\cal F}$, and therefore a Speer-Smirnov sector, identifies a particular complete cut $C_{{\cal F}}$. It is associated with the kinematic invariant $P_{C_{{\cal F}}}^2$, which is the square momentum crossing the cut.


Therefore it follows that 
\beq
D_{S}=\prod_{\c\in{\cal F}}t_{\c}^{I(\c\cap C_{{\cal F}})}\(P_{C_{{\cal F}}}^2+\sum_{C\neq C_{\cal F}}P_{C}^2\prod_{\c\in{\cal F}}t_{\c}^{\d_{C,{\cal F}}(\c)}\)=\prod_{\c\in {\cal F}}t_{\c}^{I(\c\cap C_{{\cal F}})}\,\D_S(t_{\c}),
\label{D_S}
\eeq
where
\beq \d_{C,{\cal F}}(\c)=I(\c\cap C)-I(\c\cap C_{{\cal F}}).\eeq
It is important to notice that $\d_{C,{\cal F}}(\c)$ has integer non-negative values and the condition $\d_{C,{\cal F}}(\c)\equiv 0$ for every $\c\in {\cal F}$ identifies $C$ with $C_{{\cal F}}$.

In order to obtain a  similar relation for $P_S$, we have to introduce the ordered reduced cut $\hat{C_{{\cal F}}}$. $\hat{C_{{\cal F}}}$ is obtained omitting from $C_{{\cal F}}$ the line $\widehat{l}_{{\cal F}}$ associated with the smallest link in ${\cal F}$.


The expression for $P_S$ is:
\beq
P_S=\prod_{\c\in{\cal F}}t_{\c}^{I(\c\cap \hat{C_{{\cal F}}})}\(1+\sum_{\hat{C}\neq \hat{C}_{{\cal F}}}\prod_{\c\in{\cal F}}t_{\c}^{\d_{\hat{C},{\cal F}}(\c)}\)=\prod_{\c\in{\cal F}}t_{\c}^{I(\c\cap \hat{C_{{\cal F}}})}\,\Pi_S(t_{\c})
\label{P_S}
\eeq 
where
\beq \d_{\hat{C},{\cal F}}(\c)=I(\c\cap \hat{C})-I(\c\cap \hat{C}_{{\cal F}}),\eeq 
which has non-negative integer values and it is identically null in ${\cal F}$ if and only if $\hat{C}=\hat{C}_{{\cal F}}.$

At this point let's come back to the relation $(\ref{I_Sb})$ and express it in the new variables $t_{\gamma}$. Combining the above equations together we have:
\beq I^S_{a,b,\tau}(P)=\mu^{2L\eps}\,\Gamma(I-\frac{Ld}{2}-a)\int\prod_{\substack{\c\in{\cal F}\\ \c\neq G }} dt_{\c} \, \frac{\prod_{\c}t_{\c}^{{\cal E}(\c)_{a,b,\tau}-1}}{\Pi_S^{\frac{d(L+1)}{2}-I+b}\D_S^{I-\frac{Ld}{2}-a}}
\label{I_S}
\eeq
with
\beq
\begin{array}{rl}
{\cal E}(\c)_{a,b,\tau}=&I(\c)-I(\c\cap \hat{C_{{\cal F}}})(b-a+\frac{d}{2})+I(\hat{l}_k^*\cap \c)(L\frac{d}{2}-I+a)+\sum_{i\in\c}\lambda_{i,\tau}.
\end{array}
\label{e}
\eeq
In the last expression $I(\widehat{l}_{{\cal F}}\cap \c)=I(\c\cap C_{{\cal F}})-I(\c\cap\hat{C_{{\cal F}}})$ can either be $1$, if $\widehat{l}_{{\cal F}}$ is in $\c$ or null if $\c$ does not contain $\widehat{l}_{{\cal F}}$. In other words it depends on whether $\c$ is a link or not.\\

It is important to notice at this point that our formula $(\ref{I_S})$ holds true for  dimensionally regularized unsubtracted amplitudes. Concerning the UV-subtracted amplitudes, if the UV-divergence is primitive, there appear logarithms together with rational integrands. The case of amplitudes with UV-divergent subdiagrams in the sectors where these divergences appear is discussed by Smirnov in \cite{Smirnov:2008aw}.

We shall not consider anymore UV-divergent amplitudes since we shall discuss in this thesis a one loop example in which the $UV$-divergences are primitive whenever they appear and hence do not give any contribution to the collinear divergent amplitudes.

Therefore forgetting UV-divergences we notice that, if one considers the IR-safe theories mentioned in the introduction, that is, theories in which the interaction corresponds to operators with mass dimension strictly equal to $4$, such as QCD, computing the amplitudes with non-exceptional Euclidean external momenta, one has absolutely convergent integrals which are analytic in the kinematic invariants. Indeed this follows from the known IR-power counting theorems \cite{Speer:1975dc,Zimmermann:1975gm,Lowenstein:1974qt,*Lowenstein:1975rg,*Lowenstein:1975ps,*Lowenstein:1975rf}.

This result is very important for us since it guarantees that the exponents ${\cal E}(\c)_{a,b,\tau}$ in equation $(\ref{I_S})$ are positive and hence the measure
\beq
d\bar{\mu}^S_{a,b,\tau}=\prod_{\substack{\c\in{\cal F}\\ \c\neq G }} dt_{\c}\, \frac{\prod_{\c}t_{\c}^{{\cal E}(\c)_{a,b,\tau}-1}}{\Pi_S^{\frac{d(L+1)}{2}-I+b}}
\label{mub}
\eeq
is positive and integrable in the $I-1$ dimensional hypercube $0\leq t_{\c}\leq1$. Therefore we have for the sector amplitude the expression:
\beq I^S_{a,b,\tau}(P)=\int d\bar{\mu}^S_{a,b,\tau}\,\frac{\Gamma(I-\frac{Ld}{2}-a)}{\D_S^{I-\frac{Ld}{2}-a}}.
\label{I_Sa}
\eeq
From equations $(\ref{I_Sa})$ and $(\ref{D_S})$ we see that $I^S_{a,b,\tau}(P)$ is regular and analytic in the Euclidean region if $P_{C_{{\cal F}}}^2<0$, and hence $\Delta_S<0$. $\Delta_S$ vanishes in boundary points of the $I-1$-dimensional hypercube together with $P_{C_{{\cal F}}}^2$. These are points where some $t_{\c}$ vanish. This is the starting point of our singularity study.

However, to complete this study, we have to change  our point of view. Instead of looking at a given sector-(family), we must consider one, or more than one, kinematic invariants whose vanishing together characterizes the limit under study. We have seen that once the sector-(family) is given, this identifies a complete cut and hence a kinematic invariant coinciding with the square momentum crossing the cut. 

If we start from a set of kinematic invariants with the same (or possibly proportional) negative value $-\xi^2$, we have to identify the set of families ${\cal F}^{\xi}$, the corresponding sectors $S^{\xi}$  and complete cuts $C_{{\cal F}^{\xi}}$, such that the kinematic invariants $P_{C_{{\cal F}^{\xi}}}^2$  belong to the above set and hence are equal (proportional) to $-\xi^2$. We call ${\cal C}_{\xi}$ the set of the complete cuts associated with kinematic invariants vanishing together with $-\xi^2$.

With these definitions the singular part of the amplitude appears in the sum over the sectors $S^{\xi}$ of $I^{S^{\xi}}_{a,b,\tau}(P)$:
\beq
I^{\xi}_{a,b,\tau}(P)=\sum_{{\cal F}^{\xi}}\int d\bar{\mu}_{a,b,\tau}^{S^{\xi}}\frac{\Gamma(I-\frac{Ld}{2}-a)}{(-\xi^2\, N_{S^{\xi}}(t_{\c})+M_{S^{\xi}}(t_{\c}))^{I-\frac{Ld}{2}-a}},
\label{Ixi}
\eeq
where
\beq N_{S^{\xi}}(t_{\c})=1+\sum_{\substack{C\neq C_{{\cal F}^{\xi}}\\  C\in{\cal C}_{\xi}}}\prod_{\c\in {\cal F}^{\xi}}t_{\c}^{\d_{C,{\cal F}^{\xi}}(\c)} \label{N2} \eeq
is a positive polynomial
and 
\beq
M_{S^{\xi}}(t_{\c})=\sum_{ C\notin {\cal C}_{\xi}}P_C^2\prod_{\c\in {\cal F}^{\xi}}t_{\c}^{\d_{C,{\cal F}^{\xi}}(\c)}
\label{M}
\eeq
is a negative polynomial which in general vanishes in boundary points of the hypercube.

Our purpose is to analyze the behaviour of $I^{\xi}_{a,b,\tau}$ when $\xi^2\rightarrow 0$ and this can be done employing the Mellin-Barnes transform.

However one can study the limit of the amplitude in every sector before using the Mellin Barnes to verify the presence of the divergence sector by sector. On this purpose we study the example of the two loop diagram, whose singularity families have been derived in the previous chapter. 

\subsection{An example}
Suppose we are interested in the limit $k^2\rightarrow 0$ of the amplitude corresponding to the diagram $(3.5)$. 
We consider the other kinematic invariants of the same order of magnitude: $q^2=p^2=\omega^2$.
The four complete cuts, through which $k$ flows, are:
\beq \{n, j,  i\},\,  \{n, j,  m\},\,  \{n, j,  l\},\,  \{n, m,  i\}\ 
\nonumber
\eeq 
\begin{figure}[ht]
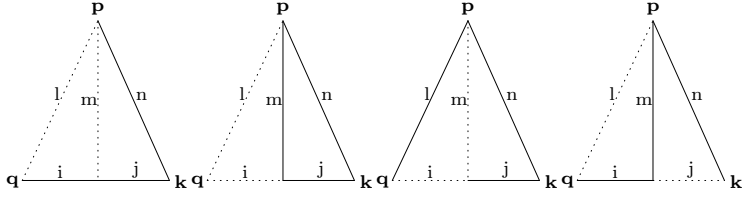

\begin{scriptsize}
\begin{center}
\begin{feynartspicture}(300,70)(4,1)
\FADiagram{}
\FAProp(1.,0.5)(10.,19.5)(0.,){/GhostDash}{0}
\FALabel(5.48,10.02)[br]{l}
\FAProp(10.,19.5)(18.5,0.5)(0.,){/Straight}{0}
\FALabel(14.52,10.02)[bl]{n}
\FAProp(10.,19.5)(10.,0.5)(0.,){/GhostDash}{0}
\FALabel(9.98,10.)[r]{m}
\FAProp(1.,0.5)(10.,0.5)(0.,){/Straight}{0}
\FALabel(5.5,0.85)[b]{i}
\FAProp(10.,0.5)(18.5,0.5)(0.,){/Straight}{0}
\FALabel(14.5,0.85)[b]{j}
\FALabel(0.6,0.35)[r]{\textbf{q}}
\FALabel(10.,20.2)[b]{\textbf{p}}
\FALabel(19.1,0.37)[l]{\textbf{k}}
\FADiagram{}
\FAProp(1.,0.5)(10.,19.5)(0.,){/GhostDash}{0}
\FALabel(5.48,10.02)[br]{l}
\FAProp(10.,19.5)(18.5,0.5)(0.,){/Straight}{0}
\FALabel(14.52,10.02)[bl]{n}
\FAProp(10.,19.5)(10.,0.5)(0.,){/Straight}{0}
\FALabel(9.98,10.)[r]{m}
\FAProp(1.,0.5)(10.,0.5)(0.,){/GhostDash}{0}
\FALabel(5.5,0.85)[b]{i}
\FAProp(10.,0.5)(18.5,0.5)(0.,){/Straight}{0}
\FALabel(14.5,0.85)[b]{j}
\FALabel(0.6,0.35)[r]{\textbf{q}}
\FALabel(10.,20.2)[b]{\textbf{p}}
\FALabel(19.1,0.37)[l]{\textbf{k}}
\FADiagram{}
\FAProp(1.,0.5)(10.,19.5)(0.,){/Straight}{0}
\FALabel(5.48,10.02)[br]{l}
\FAProp(10.,19.5)(18.5,0.5)(0.,){/Straight}{0}
\FALabel(14.52,10.02)[bl]{n}
\FAProp(10.,19.5)(10.,0.5)(0.,){/GhostDash}{0}
\FALabel(9.98,10.)[r]{m}
\FAProp(1.,0.5)(10.,0.5)(0.,){/GhostDash}{0}
\FALabel(5.5,0.85)[b]{i}
\FAProp(10.,0.5)(18.5,0.5)(0.,){/Straight}{0}
\FALabel(14.5,0.85)[b]{j}
\FALabel(0.6,0.35)[r]{\textbf{q}}
\FALabel(10.,20.2)[b]{\textbf{p}}
\FALabel(19.1,0.37)[l]{\textbf{k}}
\FADiagram{}
\FAProp(1.,0.5)(10.,19.5)(0.,){/GhostDash}{0}
\FALabel(5.48,10.02)[br]{l}
\FAProp(10.,19.5)(18.5,0.5)(0.,){/Straight}{0}
\FALabel(14.52,10.02)[bl]{n}
\FAProp(10.,19.5)(10.,0.5)(0.,){/Straight}{0}
\FALabel(9.98,10.)[r]{m}
\FAProp(1.,0.5)(10.,0.5)(0.,){/Straight}{0}
\FALabel(5.5,0.85)[b]{i}
\FAProp(10.,0.5)(18.5,0.5)(0.,){/GhostDash}{0}
\FALabel(14.5,0.85)[b]{j}
\FALabel(0.6,0.35)[r]{\textbf{q}}
\FALabel(10.,20.2)[b]{\textbf{p}}
\FALabel(19.1,0.37)[l]{\textbf{k}}
\label{cuts}
\end{feynartspicture}
\end{center}
\end{scriptsize}
\caption{Complete cuts through which $k$ flows}
\end{figure}
and are drawn in figure $(\ref{cuts})$.

We show that the amplitudes associated with the corresponding families-sectors might be singular in the $k^2\to 0$ limit. 
 
Notice that each cut corresponds to many families. We select few  examples.
Let's start considering the  type $A$ family:
\beq {\cal F}_{A,4}=G\oplus L(l,m,i,j)\oplus L(l,i,j)\oplus l \oplus i,\eeq shown in table $(\ref{tabella})$ whose corresponding complete cut $C_{{\cal F}_{A,4}}$ is $\{m, n,j\}$. 
We assign  $t_G=1$ to the element $G$ and the variables $t_i$, with $i=1,2,3,4$  to the elements of the family in the order they are written. 

Since we are only interested in the possibly singular behaviour of the amplitude in  the $k^2\to 0$ limit we replace $\Pi_S$ defined in equation $(\ref{mub})$ and $N_S$ defined in equation $(\ref{N2})$ by $1$.

Concerning $M_{S^{A,4}}(t_{\c})$, defined in equation $(\ref{M})$, we know that its dominant terms are related to the complete cuts $C$, for which $\delta_C\equiv\sum_{\c}\d_{C,{\cal F}_{A,4}^{\xi}}(\c)=I(\c\cap C)-I(\c\cap C_{{\cal F}_{A,4}})$ is minimal, for example the cut $\{m, n, l\}$ into which $p$ flows for which $\d_C=1$. 

We stop looking for other complete cuts, since for all of them $\d_C>1$ and hence they correspond to negligible corrections.
Thus the contribution of this sector to the amplitude is given by:
\beq
\int \frac{t_2\, dt_2\,dt_3}{k^2+t_3\,\omega^2}.
\eeq
Apparently it diverges in the limit $k^2\rightarrow 0$.\\

Another family of type $A$, which is worth analyzing , is
\beq {\cal F}_{A,1}=G\oplus L(l,m,i,j)\oplus L(m,i,j)\oplus m \oplus i.\eeq
corresponding to the complete cut  $C_{{\cal F}_{A,1}}=\{n,l,j\}$. 

The dominant terms in $M_{S^{A,1}}(t_{\c})$ correspond to  the contributions from the cut $\{n,i,j\}$, through which $p$ flows, and the cut $\{n,m,j\}$, through which $q$ flows.

Therefore the  amplitude is approximated by:
\beq
\int \frac{t_2\, dt_2\,dt_3\,dt_4}{k^2+(t_3+t_4)\,\omega^2},
\eeq
which doesn't diverge.\\

As a last example let's consider a family of type $B$: 
\beq
{\cal F}_{B,8}=G\oplus L(l,m,n,i)\oplus H(l,m,i)\oplus  l \oplus i
\eeq
The complete cut associated with this family is $C_{{\cal F}_{B,8}}=\{j,n,m\}$ and the dominant terms of $M_{S^{B,8}}(t_{\c})$  correspond to the complete cut $\{j,l,m\}$, through which $p$ flows, and to  $\{j,l,i\}$, through which $q$ flows. 

The scalar amplitude is approximated by:
\beq
\int \frac{dt_1\, dt_2\,dt_3\,dt_4}{k^2+t_2\,t_4\,(t_1+t_3)\,\omega^2}
\eeq
and has a stronger divergence in the limit $k^2\rightarrow 0$, since both the integral in $t_2$ and in $t_4$ diverge.\\

At the end of the analysis one finds that  the scalar amplitude diverges in the limit $k^2\rightarrow 0$ in  $20$ sectors. 

An analogous conclusion can be reached considering the limit $q^2\rightarrow 0$.

 On the contrary  in the limit $p^2\rightarrow 0$ the only families which potentially give divergent contributions are of type $A$, and are  associated with the complete cut $\{l,m,n\}$. One can verify that there is no collinear divergence in the scalar amplitude in $p^2\to 0$.

At this point we need a mathematical tool which makes us compute exactly the divergent part of the amplitudes, without approximations.

\chapter{Mellin-Barnes representation}
\textit{The Mellin-Barnes representation \cite{librosm,*Freitas:2010nx,*Allendes:2009bd,*Smirnov:2009up,*Smirnov:2008tz,*Aguilar:2008qj,*Bierenbaum:2007zza,*Gluza:2007bd,*Gluza:2007rt,*Czakon:2005rk,*Borisov:2005ka,*Friot:2009fw,*Friot:2005gh,*Friot:2005cu,*Suzuki:2003jn,*Smirnov:2004ip,*Kharchev:2000ug,*Kowalenko:1998qm,*Passare:1996db,*Elizalde:1994dm,*Bytsenko:1993cf,*Anastasiou:2005cb,*Bolzoni:2009ye,*Bolzoni:2010sp,*Drummond:2006rz} allows the analytical evaluation of the divergent infrared part of the amplitude, since it separates the singularity in the kinematic invariant from the rest of the amplitude. In this chapter we consider this important mathematical tool and in the next chapter we will apply it to a physical problem. The method we want to show is  based on Speer-Smirnov sector division combined with the Mellin-Barnes transform. It is a  successful method in the analysis of collinear divergent amplitude. }
\section{Definition}
The Mellin-Barnes transform is a useful mathematical tool based on the following formula:
\beq
\frac{\Gamma(x)}{(a+b)^{x}}=\frac{1}{2\pi i}\frac{1}{b^x}\int_{{\cal P}\[-i\infty,+i\infty\]} d\s\,\Gamma(x-\s)\,\Gamma(\s)\,\(\frac{b}{a}\)^{\s},
\label{MBd}
\eeq
\begin{center}
\begin{figure}[hh]
\scalebox{0.26}{ \input{grafico2.pstex_t}}
\scalebox{0.26}{ \input{grafico1.pstex_t}}
\end{figure}
\end{center}
where $a$ and $b$ are real non-vanishing numbers with the same sign, $x$ is a non-vanishing complex number and the path ${\cal P}\[-i\infty,+i\infty\]$ is a continuous line going from $z_i=-iT$ to $z_f=iT$ in the $T\rightarrow \infty$ limit, leaving the poles of $\Gamma(\sigma)$, $z_n=-n$ with $n\in(0,...,\infty)$, on its left-hand side and the poles of $\Gamma(x-\sigma)$, $z_m=x+m$ with $m\in(0,..,\infty)$, on its right-hand side.

This formula is easily proved if $b/a\neq 1$ since the integration path  can be closed, without further contributions, enclosing all the poles of $\Gamma(\s)$ if $b/a>1$ and all the poles of $\Gamma(x-\s)$ if $b/a<1$.

The formula can also be generalized to the case of complex non-vanishing $a$ and $b$, with $a+b\neq 0$, turning the integration path to ${\cal P}\[-e^{i\phi}\,\infty,+e^{i\phi}\,\infty\]$, with $0<\phi\leq arg\(\frac{b}{a}\)\pm\frac{\pi}{2}<\pi$ and the same conditions concerning the poles of the integrand.


\section{Employment of the MB transform in a Speer Sector}

We shall exploit $(\ref{MBd})$ in order to evaluate the behaviour of an Euclidean Feynman amplitude given in equation $(\ref{Ixi})$ when $\xi^2\rightarrow 0$. 

Formally, choosing $a=\xi^2 N_{S^{\xi}}(t_{\c})$ and $b=-M_{S^{\xi}}(t_{\c})$, we have:
\bea
 &I^{\xi}_{a,b,\tau}= \frac{1}{2\pi i}&\sum_{{\cal F}^{\xi}}\int d\bar{\mu}_{a,b,\tau}^{S^{\xi}}\ (-M_{S^{\xi}}(t_{\c}))^{a+\frac{Ld}{2}-I}\\
 &&\int_{{\cal P}\[-i\infty,+i\infty\]} d\s \Gamma(\s)\Gamma(I-\frac{Ld}{2}-a-\s)\(-\frac{M_{S^{\xi}}(t_{\c})}{\xi^2 N_{S^{\xi}}(t_{\c})}\)^{\sigma}.\nonumber
\label{MB1}
\eea
We have seen how  the integration path should be chosen depending on the value of the ratio $\(-\frac{M_{S^{\xi}}(t_{\c})}{\xi^2 N_{S^{\xi}}(t_{\c})}\)$. However, in order to profit of the Mellin-Barnes, we have to change the order of the integrals considering 
\bea
 I^{\xi}_{a,b,\tau}&=&\int_{{\cal P}\[-i\infty,+i\infty\]} \frac{d\s}{2\pi i} \,(\xi^2)^{-\s}\,\Gamma(\s)\Gamma(I-\frac{Ld}{2}-a-\s)\\
 &&\sum_{{\cal F}^{\xi}}\int d\bar{\mu}_{a,b,\tau}^{S^{\xi}}\ (-M_{S^{\xi}}(t_{\c}))^{a+\frac{Ld}{2}-I}\(-\frac{M_{S^{\xi}}(t_{\c})}{N_{S^{\xi}}(t_{\c})}\)^{\sigma}.\nonumber
 \label{MB2}
\eea
Taking into account that $N_{S^{\xi}}(t_{\c})$ is positive and bounded in the hypercube $(0\leq t_{\c}\leq 1)$, if $-M_{S^{\xi}}(t_{\c})$ were bounded from below by a positive number, for small enough $\xi^2$ the two formulae in equation $(\ref{MB1})$ and in equation $(\ref{MB2})$ would coincide. This is however not the case, since $M_{S^{\xi}}(t_{\c})$ might vanish in the boundary.

A consequence of this vanishing is the fact that the integral
\beq
f(\s)=\int d\bar{\mu}_{a,b,\tau}^{S^{\xi}}\ (-M_{S^{\xi}}(t_{\c}))^{a+\frac{Ld}{2}-I}\(-\frac{M_{S^{\xi}}(t_{\c})}{N_{S^{\xi}}(t_{\c})}\)^{\sigma},
\eeq
instead of being an analytic function of $\s$, has a finite number of poles. 

It follows that, while in the analytic case the path in equation $(\ref{MB2})$ would be closed around the poles of $\Gamma(\s),$ in the second case the path must leave not only the poles of $\Gamma(\s)$, but also those of $f(\s)$, on its left-hand side and those of $\Gamma(I-\frac{Ld}{2}-a-\s)$ on its right-hand side and the amplitude must be computed closing the path around the poles of $\Gamma(\s)$ and those of $f(\s)$.

Let us make this more clear presenting a trivial example:
\bea
&&\int_0^1\frac{1}{\xi^2+x}dx=\log\(\frac{1+\xi^2}{\xi^2}\)=\\
&&\int_{{\cal P}\[-i\infty,+i\infty\]} \frac{d\s}{2\pi i} \,(\xi^2)^{-\s}\,\Gamma(\s)\Gamma(1-\s)\int_0^1 dx\, x^{\s-1}=\nonumber\\
&&\frac{1}{2\pi i}\int d\s
\frac{(\xi^2)^{-\s}}{\s}\,\Gamma(\s)\Gamma(1-\s)=\nonumber\\&&-\log(\xi^2)-\sum_{n=1}^{\infty}\frac{(-1)^n}{n!}\xi^{2n}.\nonumber
\eea
The pole from the $x$-integral transforms that of $\Gamma(\s)$ in $\s=0$ from first to second order and hence produces the term $-\log(\xi^2)$ which gives the singular part of the original integral. The remaining power series gives $\log(1+\xi^2)$ which is the analytic part of the integral.

We shall give a less trivial example of the use of $(\ref{MBd})$ applying it to a DIS process at one loop.
\chapter{Collinear singularities in DIS}
\textit{In the first chapter we discussed the important physical process of DIS. Here we present an off-shell strategy for evaluating analytically the collinear IR-divergent part of the structure functions of DIS to first order in the strong constant $\a_S$, using the parametric representation of the Speer sectors combined with the Mellin-Barnes transform.}
\section{The computing strategy}

In the framework of the naive parton model the hadronic tensor is computed  in terms of the parton (quark) distribution function $q_0(y)$ and of the partonic tensor $W^p_{\mu,\nu}(y\,p,q)$ using the formula:
\beq
W_{\mu,\nu}(p,q)=\int_0^1dy\,q_0(y)\,W^p_{\mu,\nu}(y\,p,q),
\eeq
where $y$ is the fraction of proton momentum taken by the quark.

In QCD the naive partonic tensor is computed from the tree approximation forward quark-virtual-photon Compton scattering by means of:
\beq 
W^p_{\mu,\nu}(p,q)=-\frac{1}{4\pi}Im\(\frac{Tr\[\psl\c_{\mu}(\psl+\qsl)\c_{\nu}\]}{(p+q)^2+i\eta}\)
\eeq
\begin{figure}[ht]
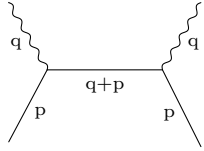

\centering
\begin{scriptsize}
\begin{feynartspicture}(100,80)(1,1)
\FADiagram{}
\FAProp(-0.,2.5)(4.,10.)(0.,){/Straight}{0}
\FALabel(2.667,6.1182)[tl]{p}
\FAProp(0.,17.)(4.,10.)(0.,){/Sine}{0}
\FALabel(1.3512,13.335)[tr]{q}
\FAProp(4.,10.)(16.,10.)(0.,){/Straight}{0}
\FALabel(10.,9.18)[t]{q+p}
\FAProp(16.,10.)(20.,2.)(0.,){/Straight}{0}
\FALabel(17.3172,5.8986)[tr]{p}
\FAProp(16.,10.)(20.,17.)(0.,){/Sine}{0}
\FALabel(18.6487,13.335)[tl]{q}
\end{feynartspicture}
\end{scriptsize}
\caption{The forward tree-approximation Compton amplitude}
\label{t-a}
\end{figure}
Thus the partonic tensor is related to the absorptive part of the tree approximation forward Compton amplitude in figure $(\ref{t-a})$, according to the optical theorem. 

\begin{figure}[ht]
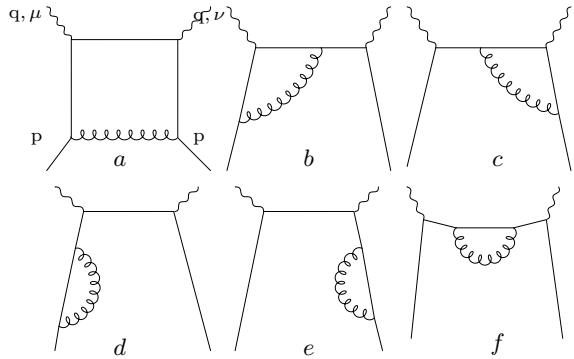

\centering
\begin{scriptsize}
\begin{feynartspicture}(250,150)(3,2.2)
\FADiagram{$a$}
\FAProp(0.,20.)(3.,16.)(0.,){/Sine}{0}
\FALabel(0.94,17.7)[tr]{q,\,$\mu$}
\FAProp(20.,20.)(16.,16.)(0.,){/Sine}{0}
\FALabel(18.52,17.48)[tl]{q,\,$\nu$}
\FAProp(3.,16.)(16.,16.)(0.,){/Straight}{0}
\FAProp(3.,16.)(3.,4.)(0.,){/Straight}{0}
\FAProp(3.,4.)(16.,4.)(0.,){/Cycles}{0}
\FAProp(16.,16.)(16.,4.)(0.,){/Straight}{0}
\FAProp(16.,4.)(20.,0.)(0.,){/Straight}{0}
\FALabel(18.52,2.02)[bl]{p}
\FAProp(3.,4.)(0.,0.)(0.,){/Straight}{0}
\FALabel(0.98,2.02)[br]{p}
\FADiagram{$b$}
\FAProp(0.,20.)(3.5,15.)(0.,){/Sine}{0}
\FAProp(3.5,15.)(1.5,6.)(0.,){/Straight}{0}
\FAProp(1.5,6.)(0.,0.)(0.,){/Straight}{0}
\FAProp(1.5,6.)(11.5,15.)(0.3023,){/Cycles}{0}
\FAProp(3.5,15.)(11.5,15.)(0.,){/Straight}{0}
\FAProp(11.5,15.)(17.,15.)(0.,){/Straight}{0}
\FAProp(17.,15.)(20.,20.)(0.,){/Sine}{0}
\FAProp(17.,15.)(20.,0.)(0.,){/Straight}{0}
\FADiagram{$c$}
\FAProp(0.,20.)(3.5,15.)(0.,){/Sine}{0}
\FAProp(3.5,15.)(0.,0.5)(0.,){/Straight}{0}
\FAProp(3.5,15.)(9.,15.)(0.,){/Straight}{0}
\FAProp(9.,15.)(17.,15.)(0.,){/Straight}{0}
\FAProp(17.,15.)(20.,20.)(0.,){/Sine}{0}
\FAProp(17.,15.)(18.5,7.)(0.,){/Straight}{0}
\FAProp(9.,15.)(18.5,7.)(0.2812,){/Cycles}{0}
\FAProp(18.5,7.)(20.,0.)(0.,){/Straight}{0}
\FADiagram{$d$}
\FAProp(1.,20.)(4.5,17.)(0.,){/Sine}{0}
\FAProp(18.5,20.)(15.5,17.)(0.,){/Sine}{0}
\FAProp(4.5,17.)(15.5,17.)(0.,){/Straight}{0}
\FAProp(4.5,17.)(3.5,12.5)(0.,){/Straight}{0}
\FAProp(15.5,17.)(20.,-0.)(0.,){/Straight}{0}
\FAProp(3.5,12.5)(1.5,3.)(-0.7745,){/Cycles}{0}
\FAProp(1.5,3.)(1.,0.)(0.,){/Straight}{0}
\FAProp(3.5,12.5)(1.5,3.)(0.,){/Straight}{0}
\FADiagram{$e$}
\FAProp(1.,20.)(4.5,17.)(0.,){/Sine}{0}
\FAProp(18.5,20.)(15.5,17.)(0.,){/Sine}{0}
\FAProp(4.5,17.)(15.5,17.)(0.,){/Straight}{0}
\FAProp(4.5,17.)(1.,-0.)(0.,){/Straight}{0}
\FAProp(15.5,17.)(16.5,12.5)(0.,){/Straight}{0}
\FAProp(16.5,12.5)(18.,4.)(0.886,){/Cycles}{0}
\FAProp(19.,-0.)(18.,4.)(0.,){/Straight}{0}
\FAProp(16.5,12.5)(18.,4.)(0.,){/Straight}{0}
\FADiagram{$f$}
\FAProp(0.,20.)(2.,16.)(0.,){/Sine}{0}
\FAProp(18.5,20.)(17.,16.)(0.,){/Sine}{0}
\FAProp(2.,16.)(0.5,1.5)(0.,){/Straight}{0}
\FAProp(17.,16.)(19.,1.5)(0.,){/Straight}{0}
\FAProp(2.,16.)(6.,15.)(0.,){/Straight}{0}
\FAProp(6.,15.)(13.,15.)(1.2307,){/Cycles}{0}
\FAProp(13.,15.)(17.,16.)(0.,){/Straight}{0}
\FAProp(6.,15.)(13.,15.)(0.,){/Straight}{0}
\end{feynartspicture}
\end{scriptsize}
\caption{The first order in $\a_S$ forward amplitudes in DIS.}
\label{gr}
\end{figure}
Taking into account the radiative corrections, one has to consider the forward virtual photon-quark scattering amplitude corresponding to the diagrams in figure $(\ref{gr})$.

In all these diagrams the initial state and the final state coincide and consist of a quark with momentum $p$ and of a virtual photon with momentum $q$.

A scattering amplitude depends on six independent kinematic invariants. These are the square momenta (masses) of the four external legs, that is, of initial and final, possibly virtual, particles, the square center of mass energy, the Mandelstam variable $S$, and the square momentum transfer, the Mandelstam variable $T$.

We denote the independent kinematic invariants of these amplitudes by $p^2$, the square momentum of the quark, $q^2$, the square momentum of the virtual photon, $S=(q+p)^2$ and of course $T=0$.

Considering the off-shell quark-virtual photon forward (Compton) amplitude, we have negative virtual photon square momentum ($q^2$), positive center of mass energy $S$ and null $T$, while we choose negative virtual quark square momentum ($p^2$).

In principle one should foresee a singularity in the amplitude at $T=0$. However this is not the case for helicity/angular momentum reasons. A vanishing angular momentum in the crossed ($T$) channel $q+\bar{q}\rightarrow \c^{*}+\c^{*}$ is excluded by helicity conservation and consequently the amplitude is regular at $T=0$. It is not analytic for $S>0$, as expected, since we are studying the absorptive part of the amplitude.

However if we complete our off-shell choice considering an unphysical negative $S$ together with the negative square momenta of the virtual scattering particles, we expect analyticity for the general reason mentioned in chapter $4$, that is, infrared power counting. Therefore the general idea is the following.
\begin{itemize}
\item We start computing the amplitudes with negative $p^2$, $q^2$ and $S$, where the amplitudes are analytic. 

\item We separate the singular part when $p^2\rightarrow 0^-$. This is expected to diverge as $\log(-p^2)$ and in fact it does. We compute the coefficient of $\log(-p^2)$ in the Mellin-Barnes expansion of the amplitude. This gives the complete $p^2$-expansion of the amplitude, however we disregard the regular contributions in the present analysis.

\item This coefficient of $\log(-p^2)$ is expected to correspond to an analytic function of $S$, in the region of negative $S$ and $q^2$. By the Mellin-Barnes formula and sector decomposition we compute explicitly the analytic coefficient of $\log(-p^2)$ (as a matter of fact it is the sum of analytic contributions) and we analytically continue it in the $Re(S)>0$ region, where we find the expected branch-cut. We compute the discontinuity, which is directly related to the singular part of the trace of the partonic tensor under study.
\end{itemize}
We repeat this analysis for each graph, noticing however that, on account of L.S.Z. formula \cite{Lehmann:1954rq}, diagrams $d$ and $e$ have to be divided by 2.

For each spinorial amplitude we select the mass-shell divergent term contributing to the trace of the hadronic tensor and we compute the trace of the spinorial matrix multiplied by $\psl$, since we want to sum over the helicities in the (logarithmically singular) mass-shell limit.


\subsection*{The box}
\begin{center}
\begin{feynartspicture}(200,150)(1,1)
\FADiagram{$a$}
\FAProp(0.,20.)(3.,16.)(0.,){/Sine}{0}
\FALabel(0.94,17.7)[tr]{q,\,$\mu$}
\FAProp(20.,20.)(16.,16.)(0.,){/Sine}{0}
\FALabel(18.52,17.48)[tl]{q,\,$\nu$}
\FAProp(3.,16.)(16.,16.)(0.,){/Straight}{1}
\FALabel(9.5,16.52)[b]{\textit{2}\ q+k}
\FAProp(3.,16.)(3.,4.)(0.,){/Straight}{-1}
\FALabel(2.1799,10.)[r]{k\ \textit{1}}
\FAProp(3.,4.)(16.,4.)(0.,){/Cycles}{0}
\FALabel(9.5,3.18)[t]{\textit{4}\ k-p}
\FAProp(16.,16.)(16.,4.)(0.,){/Straight}{-1}
\FALabel(16.52,10.)[l]{\textit{3}\ k}
\FAProp(16.,4.)(20.,0.)(0.,){/Straight}{1}
\FALabel(18.52,2.02)[bl]{p}
\FAProp(3.,4.)(0.,0.)(0.,){/Straight}{-1}
\FALabel(0.98,2.02)[br]{p}
\label{boxdiagram}
\end{feynartspicture}
\end{center}


In momentum representation and in the Feynman gauge the amplitude is
\beq
A^{(a)}_{\mu,\nu}(p,q)=i\frac{e^2\,e_s^2\,c_F}{4\,\pi}\int \frac{d^4 k}{(2\pi)^4} \,\frac{Tr[\psl\c^{\rho}\ksl\c_{\nu}(\qsl+\ksl)\c_{\mu}\ksl\c_{\rho}]}{((k^2)^2(q+k)^2(k-p)^2)},
\label{A1}
\eeq
and we have
\beq
W^{p,(a)}_{\mu,\nu}(p,q)=Im(A^{(a)}_{\mu,\nu}(p,q)).
\eeq
Let's start contracting the amplitude with the metric tensor in order to single out the collinear divergent box contribution to the trace (in Lorentz indices $\mu$ and $\nu$) of the partonic tensor. This gives us the radiative corrections to the linear combination of structure functions in the first line of equation $(\ref{c})$. The radiative corrections to the second line are given contracting the expression in equation $(\ref{A1})$ with $p^{\mu}p^{\nu}$. However we shall see that this does not contain singular contributions in $p^2\rightarrow 0^{-}$.

Computing the trace we get: 
\beq
Tr[\psl\c^{\rho}\ksl\c^{\nu}(\qsl+\ksl)\c_{\nu}\ksl\c_{\rho}]=16\(2k\cdot p\, k\cdot q+k^2(k\cdot p-p\cdot q)\).
\eeq
At this point we pass to the Schwinger parametric form following the procedure shown in chapter $2$ and disregarding the terms proportional to $p^2$ in the numerator, which give vanishing contribution in the $p^2\rightarrow 0$ limit.

We get
\beq
A_T^{(a)}=K\,p\cdot q\int_0^{\infty}\frac{\prod_{l=1}^4d\a_l}{P_a(\a)^3}\(-i\(1+\frac{3\a_2}{P_a(\a)}\)+\frac{\a_2\,D_a(\a,P)}{P_a(\a)^2}\)e^{i\frac {D_a(\a,P)}{P_a(\a)}}
\eeq 
with $K=\frac{\a_S\,e^2\,c_F}{\pi^2}$.

Then, integrating over the scale factor $t$ and hence passing from the $\a$ to the $\b$-parameters, we get:
\beq
A_T^{(a)}=K\,p\cdot q\int\frac{d\mu(\b)}{P_a(\b)^2}\(1+\frac{2\b_2}{P_a(\b)}\)\frac{1}{D_a(\b,P)}
\label{Waq}
\eeq
\beq
\text{with }\left\{
\begin{array}{rcl}
P_a(\b)&=&\b_1+\b_2+\b_3+\b_4\\
D_a(\b,P)&=&p^2\b_4(\b_1+\b_3)+q^2\b_2(\b_1+\b_3)+S\b_2\b_4+i\eta.\\
\end{array}
\right.
\label{PD1}
\eeq
We have shown explicitly the infinitesimal imaginary term $i\eta$, which accounts for the time-ordering in the Feynman amplitudes.
\normalsize
The amplitude is the sum of different terms corresponding to different Speer sectors. We want to single out the terms which diverge in the limit $p^2\rightarrow 0$. To this purpose we have already said that we can focus on a subset of sectors in which the scalar amplitude of the box is singular. This amplitude is proportional to
\beq 
A_s^{(a)}\propto\int \frac{d\mu(\b)}{D_a(\b,P)^{2}}
\nonumber
\eeq
and decomposes into 12 Speer sectors corresponding to the parametrizations:
\beq
\begin{array}{c l} 
a)&\b_1=s\leftrightarrow s\b,\,\b_2=s\, \a,\,\b_3=s\b\leftrightarrow s,\,\b_4=1;\\
b)&\b_1=1\leftrightarrow s\b,\,\b_2=s\, \a,\,\b_3=s\b\leftrightarrow 1,\,\b_4=s;\\
c)&\b_1=1\leftrightarrow s,\,\b_2=s\, \a,\,\b_3=s\leftrightarrow 1,\,\b_4=s\b;\\
d)&\b_1=1\leftrightarrow s\b,\,\b_2=s,\,\b_3=s\b\leftrightarrow 1,\,\b_4=s\a;\\
e)&\b_1=s\leftrightarrow s\b,\,\b_2=1,\,\b_3=s\b\leftrightarrow s,\,\b_4=s\a;\\
f)&\b_1=s\b,\,\b_2=1,\,\b_3=s\a,\,\b_4=s;\\
g)&\b_1=s\b,\,\b_2=s,\,\b_3=s\a,\,\b_4=1,
\end{array}
\label{abc}
\eeq
where each of the first five lines corresponds to two sectors, giving the same contribution, because of the symmetry for exchange of the parameters $\b_1$ and $\b_3$ . 

It is easy to verify that the box scalar amplitude is divergent in the limit $p^2\rightarrow 0$ only in the six sectors of  $a)$ $b)$ and $c)$:
\beq
\begin{array}{c l}
a)&\int_0^1 \frac{d\a\,ds}{(p^2+\a(q^2\,s+S))^2}\stackrel{p^2\rightarrow 0}{\rightarrow}\(\int_0^1 \frac{d\a}{\a^2}\)\int_0^1\frac{ds}{(q^2\,s+S)^2};\\
b)&\int_0^1 \frac{d\a\,ds}{(p^2+\a(q^2+S\,s))^2}\stackrel{p^2\rightarrow 0}{\rightarrow}\(\int_0^1 \frac{d\a}{\a^2}\)\int_0^1\frac{ds}{(q^2+S\,s)^2};\\
c)&\int_0^1 \frac{d\a\,ds\,d\b}{(p^2\b+\a(q^2+S\,s\b))^2}\stackrel{p^2\rightarrow 0}{\rightarrow}\(\int_0^1 \frac{d\a}{\a^2}\)\int_0^1\frac{ds\,d\b}{(q^2+S\,s\b)^2}.
\nonumber
\end{array}
\eeq
Notice that, following the method discussed at the end of section $4.1$ in order to identify the families and hence the sectors accounting for the singularities in the limit of vanishing $p^2$, that is, selecting the families for which the momentum crossing the complete cut $C_{\cal F}$ coincides with $p$, we should have selected sectors $a)$ and $b)$ forgetting sectors $c)$.

However we are not facing a counter-example to the above method; indeed  we have to consider, together with the families $a)$ and $b)$ associated with $p^2$, the families $c)$, which correspond to the kinematic invariant $T$,  since $T$ is null: $0=T<\l|p^2\r|$. 

Now, once selected the interesting sectors, we consider again the expression for the amplitude given in $(\ref{Waq})$, which is the sum of two terms. The second term, which is proportional to $\b_2$, does not contribute to the collinear divergence, since in all the considered sectors $\b_2=s\a$ vanishes with $D_a(\b,P)$ for $p^2=0$.

Thus the singularity is only due to the first term in the integrand of $(\ref{Waq})$ and hence we must evaluate the sum of the contributions to
\beq
A_T^{(a,1)}=K\,p\cdot q\int\frac{d\mu(\b)}{P_a(\b)^2}\frac{1}{D_a(\b,P)}
\nonumber
\eeq
from the six sectors $a),\ b)\ \text{and}\ c)$, that is
\bea
&&A_T^{(a,1)}=2\,K\,p\cdot q\int_0^1  \frac{ s\,ds\,d\a\,d\b}{(1+s(1+\a+\b))^{2}}\(\frac{1}{(1+\b)(p^2+q^2\,s\,\a)+S\,\a+i\eta}\right.\nonumber\\
 &&\left.+\frac{1}{(1+\b\,s)(p^2+q^2\,\a)+S\,s\,\a+i\eta}+\frac{1}{(1+s)(p^2\,\b+q^2\,\a)+S\,s\,\a\,\b+i\eta}\).\nonumber 
\eea
In order to single out the collinear divergence we apply the Mellin-Barnes formula, introduced in chapter $5$:
\bea
&&A_T^{(a,1)}=-2\,K\,p\cdot q \int_{{\cal P}\[-i\infty,+i\infty\]} \frac{d\sigma}{2\pi i}(-p^2)^{-\sigma}\frac{\Gamma(\sigma)}{\s}\Gamma(1-\sigma)\\
&&\int_0^1 \frac{s\,ds\,d\b\,\d\a}{(1+s(1+\a+\b))^{2}}\(\frac{(1+\b)^{-\s}}{\[-\a(q^2(1+\b)s+S)-i\eta\]^{1-\s}}+\right.\nonumber\\
&&\left.\frac{(1+\b\,s)^{-\s}}{\[-\a(q^2(1+\b\,s)+S\,s)-i\eta\]^{1-\s}}+\frac{(1+s)^{-\s}}{\[-\a(q^2(1+s)+S\,s\,\b)-i\eta\]^{1-\s}}\)\nonumber.
\label{abo}
\eea
Let us perform first the $\a$ integral; this has the form:
\beq\int_0^1\frac{d\a}{\a}\a^{\s}\,A(\a),
\eeq
where $A(\a)=\frac{1}{\[1+s(1+\b+\a)\]^2}$ is analytic and non-vanishing in the integration domain. It follows that the above integral is $\frac{A(0)}{\s}+{\cal R}(\s)$, where ${\cal R}(\s)$ is analytic in the whole complex $\s$-plane.

This pole in $\s=0$ is the singularity considered in chapter $5$ which must be enclosed in the path encircling the $\Gamma(\s)$ poles. Thus the $\s$-integrand has a double pole in $\s=0$, while this pole would have been simple in the absence of collinear singularities.

Notice that if we had used dimensional regularization as IR-regularization, the new pole would have been in $\s=\eps$. In this case, performing the $\s$-integral one would obtain a term $-\frac{1}{\eps}$ from the $\Gamma$-pole in $\s=0$ and $\frac{(-p^{2})^{-\s}}{\eps}$ from the new pole. Then one could "go to the limit" $p^2\rightarrow 0$. Choosing, arbitrarily, $\eps$ real and negative, the second term would vanish in the limit, leaving a $\frac{1}{\eps}$ singularity, as a memory of the collinear one.

We believe that our procedure is mathematically more sound. 

Thus, applying the above considerations to equation $(\ref{abo})$, computing the $\s$-integral and selecting the contribution of the double pole, the rest giving analytical contribution in $p^2\approx 0$, we get:
\bea
\left.A^{(a)}_T\right|_{div}&=&K\,\frac{2\,p\cdot q}{-q^2} \,\log(-p^2)\int_0^1 \frac{s\,ds\,d\b}{(1+s(1+\b))^{2}}
\(\frac{1}{(1+\b)s+\frac{S}{q^2}-i\eta}\right.\nonumber\\
&&\left.+\frac{1}{1+\b\,s+\frac{S}{q^2}\,s-i\eta}+\frac{1}{1+s+\frac{S}{q^2}\,s\,\b-i\eta}\).
\eea
This is an analytic function of the variable $\frac{S}{q^2}$ with a branch-cut on the negative real axis. We compute the discontinuity and we set
\beq
\frac{S}{q^2}=1-\frac{1}{x},
\eeq
which now is a negative quantity, since $S$ is positive\footnote{We set $x=x_B$, Bjorken variable.}, and we also set $z=1+s(1+\a+\b)$. 

The discontinuity is:
\bea
\left.W^{p,(a)}_T\right|_{div}&=&\pi\,K\,\log(-p^2)\int_0^1 ds\int_{1+s}^{1+2s}\frac{dz}{z^{2}}\\
&&\[\d(x\,z-1)+\d(x\,z-s)+\d(-z(1-x)+1+s)\].\nonumber
\eea
Notice that $S=q^2+2p\cdot q=q^2-\frac{q^2}{x}$ and that $\frac{2p\cdot q}{-q^2}=\frac{1}{x}$. Hence $\frac{2p\cdot q}{-q^2}\times x=1$.


After the integration we have
\beq K\,\log(-p^2)\,\pi\cdot
\left\{
\begin{array}{rl}
a)& \(\frac{3x-1}{2}\)\,\theta\(\frac{1}{2}-x\)\theta\(x-\frac{1}{3}\)+\(\frac{1-x}{2}\)\theta\(x-\frac{1}{2}\)\theta\(1-x\)\\
b)& x\,\theta(x)\theta\(\frac{1}{3}-x\)+(1-2x)\,\theta\(+\frac{1}{2}-x\)\theta\(x-\frac{1}{3}\)\\ 
c)&\(\frac{1-3x}{2}\)\,\theta(x)\theta\(+\frac{1}{3}-x\),\\
\end{array}
\right.
\eeq
where we see that different sectors correspond to different physical regions.

Finally the sum is
\beq \left.W^{p,(a)}_T\right|_{div}=K\,\pi\,\log(-p^2)\frac{(1-x)}{2} \label{Wa}\eeq
which is the box contribution to the collinear divergent imaginary part of the amplitude, we were looking for.\\

\bigskip

Before studying the other diagrams, we still have to contract the box amplitude $(\ref{A1})$ with $p^{\mu}p^{\nu}$.

The numerator in momentum representation is proportional to:
\beq
Tr[\psl\c^{\rho}\ksl\psl(\qsl+\ksl)\psl\ksl\c_{\rho}]=-32(k\cdot p)^2(k\cdot p+p\cdot q).\eeq
One has
\beq
p^{\mu}p^{\nu}A_{\mu\nu}^{(a)}=-2\,K(p\cdot q)^3\int d\mu(\b)\frac{\b_2^2(\b_1+\b_3+\b_4)}{P_a(\b)^2\,D_a(\b,P)^2}
\eeq
It is very easy to verify that the amplitude is not divergent in the limit $p^2\rightarrow 0$. Indeed, in analogy with the above case we have a $\b_2^2$-proportional term. $\b_2^2$ vanishes with $D_a(\b,P)^2$ in the $(p^2\rightarrow 0)$-limit.

\subsection*{The vertex correction}
\begin{center}
\begin{feynartspicture}(210,160)(2,1)
\FADiagram{$b$}
\FAProp(0.,20.)(3.5,15.)(0.,){/Sine}{0}
\FALabel(1.1649,17.2344)[tr]{q,\,$\mu$}
\FAProp(3.5,15.)(1.5,6.)(0.,){/Straight}{-1}
\FALabel(1.7109,10.782)[r]{k \textit{1}}
\FAProp(1.5,6.)(0.,0.)(0.,){/Straight}{-1}
\FALabel(-0.0311,3.3152)[r]{p}
\FAProp(1.5,6.)(11.5,15.)(0.3023,){/Cycles}{0}
\FALabel(8.2501,8.502)[tl]{\textit{3} k-p}
\FAProp(3.5,15.)(11.5,15.)(0.,){/Straight}{1}
\FALabel(7.5,15.52)[b]{q+k \textit{2}}
\FAProp(11.5,15.)(17.,15.)(0.,){/Straight}{1}
\FALabel(14.25,15.52)[b]{q+p}
\FAProp(17.,15.)(20.,20.)(0.,){/Sine}{0}
\FALabel(19.1347,17.3111)[tl]{q,\,$\nu$}
\FAProp(17.,15.)(20.,0.)(0.,){/Straight}{+1}
\FALabel(19.02,8.)[l]{p}
\FADiagram{c}
\FAProp(0.,20.)(3.5,15.)(0.,){/Sine}{0}
\FAProp(3.5,15.)(0.,0.5)(0.,){/Straight}{0}
\FAProp(3.5,15.)(9.,15.)(0.,){/Straight}{0}
\FAProp(9.,15.)(17.,15.)(0.,){/Straight}{0}
\FAProp(17.,15.)(20.,20.)(0.,){/Sine}{0}
\FAProp(17.,15.)(18.5,7.)(0.,){/Straight}{0}
\FAProp(9.,15.)(18.5,7.)(0.2812,){/Cycles}{0}
\FAProp(18.5,7.)(20.,0.)(0.,){/Straight}{0}
\end{feynartspicture}
\end{center}
Let's recall equation $(\ref{Delta})$, in which we evaluated the first order in $\a_s$ amplitude of the vertex correction:
{\small\bea
\Delta^{\mu}&=i\frac{\a_S\,e\,c_F}{2\pi}\int_0^\infty &\frac{\prod_{l}^3d\b_l\d(1-\sum_{i}^3\b_i)}{(P_G(\b))^{3}}\[\frac{\(\ \b_3\ \psl-\b_2\ \qsl\ \)\c^{\mu}\(\ \qsl\ \(\b_1+\b_3\)+\ \psl\ \b_3\)}{(p^2\,\b_1\b_3+q^2\,\b_1\b_2+S\,\b_2\b_3+i\eta)}+\right.\nonumber\\
&&-\left.\c^{\mu}\,\log\(\frac{P_G(\b)^2\mu^2}{p^2\,\b_1\b_3+q^2\,\b_1\b_2+S\,\b_2\b_3}\)\].\nonumber
\label{Delta2}
\eea}
We can write the amplitude of the diagram $b$ in terms of $\Delta^{\mu}$:
\beq
A^{(b)}_{\mu,\nu}(p,q)=-i\frac{e}{4\pi}\,\frac{Tr(\psl\,\c^{\nu}\,(\psl+\qsl)\Delta^{\mu})}{S+i\eta}.
\label{Apb}
\eeq
In much the same way as for the box amplitude, we first contract with the metric tensor and then with $p^{\mu}p^{\nu}$.

Contracting equation $(\ref{Apb})$ with the metric tensor, multiplying by two, since the diagrams in the figure give the same contribution, and finally disregarding the terms which are proportional to $p^2$, we have:
\bea
2A^{(b)}_{T}\approx& K\frac{2\,p\cdot q}{S+i\eta}\int \frac{d\mu(\b)}{P_b(\b)^3}&\[(\b_1+\b_3)\frac{(S-q^2)\b_3-(S+q^2)\b_2}{D_b(\b,P)}+\right.\nonumber\\
&&-\log\(\frac{P_b(\b)^2\mu^2}{D_b(\b,P)}\)\left.\]\nonumber
\label{tb}
\eea
\beq
\text{with }\left\{
\begin{array}{rcl}
P_b(\b)&=&\b_1+\b_2+\b_3\\
D_b(\b,P)&=&p^2\b_1\b_3+q^2\b_1\b_2+S\b_2\b_3+i\eta.\\
\end{array}
\right.
\label{PD2}
\eeq
There are 6 Speer sectors corresponding to the following parametrizations:
\bea 
&a)\,&\b_1=s,\,\b_2=s\,\a,\,\b_3=1;\nonumber\\
&b)\,&\b_1=1,\,\b_2=s\,\a,\,\b_3=s;\nonumber\\
&c)\,&\b_1=s\,\a,\,\b_2=s,\,\b_3=1;\nonumber\\
&d)\,&\b_1=s\,\a,\,\b_2=1,\,\b_3=s;\\
&e)\,&\b_1=s,\,\b_2=1,\,\b_3=s\,\a;\nonumber\\
&f)\,&\b_1=1,\,\b_2=s,\,\b_3=s\,\a\nonumber.
\eea
It is easy to verify that the scalar amplitude, corresponding to diagrams $b$ and $c$
\beq
A^{(b)}_S\propto \int d\mu(\b)\,\frac{1}{P_b(\b)D_b(\b,P)}
\nonumber
\eeq
is UV-finite. It has IR-divergences in the limit $p^2\rightarrow 0$ corresponding to the contributions from the sectors $a)$ and $b)$, since in this limit $D_b(\b,P)$ is proportional to $\a$.
Therefore we have to compute $(\ref{tb})$ in the sectors $a)$ and $b)$. 

In this equation we have put into evidence two terms. The second one comes from the subtraction of the UV-divergence. It is collinear finite, since it is a parametric integral of the logarithm of a function which vanishes on the boundary of the integration domain in the $p^2\rightarrow 0$-limit.

In the first term we pick out the only contribution which is collinear divergent in the sectors $a)$ and $b)$, which is:
\beq
2A_T^{(b,1)}=K\frac{2\,p\cdot q}{S+i\eta}\int \frac{d\mu(\b)}{P_b(\b)}(\b_1+\b_3)\frac{(S-q^2)\b_3}{D_b(\b,P)}.
\eeq
The rest is finite since it is proportional to $\b_2$, which vanishes together with $D_b(\b,P)$ in much the same way as for the box. Summing the contributions from the two sectors, we get: 
\bea
&2A_T^{(b,1)}=-K\frac{2\,p\cdot q}{S+i\eta}(S-q^2)\int \frac{(1+s)\,ds\, d\a}{(1+s(1+\a))^3}&\(\frac{1}{-p^2+\a(-q^2\,s-S)-i\eta}\quad\qquad\right.\nonumber\\
&&\left.+\frac{s}{-p^2+\a(-q^2-S\,s)-i\eta}\).
\eea
We apply the Mellin-Barnes transform and we consider the collinear divergent part.\\ This is:
\bea
&\left.2A^{(b)}_T\right|_{div}=K\frac{2\,p\cdot q}{S+i\eta}&(1-\frac{S}{q^2})\log(-p^2)\int_0^1 \frac{ds}{(1+s)^2}\nonumber\\
&&\(\frac{1}{s+\frac{S}{q^2}-i\eta}+\frac{s}{1+\frac{S}{q^2}\,s-i\eta}\).
\label{newone}
\eea
Again this is an analytic function of the variable $\frac{S}{q^2}$ with a branch-cut in the negative real axis and a pole in $-i\eta$. 

The novelty of the vertex correction contribution lies in the presence of the pole superimposed on the branch cut. Thus computing the absorptive part we should take into account the branch-cut discontinuity and the contribution proportional to $-i\pi\delta(S)$ coming from the pole. This corresponds to a single scattered parton final state which is also present in the collinear-divergent contributions from diagrams $d$ and $e$.

However the term $i\pi\delta(S)$ multiplying the vertex  
correction in diagrams $b$ and $c$ is ill-defined, since we see from  
equation $(\ref{newone})$ that the coefficient of the Dirac delta contains the  
ill-defined integral $\int_0^1ds/[(1+s)^2(s-i\eta)]\ .$

  A general remark is here in order. Even if the Fourier transformed  
Feynman amplitude corresponding to a given diagram is the product of  
the contributions from the parts of the diagram, this product  
structure is not suitable for spectral analyses. Indeed for this  
purpose one has to introduce a spectral representation for the whole  
amplitude and compute, e.g. its imaginary part, on the basis of this  
representation.

In the light of this comment it is easy to verify that the two above mentioned inconsistencies compensate each other. Indeed let us look more carefully at equation $(\ref{newone})$. It can be written as follows:
\bea
\left.2A^{(b)}_T\right|_{div}&=&K\,\frac{log(-p^2)}{x}\,\[\int_0^1 \frac{ds}{s\,\(1+s\)^2}\(\frac{1}{s+\frac{S}{q^2}-i\eta}-\frac{1}{\frac{S}{q^2}-i\eta}\)+\right.\nonumber\\
&&-\frac{1}{S/q^2-i\eta}\int_0^1 \frac{s\,ds}{\(1+s\)^2}\frac{1}{\frac{S}{q^2}\,s-i\eta}+\int_0^1\frac{ds}{\(1+s\)^2}\cdot\nonumber\\
&&\left. \(\frac{1}{s+\frac{S}{q^2}-i\eta}-\frac{s}{1+s\,\frac{S}{q^2}-i\eta}\)\right].
\label{Apb3}
\eea
Now, considering the imaginary part and setting 
\beq
\frac{S}{q^2}=1-\frac{1}{x},
\nonumber\eeq
we have:
\bea
\left.2W^{p,(b)}_T\right|_{div}&=&K\,\pi\,\log(-p^2)\,x\,\[\int_{1/2}^1 \frac{dt}{1-t}t\(\delta(x-t)-\delta(x-1)\)\right.\nonumber\\
&&+\int_{1/2}^1 dt\(\frac{x}{1-x}\delta(x-(1-t))+\delta(x-t)+\delta(x-(1-t))\)\nonumber\\
&&\left.-\delta(x-1)\(\log(2)-\frac{1}{2}\)\],\nonumber
\eea
where we have changed the integration variable $t=\frac{1}{1+s}.$

The question is if this is a distribution in $0\leq x\leq 1$. To check this point, we multiply $2W^{p,(b)}_{T}\left.\right|_{div}$ by a $C_{\[0,1\]}^{\infty}$ function $\phi(x)$ and integrate over $x$. \\
We get:
\bea
\int_0^1 dx\left.2W^{p,(b)}_T\right|_{div}&=&K\,\pi\,\log(-p^2)\[\int_{1/2}^1 dx\,\frac{x}{1-x}\(\phi(x)-\phi(1)\)\right.\nonumber\\
&&+\left.\int_0^{1/2}dx\,\frac{x}{1-x}\phi(x) -(\log(2-\frac{1}{2}))\phi(1)\]=\nonumber\\
&&\int_0^1 dx\frac{x\phi(x)-\phi(1)}{1-x}+\phi(1).
\label{Wpb}
\eea
Defining as in literature the distribution $\frac{1}{(1-x)_+}$ as
\beq
\int_0^1 dx\frac{f(x)}{(1-x)_+}=\int_0^1dx\frac{f(x)-f(1)}{1-x},
\eeq
we identify equation $(\ref{Wpb})$ with:
\beq
\left.2W^{p,(b)}_T\right|_{div}=K\,\pi\,\log(-p^2)\(\frac{x}{(1-x)_+}+\delta(1-x)\).
\eeq
We sum this result with $(\ref{Wa})$ getting
\beq
\left.(W^{p,(a)}_T+2W^{p,(b)}_T)\right|_{div}=\frac{\a_s\,e^2\,c_F}{2\,\pi}\,\log(-p^2)\(\frac{1+x^2}{(1-x)_+}+2\,\delta(1-x)\).
\eeq
From $(\ref{F})$ and $(\ref{G})$ one computes the splitting function $P(x)$
\beq
P(x)=c_F\(\frac{1+x^2}{(1-x)_+}+2\,\delta(1-x)+C\,\delta(1-x)\),
\eeq
where $C$ is the coefficient of term from the diagrams $d$ and $e$. Notice that diagram $f$ does not contribute to the collinear-divergent part of $g^{\a,\b}W_{\a,\b}$ since it is proportional to the tree diagram term multiplied by $\log(S)$.

\begin{center}
\begin{scriptsize}
\begin{feynartspicture}(350,120)(3,1)
\FADiagram{$d$}
\FAProp(1.,20.)(4.5,17.)(0.,){/Sine}{0}
\FALabel(2.3839,17.9929)[tr]{q,$\mu$}
\FAProp(18.5,20.)(15.5,17.)(0.,){/Sine}{0}
\FALabel(17.52,17.98)[tl]{q,$\nu$}
\FAProp(4.5,17.)(15.5,17.)(0.,){/Straight}{1}
\FALabel(10.,17.52)[b]{p+q}
\FAProp(4.5,17.)(3.5,12.5)(0.,){/Straight}{-1}
\FALabel(3.2109,15.032)[r]{p}
\FAProp(15.5,17.)(20.,-0.)(0.,){/Straight}{1}
\FALabel(18.2367,8.7558)[l]{p}
\FAProp(3.5,12.5)(1.5,3.)(-0.7745,){/Cycles}{0}
\FALabel(6.8168,6.9445)[l]{k-p}
\FAProp(1.5,3.)(1.,0.)(0.,){/Straight}{-1}
\FALabel(0.4476,1.7137)[r]{p}
\FAProp(3.5,12.5)(1.5,3.)(0.,){/Straight}{-1}
\FALabel(1.7078,8.0178)[r]{k}
\FADiagram{$e$}
\FAProp(1.,20.)(4.5,17.)(0.,){/Sine}{0}
\FAProp(18.5,20.)(15.5,17.)(0.,){/Sine}{0}
\FAProp(4.5,17.)(15.5,17.)(0.,){/Straight}{0}
\FAProp(4.5,17.)(1.,-0.)(0.,){/Straight}{0}
\FAProp(15.5,17.)(16.5,12.5)(0.,){/Straight}{0}
\FAProp(16.5,12.5)(18.,4.)(0.886,){/Cycles}{0}
\FAProp(19.,-0.)(18.,4.)(0.,){/Straight}{0}
\FAProp(16.5,12.5)(18.,4.)(0.,){/Straight}{0}
\FADiagram{$f$}
\FAProp(0.,20.)(2.,16.)(0.,){/Sine}{0}
\FALabel(0.3172,17.8986)[tr]{q,$\mu$}
\FAProp(18.5,20.)(17.,16.)(0.,){/Sine}{0}
\FALabel(18.5941,17.5034)[l]{q,$\nu$}
\FAProp(2.,16.)(0.5,1.5)(0.,){/Straight}{-1}
\FALabel(0.4369,8.8837)[r]{p}
\FAProp(17.,16.)(19.,1.5)(0.,){/Straight}{1}
\FALabel(18.5106,8.8866)[l]{p}
\FAProp(2.,16.)(6.,15.)(0.,){/Straight}{1}
\FALabel(3.6847,14.7188)[t]{p+q}
\FAProp(6.,15.)(13.,15.)(1.2307,){/Cycles}{0}
\FALabel(9.5,9.8723)[t]{k-p-q}
\FAProp(13.,15.)(17.,16.)(0.,){/Straight}{1}
\FALabel(15.3152,14.7188)[t]{p+q}
\FAProp(6.,15.)(13.,15.)(0.,){/Straight}{1}
\FALabel(9.5,14.18)[t]{k}
\end{feynartspicture}
\end{scriptsize}
\end{center}

Considering that, on account of L.S.Z formula, diagrams $d$ and $e$ have to be divided by 2, one has 
\beq
\left.(W^{p,(d)}_T+W^{p,(e)}_T)\right|_{div}=-\frac{1}{4}\frac{\a_S\,e^2 c_F}{\pi}\,\log(-p^2) \,x\, \delta(1-x).
\eeq
Hence $C=-\frac{1}{2}$ and we get the known value of the splitting function:
\beq
P(x)=c_F\(\frac{1+x^2}{(1-x)_+}+\frac{3}{2}\delta(1-x)\).
\eeq

We have still to contract $2A^{(b)}_{\mu,\nu}$ with $p^{\mu}p^{\nu}$. One easily sees that the result vanishes in the collinear limit. 
\chapter*{Conclusions}
We have seen in the last chapter how one can study collinear
divergences in DIS cross sections by considering the quark mass-shell
limit of Feynman amplitudes and exploiting the simplifying power of
the Speer-Smirnov sector decomposition of the parametric integral
expressions for the amplitudes together with that of the Mellin-Barnes technique.

In particular, we have shown that the off-shell regularization is a
natural tool for the analysis of collinear divergences in inclusive
cross sections, and that the use of this regularization is strongly
simplified by the Speer-Smirnov sector decomposition. The recourse to
Mellin-Barnes formula is the obvious final tool for the computation of
singular parts of the amplitudes.

The suggestion of initial state off-shell regularization of inclusive
cross sections as a reasonable "physical option", as an alternative 
to the widely adopted dimensional scheme,
is one of the results of the present thesis. 
However the principal goal of this thesis was to identify a general
method of analysis of mass singularities in the Feynman amplitudes of
massless field theories, which could be automatically applied to
multi-loop and multi-external-vertex diagrams. For this reason we have
described with great care the role of different kinds of cuts in the
construction of the parametric form of a Feynman integral. These cuts
are sets of lines in Feynman diagrams whose deletion either breaks the
original diagram in two parts, or transforms it into a tree diagram,
or else does both things together (chapter 2). We have also analyzed
the role of the cuts in the identification of the Speer-Smirnov
singularity families (chapter 3) and sectors (chapter 4). The idea was
that the identification of sectors contributing to the IR-singular
parts associated with the vanishing of kinematic invariants is a
central step in the study of these singularities, and that the method
could be extended to multi-loop and multi-external-vertex diagrams
developing suitable software tools. As a matter of fact, in the last
year an example of such tools (Fiesta \cite{Smirnov:2008py}) was
developed by Smirnov and Tentyukov. However, to our knowledge, the
method has not yet been used in the explicit analysis of IR
divergences.

A systematic use of the techniques presented in this thesis may in principle
prove useful in different contexts of phenomenological
interest. High-order perturbative calculations
in QCD, for example, may take advantage of a systematic separation of
singular contributions.
From a different point of view, an efficient way of identifying IR divergent
terms to all orders in perturbation theory can be useful when one is faced
with the problem of resumming the whole perturbative series in
special kinematic regimes, where powers of large logarithms of ratios of
invariants spoil the reliability of fixed-order calculations.

There are, of course, many points that need a further and deeper discussion. 
For example the gauge independence of the singular parts
of the forward amplitudes \cite{Gabrieli:1997rx,*Dorn:2008dz,*Carlitz:1988ab}, has been
mentioned in short occasional remarks. 
This could be better justified extending the basic argument proving
gauge independence on the charged particle mass shell of massive QED
amplitudes \cite{Lowenstein:1972pr}. Indeed in any gauge theory there
is a general connection between the gauge independence of the
"physical" amplitudes and Ward-Slavnov-Taylor identities \cite{Becchi:1974md}. In the case of amplitudes involving only local
physical (BRS invariant) operators one finds directly that their
partial derivative with respect to any gauge fixing parameter
vanishes. The case of amplitudes involving charged fields, which are
not BRS invariant, is more difficult, indeed one finds that the same
derivatives are less singular on the mass-shell than the original
amplitudes; poles are replaced by branch cuts. The idea which must be
verified in a general situation is that in mass-less theories the same
thing happens since any gauge parameter derivative of a collinear
divergent amplitude is less singular, or even null in the mass-shell
limit. This is fairly clear in our DIS example, indeed, using
Slavnov-Taylor identities, one shows that the Feynman-gauge parameter
derivative of the non-amputated forward scattering amplitude is
proportional to the box diagram amplitude in which the gluon
propagator has been replaced by a scalar propagator. The result is
apparently proportional to the box amplitude (there is a $-2$ factor),
therefore it is proportional to $\log (-p^2)$. However to get the
derivative of the forward scattering amplitude one has to amputate the
amplitude, which means multiply it by $p^2$, and hence the collinear
divergence disappears and the gauge derivative vanishes in the
mass-shell limit.

A second point that we have not discussed with sufficient completeness is the
effect of UV-divergent parts on the IR analysis. This is essentially
the consequence of the choice of the Smirnov version of Speer's
construction, which, due to the crucial role given to the links,
particular subdiagrams containing all the external vertices, is more
suited for the study of IR divergences.

In the original Speer construction IR and UV divergent subdiagrams
participated in the singularity families on the same footing. A
further study on this point should recover this advantage of Speer's
construction which is crucial if one tries to work with high-order
diagrams. As a matter of fact one should compare Speer-Smirnov
singularity families with Hepp-Zimmermann forests and show that the UV
divergent parts correspond to singularity families and hence sectors
containing one-particle-irreducible UV divergent components
consistently with UV power counting. Therefore, being clear that
Speer's sector decomposition is useful in the singularity analysis,
while it does not seem particularly suited for regular diagram
calculations for which other sector decomposition have been developed
together with a suitable software, we believe that our work has been
convincing enough concerning the advantages of the Speer-Smirnov
sector decomposition combined with the Mellin-Barnes formula in the
IR-singularity analysis, also in view of the study of multi-loop
multi-leg amplitude. This point is also stressed by A.V. Smirnov,
V.A. Smirnov and M. Tentyukov in few recent works
\cite{Smirnov:2009pb,*Smirnov:2010hd,*Tentyukov:2010qm}.

\nocite{*}
\addcontentsline{toc}{chapter}{Bibliography}
\bibliographystyle{unsrt}
\bibliography{biblionuova}

\begin{thebibliography}{10}

\bibitem{Feynman:1969}
R.~P. Feynman.
\newblock {Proceedings of the 3rd Topical Conference on High Energy Collision
  of Hadrons}.
\newblock Stony Brook, N. Y. (1969).

\bibitem{Bjorken:1969ja}
J.~D. Bjorken and Emmanuel~A. Paschos.
\newblock {Inelastic Electron Proton and $\gamma$ Proton Scattering, and the
  Structure of the Nucleon}.
\newblock {\em Phys. Rev.}, 185:1975--1982, 1969.

\bibitem{Gross:1973ju}
D.~J. Gross and Frank Wilczek.
\newblock {Asymptotically Free Gauge Theories. 1}.
\newblock {\em Phys. Rev.}, D8:3633--3652, 1973.

\bibitem{Gross:1974cs}
D.~J. Gross and Frank Wilczek.
\newblock {Asymptotically Free Gauge Theories. 2}.
\newblock {\em Phys. Rev.}, D9:980--993, 1974.

\bibitem{unp:1972}
G.~'t~Hooft.
\newblock {Unpublished Talk at the Marseille Conference on Renormalization of
  Yang-Mills Fields and Applications to Particle Physics (June 1972)}.

\bibitem{Politzer:1973fx}
H.~David Politzer.
\newblock {Reliable Perturbative Results for Strong Interactions?}
\newblock {\em Phys. Rev. Lett.}, 30:1346--1349, 1973.

\bibitem{Bjorken:1968dy}
J.~D. Bjorken.
\newblock {Asymptotic Sum Rules at Infinite Momentum}.
\newblock {\em Phys. Rev.}, 179:1547--1553, 1969.

\bibitem{'tHooft:1972fi}
Gerard 't~Hooft and M.~J.~G. Veltman.
\newblock {Regularization and Renormalization of Gauge Fields}.
\newblock {\em Nucl. Phys.}, B44:189--213, 1972.

\bibitem{Bollini:1995pp}
C.~G. Bollini and J.~J. Giambiagi.
\newblock {Dimensional Renormalization: The Number of Dimensions as a
  Regularizing Parameter}.
\newblock {\em Nuovo Cim.}, B12:20--25, 1972.

\bibitem{Breitenlohner:1977hr}
P.~Breitenlohner and D.~Maison.
\newblock Dimensional renormalization and the action principle.
\newblock {\em Commun. Math. Phys.}, 52:11--38, 1977.

\bibitem{Altarelli:1977zs}
Guido Altarelli and G.~Parisi.
\newblock {Asymptotic Freedom in Parton Language}.
\newblock {\em Nucl. Phys.}, B126:298, 1977.

\bibitem{Altarelli:1978id}
Guido Altarelli, R.~Keith Ellis, and G.~Martinelli.
\newblock {Leptoproduction and Drell-Yan Processes Beyond the Leading
  Approximation in Chromodynamics}.
\newblock {\em Nucl. Phys.}, B143:521, 1978.

\bibitem{Curci:1980uw}
G.~Curci, W.~Furmanski, and R.~Petronzio.
\newblock {Evolution of Parton Densities Beyond Leading Order: the Nonsinglet
  Case}.
\newblock {\em Nucl. Phys.}, B175:27, 1980.

\bibitem{Speer:1970ss}
E.~R. Speer and M.~J. Westwater.
\newblock {Generic Feynman Amplitudes}.
\newblock {PRINT-70-1319 (IAS,-PRINCETON), Jun 1970. 79pp}.

\bibitem{Speer:1975dc}
E.~R. Speer.
\newblock {Ultraviolet and Infrared Singularity Structure of Generic Feynman
  Amplitudes}.
\newblock {\em Annales Poincare Phys. Theor.}, 23:1--21, 1975.

\bibitem{Carter:2010hi}
Jonathon Carter and Gudrun Heinrich.
\newblock {SecDec: A General Program for Sector Decomposition}.
\newblock 2010.

\bibitem{Smirnov:2008aw}
A.~V. Smirnov and V.~A. Smirnov.
\newblock {Hepp and Speer Sectors within Modern Strategies of Sector
  Decomposition}.
\newblock {\em JHEP}, 05:004, 2009.

\bibitem{1961AnPhy..13..379Y}
D.~R. {Yennie}, S.~C. {Frautschi}, and H.~{Suura}.
\newblock {The Infrared Divergence Phenomena and High-Energy Processes}.
\newblock {\em Annals of Physics}, 13:379--452, June 1961.

\bibitem{Kinoshita:1962ur}
T.~Kinoshita.
\newblock {Mass singularities of Feynman Amplitudes}.
\newblock {\em J. Math. Phys.}, 3:650--677, 1962.

\bibitem{Lee:1964is}
T.~D. Lee and M.~Nauenberg.
\newblock {Degenerate Systems and Mass Singularities}.
\newblock {\em Phys. Rev.}, 133:B1549--B1562, 1964.

\bibitem{Zimmermann:1975gm}
W.~Zimmermann.
\newblock {The Power Counting Theorem for Feynman Integrals with Massless
  Propagators}.
\newblock In *Erice 1975, Proceedings, Renormalization Theory*, Dordrecht 1976,
  171-184.

\bibitem{Gabrieli:1997rx}
Andrea Gabrieli and Giovanni Ridolfi.
\newblock {The Gluon Contribution to the Polarized Structure Function g2}.
\newblock {\em Phys. Lett.}, B417:369--373, 1998.

\bibitem{Dorn:2008dz}
Harald Dorn and Charlotte~Grosse Wiesmann.
\newblock {Matching Gluon Scattering Amplitudes and Wilson Loops in of-shell
  Regularization}.
\newblock {\em Phys. Lett.}, B668:429--431, 2008.

\bibitem{Carlitz:1988ab}
Robert~D. Carlitz, John~C. Collins, and Alfred~H. Mueller.
\newblock {The Role of the Axial Anomaly in Measuring Spin Dependent Parton
  Distributions}.
\newblock {\em Phys. Lett.}, B214:229, 1988.

\bibitem{librosm}
V.A. Smirnov.
\newblock {Feynman Integral Calculus}.
\newblock Berlin, Germany: Springer (2006) 283 p.

\bibitem{Freitas:2010nx}
Ayres Freitas and Yi-Cheng Huang.
\newblock {On the Numerical Evaluation of Loop Integrals With Mellin- Barnes
  Representations}.
\newblock {\em JHEP}, 04:074, 2010.

\bibitem{Allendes:2009bd}
Pedro Allendes, Natanael Guerrero, Igor Kondrashuk, and Eduardo~A.
  Notte~Cuello.
\newblock {New Four-Dimensional Integrals by Mellin-Barnes Transform}.
\newblock {\em J. Math. Phys.}, 51:052304, 2010.

\bibitem{Smirnov:2009up}
A.~V. Smirnov and V.~A. Smirnov.
\newblock {On the Resolution of Singularities of Multiple Mellin-Barnes
  Integrals}.
\newblock {\em Eur. Phys. J.}, C62:445--449, 2009.

\bibitem{Smirnov:2008tz}
Alexander~V. Smirnov, Vladimir~A. Smirnov, and Matthias Steinhauser.
\newblock {Applying Mellin-Barnes Technique and Groebner Bases to the
  Three-Loop Static Potential}.
\newblock {\em PoS}, RADCOR2007:024, 2007.

\bibitem{Aguilar:2008qj}
Jean-Philippe Aguilar, David Greynat, and Eduardo De~Rafael.
\newblock {Muon Anomaly from Lepton Vacuum Polarization and The Mellin-Barnes
  Representation}.
\newblock {\em Phys. Rev.}, D77:093010, 2008.

\bibitem{Bierenbaum:2007zza}
Isabella Bierenbaum, Johannes Blumlein, and Sebastian Klein.
\newblock {Calculating Massive Two-Loop Two-Point Functions and Operator Matrix
  Elements with Mellin-Barnes Integrals}.
\newblock {\em Nucl. Phys. Proc. Suppl.}, 174:75--78, 2007.

\bibitem{Gluza:2007bd}
J.~Gluza, F.~Haas, K.~Kajda, and T.~Riemann.
\newblock {Automatizing the Application of Mellin-Barnes Representations for
  Feynman Integrals}.
\newblock {\em PoS}, ACAT2007:081, 2007.

\bibitem{Gluza:2007rt}
J.~Gluza, K.~Kajda, and T.~Riemann.
\newblock {AMBRE - A Mathematica Package for the Construction of Mellin-Barnes
  Representations for Feynman Integrals}.
\newblock {\em Comput. Phys. Commun.}, 177:879--893, 2007.

\bibitem{Czakon:2005rk}
M.~Czakon.
\newblock {Automatized Analytic Continuation of Mellin-Barnes Integrals}.
\newblock {\em Comput. Phys. Commun.}, 175:559--571, 2006.

\bibitem{Borisov:2005ka}
Lev~A. Borisov and R.~Paul Horja.
\newblock {Mellin-Barnes Integrals as Fourier-Mukai Transforms}.
\newblock 2005.

\bibitem{Friot:2009fw}
Samuel Friot and David Greynat.
\newblock {Non-Perturbative Asymptotic Improvement of Perturbation Theory and
  Mellin-Barnes Representation}.
\newblock {\em SIGMA}, 6:079, 2010.

\bibitem{Friot:2005gh}
Samuel Friot and David Greynat.
\newblock {Asymptotic Expansions of Feynman Diagrams and the Mellin- Barnes
  Representation}.
\newblock {\em Nucl. Phys. Proc. Suppl.}, 164:199--202, 2007.

\bibitem{Friot:2005cu}
Samuel Friot, David Greynat, and Eduardo De~Rafael.
\newblock {Asymptotics of Feynman Diagrams and the Mellin-Barnes
  Representation}.
\newblock {\em Phys. Lett.}, B628:73--84, 2005.

\bibitem{Suzuki:2003jn}
A.~T. Suzuki, E.~S. Santos, and A.~G.~M. Schmidt.
\newblock {One-Loop N-Point Equivalence among Negative-Dimensional
  Mellin-Barnes and Feynman Parametrization Approaches to Feynman Integrals}.
\newblock {\em J. Phys.}, A36:11859--11872, 2003.

\bibitem{Smirnov:2004ip}
Vladimir~A. Smirnov.
\newblock {Evaluating Multiloop Feynman Integrals by Mellin-Barnes
  Representation}.
\newblock {\em Nucl. Phys. Proc. Suppl.}, 135:252--256, 2004.

\bibitem{Kharchev:2000ug}
S.~Kharchev and D.~Lebedev.
\newblock {Eigenfunctions of $GL(N,R)$ Toda Chain: The Mellin- Barnes
  Representation}.
\newblock {\em Pisma Zh. Eksp. Teor. Fiz.}, 71:338--343, 2000.

\bibitem{Kowalenko:1998qm}
V.~Kowalenko and A.~A. Rawlinson.
\newblock {Mellin-Barnes Regularization, Borel Summation and the Bender
  Asymptotics for the Anharmonic Oscillator}.
\newblock {\em J. Phys.}, A31:L663--L670, 1998.

\bibitem{Passare:1996db}
M.~Passare, A.~K. Tsikh, and A.~A. Cheshel.
\newblock {Multiple Mellin-Barnes Integrals as Periods of Calabi-Yau Manifolds
  with Several Moduli}.
\newblock {\em Theor. Math. Phys.}, 109:1544--1555, 1997.

\bibitem{Elizalde:1994dm}
E.~Elizalde, K.~Kirsten, and S.~Zerbini.
\newblock {Applications of the Mellin-Barnes Integral Representation}.
\newblock {\em J. Phys.}, A28:617--630, 1995.

\bibitem{Bytsenko:1993cf}
A.~A. Bytsenko, E.~Elizalde, S.~D. Odintsov, and S.~Zerbini.
\newblock {Mellin-Barnes Representation for the Genus g Finite Temperature
  String Theory}.
\newblock {\em Phys. Lett.}, B311:87--92, 1993.

\bibitem{Anastasiou:2005cb}
Charalampos Anastasiou and Alejandro Daleo.
\newblock {Numerical Evaluation of Loop Integrals}.
\newblock {\em JHEP}, 10:031, 2006.

\bibitem{Bolzoni:2009ye}
Paolo Bolzoni, Sven-Olaf Moch, Gabor Somogyi, and Zoltan Trocsanyi.
\newblock {Analytic Integration of Real-Virtual Counterterms in NNLO Jet Cross
  Sections II}.
\newblock {\em JHEP}, 08:079, 2009.

\bibitem{Bolzoni:2010sp}
Paolo Bolzoni and Gabor Somogyi.
\newblock {IR Subtraction Schemes: Integrating the Counterterms at NNLO in
  QCD}.
\newblock 2010.

\bibitem{Drummond:2006rz}
J.~M. Drummond, J.~Henn, V.~A. Smirnov, and E.~Sokatchev.
\newblock {Magic Identities for Conformal Four-Point Integrals}.
\newblock {\em JHEP}, 01:064, 2007.

\bibitem{Breitenlohner:1975hg}
P.~Breitenlohner and D.~Maison.
\newblock {Dimensionally Renormalized Green's Functions for Theories with
  Massless Particles. 1}.
\newblock {\em Commun. Math. Phys.}, 52:39, 1977.

\bibitem{mia:tesi}
A.Repetto C.M~Becchi.
\newblock {Studio dell'Andamento Asintotico delle Ampiezze di Feynman}.
\newblock Tesi di Laurea in fisica presso Università di Genova, 2007.

\bibitem{Weinberg:1996kr}
Steven Weinberg.
\newblock {The Quantum Theory of Fields. Vol. 2: Modern Applications}.
\newblock Cambridge, UK: Univ. Pr. (1996) 489 p.

\bibitem{Weinberg:1995mt}
Steven Weinberg.
\newblock {The Quantum theory of fields. Vol. 1: Foundations}.
\newblock Cambridge, UK: Univ. Pr. (1995) 609 p.

\bibitem{Becchi:2006sk}
C.~M. Becchi and G.~Ridolfi.
\newblock {An Introduction to Relativistic Processes and the Standard Model of
  Electroweak Interactions}.
\newblock Milan, Italy: Springer (2006) 139 p.

\bibitem{Itzykson:1980rh}
C.~Itzykson and J.~B. Zuber.
\newblock Quantum field theory.
\newblock New York, Usa: Mcgraw-hill (1980) 705 P.(International Series In Pure
  and Applied Physics).

\bibitem{Ellis:1991qj}
R.~Keith Ellis, W.~James Stirling, and B.~R. Webber.
\newblock {QCD and Collider Physics}.
\newblock {\em Camb. Monogr. Part. Phys. Nucl. Phys. Cosmol.}, 8:1--435, 1996.

\bibitem{Peskin:1995ev}
Michael~E. Peskin and Daniel~V. Schroeder.
\newblock {An Introduction to Quantum Field Theory}.
\newblock Reading, USA: Addison-Wesley (1995) 842 p.

\bibitem{jauch:r}
F.~Rohrlich J.~M.~Jauch.
\newblock {The Theory of Photons and Electrons}.
\newblock {Springer-Verlag, 1980-06}.

\bibitem{Doria:1980ak}
R.~Doria, J.~Frenkel, and J.~C. Taylor.
\newblock {Counter Example to Non-Abelian Bloch-Nordsieck Theorem}.
\newblock {\em Nucl. Phys.}, B168:93, 1980.

\bibitem{Di'Lieto:1980dt}
C.~Di'Lieto, S.~Gendron, I.~G. Halliday, and Christopher~T. Sachrajda.
\newblock {A Counter Example to the Bloch-Nordsieck Theorem in Non-Abelian
  Gauge Theories}.
\newblock {\em Nucl. Phys.}, B183:223, 1981.

\bibitem{Ridolfi:2002}
Giovanni Ridolfi.
\newblock {Spin Structure Functions}.
\newblock Lections given at the Dixième Séminaire Rhodanien de Physique. Le
  Spin en Physique. Torino, 3-8 March 2002.

\bibitem{Altarelli:2002wg}
Guido Altarelli.
\newblock {A QCD Primer}.
\newblock 2002.

\bibitem{Lehmann:1954rq}
H.~Lehmann, K.~Symanzik, and W.~Zimmermann.
\newblock {On the Formulation of Quantized Field Theories}.
\newblock {\em Nuovo Cim.}, 1:205--225, 1955.

\bibitem{Lowenstein:1974qt}
J.~H. Lowenstein and W.~Zimmermann.
\newblock {On the Formulation of Theories with Zero Mass Propagators}.
\newblock {\em Nucl. Phys.}, B86:77, 1975.

\bibitem{Lowenstein:1975rg}
J.~H. Lowenstein and W.~Zimmermann.
\newblock {The Power Counting Theorem for Feynman Integrals with Massless
  Propagators}.
\newblock {\em Commun. Math. Phys.}, 44:73--86, 1975.

\bibitem{Lowenstein:1975ps}
J.~H. Lowenstein.
\newblock {Convergence Theorems for Renormalized Feynman Integrals with
  Zero-Mass Propagators}.
\newblock {\em Commun. Math. Phys.}, 47:53--68, 1976.

\bibitem{Lowenstein:1975rf}
J.~H. Lowenstein and W.~Zimmermann.
\newblock {Infrared Convergence of Feynman Integrals for the Massless a**4
  Model}.
\newblock {\em Commun. Math. Phys.}, 46:105--118, 1976.

\bibitem{Smirnov:2008py}
A.~V. Smirnov and M.~N. Tentyukov.
\newblock {Feynman Integral Evaluation by a Sector Decomposition Approach
  (FIESTA)}.
\newblock {\em Comput. Phys. Commun.}, 180:735--746, 2009.

\bibitem{Lowenstein:1972pr}
J.~H Lowenstein and B.~Schroer.
\newblock {Gauge Invariance and Ward Identities in a Massive Vector Meson
  Model}.
\newblock {\em Phys. Rev.}, D6:1553--1571, 1972.

\bibitem{Becchi:1974md}
C.~Becchi, A.~Rouet, and R.~Stora.
\newblock {Renormalization of the Abelian Higgs-Kibble Model}.
\newblock {\em Commun. Math. Phys.}, 42:127--162, 1975.

\bibitem{Smirnov:2009pb}
A.~V. Smirnov, V.~A. Smirnov, and M.~Tentyukov.
\newblock {FIESTA 2: Parallelizeable Multiloop Numerical Calculations}.
\newblock {\em Comput. Phys. Commun.}, 182:790--803, 2011.

\bibitem{Smirnov:2010hd}
A.~V. Smirnov and M.~Tentyukov.
\newblock {Four Loop Massless Propagators: a Numerical Evaluation of All Master
  Integrals}.
\newblock {\em Nucl. Phys.}, B837:40--49, 2010.

\bibitem{Tentyukov:2010qm}
M.~Tentyukov and A.~V. Smirnov.
\newblock {Applications of FIESTA}.
\newblock 2010.

\bibitem{Bogoliubov:1957gp}
N.~N. Bogoliubov and O.~S. Parasiuk.
\newblock {On the Multiplication of the Causal Function in the Quantum Theory
  of Fields}.
\newblock {\em Acta Math.}, 97:227--266, 1957.

\bibitem{Hepp:1966eg}
Klaus Hepp.
\newblock {Proof of the Bogolyubov-Parasiuk Theorem on Renormalization}.
\newblock {\em Commun. Math. Phys.}, 2:301--326, 1966.

\bibitem{Speer:1977uf}
E.~R. Speer.
\newblock {Mass Singularities of Generic Feynman Amplitudes}.
\newblock {\em Annales Poincare Phys. Theor.}, 26:87--105, 1977.

\bibitem{Smirnov:1990rz}
Vladimir~A. Smirnov.
\newblock {Asymptotic Expansions in Limits of Large Momenta and Masses}.
\newblock {\em Commun. Math. Phys.}, 134:109--137, 1990.

\bibitem{Smirnov:2002pj}
Vladimir~A. Smirnov.
\newblock {Applied Asymptotic Expansions in Momenta and Masses}.
\newblock {\em Springer Tracts Mod. Phys.}, 177:1--262, 2002.

\end{thebibliography}
\newpage 
\pagestyle{plain}
\mbox{}


\end{document}